\documentclass[]{jfm}

\usepackage{amsmath}
\usepackage{graphicx}
\usepackage{epstopdf, epsfig, xcolor}
\usepackage[lofdepth,lotdepth]{subfig}
\usepackage[colorlinks,urlcolor=blue,linkcolor=blue,citecolor=blue]{hyperref}
\usepackage{tikz}
\usepackage{multirow,multicol}
\usepackage{lscape}
\usepackage{natbib}
\usepackage[normalem]{ulem}
\usepackage{lmodern,textcomp}
\usepackage{microtype}
\usepackage[symbol]{footmisc}

\DeclareRobustCommand\Ltransport  {\tikz[baseline=-0.6ex]\draw[thick] (0,0)--(0.9,0);}
\DeclareRobustCommand\Lpressure  {\tikz[baseline=-0.6ex]\draw[thick,dash pattern={on 8pt off 1pt on 3pt off 1pt on 3pt off 1pt}] (0,0)--(0.9,0);}
\DeclareRobustCommand\Lsurface {\tikz[baseline=-0.6ex]
\draw [thick,dash pattern={on 4pt off 1pt on 1pt off 1pt}] (0,0) -- (0.9,0);}
\DeclareRobustCommand\Lviscous {\tikz[baseline=-0.6ex]
\draw [thick,dash pattern={on 2pt off 2pt}] (0,0) -- (0.9,0);}
\DeclareRobustCommand\Lforcing {\tikz[baseline=-0.6ex]
\draw [thick,dash pattern={on 6pt off 2pt}] (0,0) -- (0.9,0);}

\renewcommand{\vec}[1]{\boldsymbol{#1}}
\newcommand{\vect}[1]{\boldsymbol{#1}}

\makeatletter
\newcommand*\bigcdot{\mathpalette\bigcdot@{.5}}
\newcommand*\bigcdot@[2]{\mathbin{\vcenter{\hbox{\scalebox{#2}{$\m@th#1\bullet$}}}}}
\makeatother

\graphicspath{{fig/}}

\shorttitle{Energy budget for multiphase turbulence}

\title{Scale-by-scale kinetic energy budgets in multiphase turbulence}
\author{F. Thiesset\aff{1}\corresp{\email{fabien.thiesset@cnrs.fr}}, J. Vah{\'e}\aff{1}}
\affiliation{
  \aff{1} CNRS, CORIA, UMR 6614, Normandy Univ., UNIROUEN, INSA Rouen,
  76000 Rouen, France}

\begin{document}
\maketitle
\begin{abstract}
  The present work aims at exploring the scale-by-scale kinetic energy exchanges in multiphase turbulence. For this purpose, we derive the K{\'a}rm{\'a}n-Howarth-Monin equation which accounts for the variations of density and viscosity across the two phases together with the effect of surface tension. We consider both conventional and phase conditional averaging operators. This framework is applied to numerical data from detailed simulations of forced homogeneous and isotropic turbulence covering different values for the liquid volume fraction, the liquid/gas density ratio, the Reynolds, and Weber numbers. 
  We confirm the existence of an additional transfer term due to surface tension. Part of the kinetic energy injected at large scales is transferred into kinetic energy at smaller scales by classical non-linear transport while another part is transferred to surface energy before being released back into kinetic energy, but at smaller scales. The overall kinetic energy transfer rate is larger than in single phase flows.  
  Kinetic energy budgets conditioned in a given phase show that the scale-by-scale transport of turbulent kinetic energy due to pressure is a gain (loss) of kinetic energy for the lighter (heavier) phase. Its contribution can be dominant when the gas volume fraction becomes small or when the density ratio increases. 
  Building on previous work, we hypothesize the existence of a pivotal scale above which kinetic energy is stored into surface deformation and below which the kinetic energy is released by interface restoration. Some phenomenological predictions for this scale are discussed.
\end{abstract}
   

% \tableofcontents

% ---------------------------------------------------------
% ---------------------------------------------------------
% ---------------------------------------------------------

% ---------------------------------------------------------
% ---------------------------------------------------------
% ---------------------------------------------------------

\section{Introduction}

Turbulence is a state of fluids characterized by some seemingly erratic motions, although not Brownian, of many sized, yet hierarchized, eddies. It is ubiquitous in our everyday life and recognized as one of the most important open problems of modern physics. The study of single fluid turbulence has concentrated most of the researchers’ efforts over the last century and remains even nowadays a field of intense research. The case of turbulent flows involving a mixture of immiscible fluids separated by an interface have received much less attention. 

Most of available fundamental knowledge on multiphase fluid turbulence applies to the dispersed regime \citep{Balachandar2010}, i.e. to cases where one phase, the carrier phase, is much more abundant than the other, the dispersed phase. When the minority phase represents more than say 5\% of total volume, we talk about multiphase fluid turbulence in the dense regime. Investigating this regime of multiphase flows has been made possible only recently with the aid of high fidelity numerical simulations \citep[see e.g.][among others]{Duret2012,Dodd2016,Lu2018,Lu2019,Rosti2019,CrialesiEsposito2022}. 

Multiphase turbulence distinguishes from its single-phase counterpart by several aspects. First, in most situations, the two phases have different density and different viscosity. Despite fundamental differences in their physical mechanisms, multiphase turbulence shares key features with compressible, reacting, and gas mixture flows, notably spatial and temporal variations in density and viscosity. There is potential for valuable cross-fertilization, particularly through the use of analytical tools that have been independently developed for these distinct classes of flow. Density/viscosity variations are known to influence the whole flow dynamics and in particular the transfer of kinetic energy between the different scales of the turbulent spectrum \citep[see e.g.][]{Taguelmimt2016,Gauding2018,Lai2018,Galtier2011,Aluie2011,Eyink2018,Whitman2019,Sabelnikov2019,Sabelnikov2019a,Kolla2014}. Secondly, the other salient feature of multiphase turbulence is the presence of interfacial forces associated to surface tension. The presence of the interface couples the interface geometry to the flow dynamics, and the interface may either store energy or release energy from/to the rest of the domain \citep[e.g.][]{Dodd2016,TrefftzPosada2023}. The interface is also known to increase internal intermittency \citep{CrialesiEsposito2023}.

As for its single-phase counterpart, multiphase turbulence gives rise to a wide and continuous spectrum of eddies. According to the classical picture of the turbulent cascade of \cite{Richardson1922}, velocity fluctuations are injected at large scales, then transferred between scales of decreasing sizes, down to the small scales where energy is dissipated. The greatest challenge in the fluid turbulence community is thus to describe how the velocity fluctuations vary with eddy size. This requires what is referred to as a scale-by-scale theory that explicitly specifies that notion of scale. This is traditionally done using Fourier transforms \citep{Lance1991,Bunner2002,Dodd2016,Risso2018,Hellinger2021,Hellinger2021a,CrialesiEsposito2022,Innocenti2021,Zamansky2024,Ramirez2024}, structure (or correlation) functions \citep{Galtier2011,Banerjee2013,Lai2018,Gauding2018,Ferrand2020,Hellinger2021,Hellinger2021a,Kritsuk2013,Wagner2012,Bunner2002,Lu2018,Lu2019,Trautner2021}, coarse-grained approaches \citep{Aluie2011,Aluie2013,Eyink2018,Wang2018,Pandey2020,Innocenti2021,Pandey2023,Narula2025}, wavelet decomposition \citep{Freund2019}. Note that the above cited literature is not exhaustive and relates to either multiphase flows or to flows with density/viscosity variations. 

As far as multiphase flows are concerned, the aforementioned analyses of the scale-by-scale kinetic energy budgets have revealed some crucial departures from single-phase flows. One of the most important departures is that surface tension acts as an additional kinetic energy transfer between the different turbulent eddies. This has been shown for both liquid-gas flows \citep{Pandey2020,Pandey2023,Ramirez2024} and liquid-liquid emulsions \citep{Perlekar2019,CrialesiEsposito2022,CrialesiEsposito2023a,Cannon2024}. The interface was found to absorb kinetic energy from the large scales and release it at smaller scales. For turbulent emulsions, \cite{Perlekar2012,CrialesiEsposito2022,CrialesiEsposito2023a,Cannon2024} explain this behaviour under the framework proposed by \cite{Kolmogorov1949} and \cite{Hinze1955} \citep[see also][]{Ni2024}. They suggest that the large turbulent eddies are intense enough to overcome the cohesive force due to surface tension, thereby leading to the fragmentation of the fluid parcels. Through this fragmentation process, the interface area increases, and kinetic energy is transferred into surface energy through the process identified by \cite{Dodd2016}. In this range of scales, the interface thus absorbs kinetic energy and stores it as surface energy. On the contrary, at smaller scales, the fluid parcels remain stable under the effect of surface tension. These parcels can however coalesce, which leads to a decrease of the interface area. In this range of scales, the surface energy is thus released into kinetic energy. As per \cite{CrialesiEsposito2022,CrialesiEsposito2023a,Cannon2024}, the pivotal scale between kinetic energy absorption and kinetic energy release is thus expected to correspond to the pivotal scale between breakup and coalescence which is nothing but the Kolmogorov-Hinze scale. However, there are situations with no break-up nor coalescence, in particular in gravity driven bubbly flows \citep{Pandey2020,Pandey2023}, where the scale-by-scale contribution of the surface tension term exhibits a similar pivotal scale. Hence, the Kolmogorov-Hinze framework and its underlying mechanisms cannot explain the effect of surface tension in general situations. In bubbly flows, \cite{Pandey2020} interpret the role of the surface tension term as a mechanism by which large-scale kinetic energy is absorbed through bubble deformation and stretching. When relaxing, this stored energy is released, but primarily at smaller scales, effectively acting as a scale-conversion process mediated by surface tension. Building on this, we also hypothesize that the breakup/coalescence range could likely be replaced by the more general mechanisms of interface deformation at large scales and interface restoration at smaller scales. This idea originates from \cite{Perlekar2019} who conjectured that the interface may alter the turbulent kinetic energy transfer similarly to some situations of viscoelastic fluid turbulence \citep[see e.g.][]{Valente2014,Nguyen2016} where polymers store kinetic energy when being stretched and release kinetic energy when relaxing, these two processes occurring in different ranges of scales. A similar idea was also conjectured by \cite{Pandey2020}.

% The contribution of the surface tension term in the scale-by-scale energy budget can thus be possibly be explained by a slightly more general scenario:  interface deformation/kinetic energy pumping at large scales, and an interface restoration/kinetic energy release at smaller scales.

Many studies on multiphase turbulence focus on the case of equal density between the two phases \citep[e.g.][among others]{Mukherjee2019,CrialesiEsposito2022,Cannon2024}. However, the presence of density difference leads to an additional contribution in the scale-by-scale kinetic energy budget term due to pressure \citep{Pandey2020,Pandey2023,Narula2025} even when the flow is homogeneous, which is generally referred to as the baropycnal work \citep{Aluie2013}. Contrasts of density may also change the whole flow dynamics as the phase with higher density carries the flow momentum. The role played by density ratio remains yet only partly understood, and more dedicated studies are required.

Here, the general objective is to explore the scale-by-scale kinetic energy exchanges in multiphase turbulence by use of a detailed analysis of the structure function transport equation (known as the K{\'a}rm{\'a}n-Howarth-Monin equation, abbreviated KHM). Only very few studies of this type exist \citep{Pan2022,Pan2025,Narula2025} which motivates the present work. In particular, we aim at using this framework to unravel the effect of density and/or viscosity variations, together with surface tension on the kinetic energy scale distribution. Another objective of the present study is to infer whether the turbulence properties taken in one given phase differ from classical single-phase constant-density, constant-viscosity turbulence. This requires conditioning the structure functions by the phase, in the lines of \cite{Lu2018,Lu2019,Trautner2021} and derive their associated budget equation in a somewhat similar fashion as \cite{Freund2019} who used the wavelet approach. In other words, we aim at extending the conditionally averaged kinetic energy budgets derived by \cite{Dodd2016,TrefftzPosada2023} to the two-point statistics. We will investigate the peculiar case of homogeneous and isotropic turbulence using numerical simulations data obtained by a standard front capturing code named \texttt{archer}. 

The paper is organized as follows. First, the general KHM equation applying to multiphase flows is derived in section \S \ref{sec:KHM}. We will consider both unconditional and conditional averaged equations. All terms of the KHM equation are also analysed in the asymptotic limit of large separations. The numerical setup is detailed in section \S \ref{sec:DNS}. Results are presented in section \S \ref{sec:results} and conclusions are drawn in a last section \S \ref{sec:conclusions}. Technical aspects regarding the derivation of the KHM equation and the numerical simulations are gathered in appendices \ref{app:derivation} and \ref{app:dns}, respectively.

\section{K{\'a}rm{\'a}n-Howarth-Monin equation for multiphase turbulence} \label{sec:KHM}

\subsection{Derivation}

We seek for a generalized K{\'a}rm{\'a}n-Howarth-Monin (KHM) equation (sometimes also referred to as the generalized Kolmogorov equation) that applies to two-phase flows. The latter should account for the variations of density and viscosity together with surface tension. As per \cite{Galtier2011,Lai2018,Hellinger2021a} and reference therein, one possible definition of the scale-by-scale kinetic energy in variable-density flows is: 
\begin{eqnarray}
  |\delta \vect{u}|_\rho^2 := (\delta (\rho \vect{u})) \cdot (\delta \vect{u}), \label{eq:dq2}
\end{eqnarray}
where $(\delta (\rho \vect{u})) := (\rho \vect{u})^+ - (\rho \vect{u})^-$ and $(\delta \vect{u}) :=  \vect{u}^+ -  \vect{u}^-$ are the increment of $\rho \vect{u}$ and $\vect{u}$, respectively, between two points $\vect{x}^+$ and $\vect{x}^-$ arbitrarily separated in space by a distance $\vect{r} := \vect{x}^+ - \vect{x}^-$. The superscript $+(-)$ is used to denote that quantities are taken at point $\vect{x^+}(\vect{x^-})$. It is also convenient to define $(\overline{\delta} \bullet ) := (\bullet^+ + \bullet^-)/2$, i.e. the arithmetic mean of $\bullet$ between point $\vect{x^+}$ and $\vect{x^-}$. Note that some alternative definitions for the scale-by-scale kinetic energy in variable-density flows can be found in the literature \citep{Ferrand2020,Brahami2020}. This is discussed in more details in Appendix \ref{app:dq2_def} and in subsection \S\ref{sec:discussion}.

The transport equation for $|\delta \vect{u}|_\rho^2$ is obtained directly from the one-fluid formulation for the two-phase incompressible Navier--Stokes (NS) equation. Each term can be derived from the following general expression:
\begin{eqnarray}
  (\delta (\rho \vect{u})) \cdot (\delta \vect{T}_{u}) + (\delta \vect{u}) \cdot (\delta \vect{T}_{\rho u}), \label{eq:KHM_general}
\end{eqnarray}
where $\vect{T}_u$ and $\vect{T}_{\rho u}$ denote the transport equation for $\vect{u}$ and $\rho \vect{u}$, respectively. The one-fluid formulation of the two-phase incompressible NS-equation writes:
% \begin{subequations}
\begin{eqnarray}
  \vect{T}_{\rho u} :=~& \partial_t \rho \vect{u} &= - \vect{\nabla} \cdot \rho \vect{u} \vect{u} - \vect{\nabla} P + \vect{\nabla} \cdot \mathsfbi{t} + \vect{S} + \vect{F}. \label{eq:nsa} 
\end{eqnarray}
The incompressibility condition further implies $\vect{\nabla}\cdot \vect{u} =0$. In Eq. \eqref{eq:nsa}, the mechanical pressure is denoted $P$, $\mathsfbi{t} := \mu 2\mathsfbi{S}$ is the viscous stress tensor with $\mathsfbi{S} := (\vect{\nabla} \vect{u} + \vect{\nabla}\vect{u}^T)/2$ the strain rate tensor, $\mu$ the dynamic viscosity. The term noted $\vect{F}$ corresponds to a generic forcing, while $\vect{S} := 2\sigma H \delta_\Gamma(\vect{x}-\vect{x}_\Gamma) \vect{n}_\Gamma$ is the surface tension term, where $\sigma$ is the surface tension coefficient, $H$ is the mean curvature of the interface, $\vect{n}_\Gamma$ is the normal to the interface. The surface tension term acts only at the interface, hence the presence of the Dirac $\delta_\Gamma(\vect{x}-\vect{x}_\Gamma)$ function allowing the interface position $\vect{x}_\Gamma$ to be tracked. Using the continuity equation, one then obtains the transport equation for $\vect{u}$:
\begin{eqnarray}
    \vect{T}_u := ~& \partial_t \vect{u} &= - (\vect{u} \cdot \vect{\nabla}) \vect{u} - v \vect{\nabla} P + v \vect{\nabla} \cdot \mathsfbi{t} + v \vect{S} + v \vect{F}, \label{eq:nsb}
\end{eqnarray}
where $v := 1/\rho$ is the specific volume.

The quantity $|\delta \vect{u}|_\rho^2$ is a fluctuating field variable that {\it a priori} depends on time $t$, the position $\vect{X}:=(\vect{x}^+ + \vect{x}^-)/2$ of the midpoint between $\vect{x^+}$ and $\vect{x^-}$ and the separation vector $\vect{r} = \vect{x}^+ - \vect{x}^-$ (See Appendix \ref{app:derivation}). The scale-by-scale kinetic energy $|\delta \vect{u}|_\rho^2$ is better assessed in a statistical sense, taking advantage of the statistical symmetry of the flow (if any). Here, we first employ an ensemble average over many statistically independent samples of the flow. These averages will be noted $\langle \bullet \rangle_\mathbb{E}$, where $\bullet$ denotes any quantity (in practice, the flow being at statistically stationary state, time averages are computed in place of ensemble average, supposing ergodicity). This ensemble average is then supplemented by a spatial average over a set of points $\mathbb{C}$: 
\begin{eqnarray}
  \langle \bullet \rangle_\mathbb{C} := \frac{\int_{\mathbb{C}} \langle \bullet \rangle_\mathbb{E}~ {\rm d}\vect{V}}{\int_{\mathbb{C}} {\rm d}\vect{V}}.
\end{eqnarray} 
We will consider four different sets of averaging points $\mathbb{C}$ as detailed below:
\begin{enumerate}
  \vspace{4pt}
  \item The first average covers the ensemble $\mathbb{T}$ of all points of the simulation domain. The symbol $\mathbb{T}$ stands for “total”. This is for instance the classical volume average in Direct Numerical Simulation of the NS equations in a periodic box. 
  \vspace{4pt}
  \item We also consider the subset $\mathbb{L}$ of the simulation domain which covers the points $\vect{X}$ such as both $\vect{x}^+$ and $\vect{x}^-$ lie in the liquid phase, i.e. $\mathbb{L} := \{\vect{X}:(\vect{x}^+ \in {\rm liquid})\land (\vect{x}^- \in {\rm liquid} )\}$. The symbol $\mathbb{L}$ stands for “liquid”. 
  \vspace{4pt} 
  \item The complementary subset for the gas phase is further considered, i.e. $\mathbb{G} := \{\vect{X}:(\vect{x}^+ \in {\rm gas})\land (\vect{x}^- \in {\rm gas} )\}$. The symbol $\mathbb{G}$ stands for “gas”. 
  \vspace{4pt}
  \item We finally define conditional averages over the set of points $\mathbb{M}$ such as $\vect{x}^+$ and $\vect{x}^-$ lie in different phases, i.e. $\mathbb{M} := \{\vect{X}:(\vect{x}^+ \in {\rm liquid})\land (\vect{x}^- \in {\rm gas} )\}$ (and {\it vice versa}). The symbol $\mathbb{M}$ stands for “mixed”.
  \vspace{4pt}
\end{enumerate}

More details on conditional averages are given in Appendix \ref{app:derivation}, see Eqs. \eqref{eq:conditional_average}. Note that for conditional averages the averaging volume depends on $\vect{r}$. Its boundary, noted $\partial \mathbb{C}$, does not necessarily coincide with the liquid-gas interface. 

Once spatially averaged, two-point statistics depend on time $t$ and the separation vector $\vect{r}$. In case of isotropic flows, two-point statistics are only function of the modulus of the vector $\vect{r}$, thereby reducing the problem complexity to 2-dimensions ($|\vect{r}|,t$). When the flow is anisotropic, one could keep the dependence to $\vect{r}$ (a 3-dimensional vector), thus paying the price of a 4-dimensional problem. The other solution to cope with anisotropy is to apply an angular average over all orientations of the separation vector $\vect{r}$ \citep{Nie1999, Hill2002}. 
\begin{eqnarray}
  \langle \bullet \rangle_{\mathbb{C},\Omega} := \frac{1}{4\pi} \iint_\Omega \langle \bullet \rangle_\mathbb{C} ~\sin \theta {\rm d}\theta {\rm d} \varphi , \label{eq:angle_av} \,
\end{eqnarray}
where the set of solid angles is given by $\Omega = \lbrace \varphi , \theta ~|~ 0 \leq \varphi \leq \pi, 0 \leq \theta \leq 2 \pi \rbrace$ with $\varphi = \arctan(r_y/r_x)$ and $\theta = \arccos(r_z/|\vect{r}|)$ ($r_x$, $r_y$, and $r_z$ are the components of the $\vect{r}$ vector in $x$, $y$, and $z$ directions, respectively). Since angularly averaged two-point statistics depend on $|\vect{r}|$ only (and not its orientation anymore), the effect of anisotropy is concealed. However, it has the advantage of lightening the analysis since $\langle \bullet \rangle_{\mathbb{C},\Omega}$ depends only on time $t$ and $r = |\vect{r}|$, i.e. a 2-dimensional problem, as for isotropic flows. In what follows, an angular average will be applied to the two-point statistics, although for the sake of clarity, the subscript $\Omega$ will be dropped from the notations. Note that the flow under consideration here is isotropic, and hence angular average are only used to increase the statistical convergence \citep{Taylor2003,Thiesset2020,Gauding2022,Federrath2021}. 

The transport equation for $\langle |\delta \vect{u}|_\rho^2 \rangle_\mathbb{C}$ is referred to as the KHM equation. Its detailed derivation is given in Appendix \ref{app:derivation}, and we summarize here the main results. The general KHM equation can be written formally as 
\begin{equation}
  d_t \langle |\delta \vect{u}|_\rho^2 \rangle_\mathbb{C} = \langle \mathcal{T} \rangle_\mathbb{C} + \langle \mathcal{P} \rangle_\mathbb{C}+ \langle \mathcal{V} \rangle_\mathbb{C} + \langle \mathcal{S} \rangle_\mathbb{C} + \langle \mathcal{F} \rangle_\mathbb{C} \label{eq:KHM_symbolic}
\end{equation}
This equation is one of the key result of the present paper. It is a generalization of the KHM equation to account for variable density and the presence of surface tension. In this formulation, it applies irrespectively of the averaging volume $\mathbb{C} \in \{\mathbb{T}, \mathbb{L}, \mathbb{G}, \mathbb{M}\}$ which is another strength of the present work. The newly derived KHM equation contains different terms which are now detailed.
\begin{itemize}
  
  \vspace{4pt}
  \item $d_t \langle |\delta \vect{u}|_\rho^2 \rangle_\mathbb{C}$ denotes the time variations of $\langle |\delta \vect{u}|_\rho^2 \rangle_\mathbb{C}$. In statistically stationary flows, this term is zero.
  
  \vspace{4pt}
  \item $\langle \mathcal{T} \rangle_\mathbb{C}$ represents the “transport” of $\langle |\delta \vect{u}|_\rho^2 \rangle_\mathbb{C}$. In incompressible flows, the latter can be decomposed into three contributions:
  \begin{align}
    \langle \mathcal{T} \rangle_\mathbb{C} = - \langle \vect{\nabla_X} \cdot (\overline{\delta} \vect{u}) |\delta \vect{u}|^2_\rho \rangle_\mathbb{C} - \langle \vect{\nabla_r} \cdot (\delta \vect{u}) |\delta \vect{u}|^2_\rho \rangle_\mathbb{C} + \int_{\partial \mathbb{C}} |\delta \vect{u}|^2_\rho ~\vect{u}_b \cdot \vect{n} \textrm{d}S. \label{eq:transport} 
  \end{align}   
  The leftmost term on RHS of Eq. \eqref{eq:transport} is the averaged transfer of $|\delta \vect{u}|_\rho^2$ in flow position space $\vect{X}$. Using the divergence theorem, it can be rewritten as 
  \begin{align}
    - \langle \vect{\nabla_X} \cdot (\overline{\delta} \vect{u}) |\delta \vect{u}|^2_\rho \rangle_\mathbb{C} = - \int_{\partial \mathbb{C}} |\delta \vect{u}|^2_\rho ~ (\overline{\delta} \vect{u})\cdot \vect{n}  \textrm{d}S.
  \end{align}   
  This term can thus be interpreted as a flux of the quantity $|\delta \vect{u}|^2_\rho$ which passes through the averaging volume boundary $\partial \mathbb{C}$ with a velocity $\overline{\delta} \vect{u}$, with $\vect{n}$ being the outwardly pointing normal vector to $\partial \mathbb{C}$. If the averaging volume is a periodic box ($\mathbb{C}=\mathbb{T}$ is periodic), then this term is zero since all the fluxes coming in and out from the domain cancels out \citep{Hill2002}. In contrast, when $\mathbb{C} \in \{\mathbb{L},\mathbb{G},\mathbb{M}\}$, the transfer in $\vect{X}$-space cannot be dropped out. 

  The second term on RHS of Eq. \eqref{eq:transport} represents a flux of $|\delta \vect{u}|_\rho^2$ in scale-space $\vect{r}$, thus corresponding to the transfer of kinetic energy among the different scales of the flow. 
  
  The rightmost term on RHS of Eq. \eqref{eq:transport} arises when applying the Reynolds transport theorem to the volume-averaged time-derivative of $|\delta \vect{u}|^2_\rho$, viz.
  \begin{align}
    \langle \partial_t |\delta \vect{u}|^2_\rho \rangle_\mathbb{C} = d_t \langle |\delta \vect{u}|_\rho^2 \rangle_\mathbb{C} - \int_{\partial \mathbb{C}} |\delta \vect{u}|^2_\rho ~\vect{u}_b \cdot \vect{n} \textrm{d}S.
  \end{align} 
  Here, $\vect{u}_b$ is the velocity of the control volume boundary $\partial \mathbb{C}$. When the averaging volume is fixed, for instance when $\mathbb{C}=\mathbb{T}$, one has $\vect{u}_b \equiv 0$ and this term vanishes. Otherwise, it represents the averaged flux of $|\delta \vect{u}|^2_\rho$ which flows through the control volume boundary at a velocity $\vect{u}_b$.

  \vspace{4pt}
  \item $\langle \mathcal{P} \rangle_\mathbb{C}$ represents the effect of pressure on the evolution of the scale-by-scale kinetic energy. In incompressible flows, it can be decomposed into two contributions:
  \begin{align} 
    \langle \mathcal{P} \rangle_\mathbb{C} = - 2 \langle \vect{\nabla_X} \cdot (\delta \vect{u}) (\delta P)\rangle_\mathbb{C} - \langle \mathcal{C}(-\vect{\nabla}P)\rangle_\mathbb{C}  \label{eq:pressure},
  \end{align}  
  The leftmost term on RHS of Eq. \eqref{eq:pressure} is the scale-by-scale transport of turbulent kinetic energy due to pressure, hereafter simply referred to as pressure transport. It can be rewritten using the divergence theorem
  \begin{align} 
    - 2 \langle \vect{\nabla_X} \cdot (\delta \vect{u}) (\delta P)\rangle_\mathbb{C} = - \int_{\partial \mathbb{C}} (\delta P) (\delta \vect{u}) \cdot \vect{n} {\rm d}S
  \end{align}   
  This term accounts for the pressure transport of scale-by-scale kinetic energy through the averaging volume boundary $\partial \mathbb{C}$. Hence, if $\mathbb{C} = \mathbb{T}$ is periodic, this term is zero and the contribution due to pressure reduces to $\langle \mathcal{P} \rangle_\mathbb{T} = -\langle \mathcal{C}(-\vect{\nabla}P) \rangle_\mathbb{T}$ (the meaning of term $\mathcal{C}$ will be described very shortly). It is also sometimes referred to as the scale-by-scale pressure diffusion as it acts in redistributing the scale-by-scale turbulent kinetic energy across position $\vect{X}$ and scales $\vect{r}$. 

  The second contribution of pressure is due to the term $- \langle \mathcal{C}(-\vect{\nabla}P)\rangle_\mathbb{C}$ which arises due to variations of the density (see Eq. \eqref{eq:Cab} in Appendix \ref{app:derivation}). It is likely to be analogous to what \cite{Aluie2013} identifies to the baropycnal work. By coarse-graining the compressible NS-equation, Aluie demonstrated that a term similar to $- \mathcal{C}(-\vect{\nabla}P)$ should operate even in incompressible flows, provided there are density variations. The analyses by \cite{Pandey2020,Pandey2023,Narula2025} further support the presence of such a contribution in the energy budget of two-phase incompressible flows. The presence of this term is discussed in more details in subsection \S\ref{sec:discussion}.
  
  Note that since the density is constant in either the liquid or gas phase, the term $- \langle \mathcal{C}(-\vect{\nabla}P)\rangle_\mathbb{C}$ should be accounted for only if $\mathbb{C} \in \{\mathbb{T},\mathbb{M}\}$ and if the density is different in the liquid and gas phase. In summary, when $\mathbb{C} \equiv \mathbb{T}$, only the term $- \langle \mathcal{C}(-\vect{\nabla}P)\rangle_\mathbb{C}$ contributes to the budget since the pressure transport is zero by homogeneity. In contrast, when $\mathbb{C} \in \{\mathbb{L},\mathbb{G}\}$, we have $- \langle \mathcal{C}(-\vect{\nabla}P)\rangle_\mathbb{C}=0$ because the density is constant in each phase and only the pressure transport term contributes. 
    
  \vspace{4pt}
  \item The terms $\langle \mathcal{V}\rangle_\mathbb{C}$, $\langle \mathcal{S}\rangle_\mathbb{C}$ and $\langle \mathcal{F}\rangle_\mathbb{C}$ represent the scale-by-scale contribution of viscous diffusion, surface tension and forcing, respectively. Their general compact expression is given in Eq. \eqref{eq:VST} of Appendix \ref{app:derivation}. The expanded version of the viscous term has been derived by \cite{Lai2018}. As with the pressure term, each of these terms can be decomposed into a contribution that looks similar to its constant-density counterpart, to which is added the correction, noted $\mathcal{C}$, which accounts for variations of density within the averaging volume. For the same reason as above, these correction terms vanish when $\mathbb{C} \in \{\mathbb{L},\mathbb{G}\}$ because density is constant per phase. Note also that the term due to surface tension $\langle \mathcal{S}\rangle_\mathbb{C}$ contributes to the budget only if $\mathbb{C} = \mathbb{T}$.
  
  {It is worth stressing that the viscous term $\langle \mathcal{V}\rangle_\mathbb{C}$ can be decomposed into different contributions: a transport term in scale space $r$, a transport term in position space $\vect{X}$ and a contribution due to the kinetic energy dissipation rate \citep[see for instance][]{Galtier2011,Lai2018,Hellinger2021a}. Such a decomposition will not be carried out here, although it may provide many important insights into the role of viscosity in the scale-by-scale kinetic energy budget. This aspect could be addressed in future work.}
\end{itemize}

In appendix \ref{app:derivation}, we also address the asymptotic behaviour of all terms in the KHM equation when the separation $r$ tends to infinity. Under the condition of statistical homogeneity, it is proved that all terms of the KHM equation tend towards (four times) their counterpart in the one-point scale-integrated kinetic energy budgets.
For the flow setup studied here, they write:
\begin{subequations}
  \begin{eqnarray}
    d_t \langle k \rangle_\mathbb{T}=\langle F \rangle_\mathbb{T}  -  \langle \epsilon \rangle_\mathbb{T} + \langle S \rangle_\mathbb{T}, \\
    d_t \langle k \rangle_\mathbb{L} = \langle  F\rangle_\mathbb{L} - \langle \epsilon \rangle_\mathbb{L} + \langle T_\nu \rangle_\mathbb{L} + \langle T_p \rangle_\mathbb{L},\\
    d_t \langle k \rangle_\mathbb{G} = \langle  F\rangle_\mathbb{G} - \langle \epsilon \rangle_\mathbb{G} + \langle T_\nu \rangle_\mathbb{G} + \langle T_p \rangle_\mathbb{G},
  \end{eqnarray} \label{eq:1pt_budget}
\end{subequations}
The turbulent kinetic energy is given by $k=\frac{1}{2}\rho|\vect{u}|^2$. The terms $F$, $\epsilon$, $S$ represent the contribution due to forcing, kinetic energy dissipation and surface tension, respectively. The terms $T_\nu$ and $T_p$ correspond to the transport of kinetic energy due to viscous diffusion and pressure. In absence of forcing $\vect{F} \equiv 0$, one recovers from Eqs. \eqref{eq:1pt_budget} the equations firstly derived by \cite{Dodd2016}. The recent work by \cite{TrefftzPosada2023} extended these equations to homogeneous shear two-phase flows. The explicit expression for $T_\nu$ and $T_p$ is given by \cite{Dodd2016} and in Appendix \ref{app:derivation}. It is important to notice that for the total fluctuating field, the pressure term is absent from the one-point kinetic energy budget Eq. \eqref{eq:1pt_budget}(a) although the baropycnal work is {\it a priori} non-zero in the KHM equation. This means that this term is likely to represent a transfer or conversion of energy between scales although its net contribution when integrated over all scales is zero. This was already anticipated by \cite{Aluie2013} for the coarse-grained kinetic energy budget in compressible or variable-density flows.

Compared to the analysis of \cite{Lai2018} or \cite{Yao2023}, we consider here the KHM equation for the total velocity field without carrying any decomposition between its (Reynolds or Favre) mean and its (Reynolds or Favre) fluctuation. The reason is that {\it (i)} we want these equations to remain as compact as possible and {\it (ii)} the flow we will consider hereafter has zero mean. In flows with strong inhomogeneities, it may however be relevant to incorporate such a decomposition in order to reveal some mechanisms such as the transport and production of turbulent fluctuations by the mean flow. 

The KHM equation for variable density multiphase flows Eq. \eqref{eq:KHM_symbolic} provides an explicit representation of the effects of unsteadiness, inertia, pressure, surface tension or viscous effect on the evolution of turbulent kinetic energy at different scales of the turbulent spectrum. Similar to the wavelet approach presented by \cite{Freund2019} or the coarse-grained methodology followed by \cite{Pandey2020}, this framework allows the evolution of turbulent kinetic energy to be represented in the compound position($\vect{X}$)/scale($\vect{r}$) space. Hence, it can easily be conditionally averaged in the same way as \cite{Freund2019}, i.e. by considering 3 different situations where the two points lie within either the liquid or within the gas phase, or where the two points separation crosses an interface. 

Note that here the evolution for the total kinetic energy is recovered by applying the limit to infinite separations while for the wavelet approach or with the spectral analysis, the latter is obtained by integrating over all scales. Hence, the structure functions should rather be interpreted as a cumulative distribution of energy rather than an energy density distribution. Some authors proposed some definitions for the energy density using scale derivatives of the structure functions \citep{Davidson2005,Danaila2012a,Hamba2015,Hamba2018,Arun2021}. This could be addressed in a follow-up study.

\subsection{Discussions} \label{sec:discussion}

At this stage of the theoretical developments, it seems important to draw attention to a few points requiring careful consideration.

First-of-all, in several studies, the analysis of scale-by-scale kinetic energy budget is performed in spectral space \citep{CrialesiEsposito2022,CrialesiEsposito2023a,Ramirez2024,Cannon2024}. It is worth stressing that the use of Fourier transforms may raise some issues. Firstly, spectra can be computed only over homogeneous directions with periodic boundary conditions. This is a very stringent restriction that makes spectra inapplicable to all flow situations, in particular in statistically inhomogeneous flows. In addition, as discussed by notably \cite{Lucci2010,Duret2012,Ramirez2024}, two-phase flows reveal local discontinuities associated with the presence of the interface. This can result in oscillations in the kinetic energy spectra thereby corrupting the physical interpretation of the associated budget. Note though that \citep{Ramirez2024} succeeded in distinguishing the spectral contribution due to discontinuities from the one due to the bulk phases allowing them to be interpreted separately. Another issue arises due to the spatial non-locality of Fourier transforms that precludes the definition of phase-conditioned spectral quantities. \cite{Ramirez2024} showed however that spectral approaches might remain valuable to explore the dynamics of the two-phase mixture, i.e. the case $\mathbb{C}=\mathbb{T}$, once the spectral signatures of discontinuities properly interpreted. If one aims at inferring the influence of the interface on the dynamics of each phase separately, analysis in physical space using either coarse-graining approaches, wavelet decompositions or point-splitting methods based on correlation or structure functions should be preferred.

Secondly, the derivation of the present form of the KHM equation requires the transport equation for both velocity $\vect{u}$ and momentum $\rho\vect{u}$. The latter is obtained rigorously from the two-fluid formulation of the two-phase incompressible Navier-Stokes equation and is valid in the sense of generalized function. On the other hand, the equation for $\vect{u}$ derives from the additional application of the incompressibility condition. As stressed by \cite{Ramirez2024}, this equation is mathematically ill-defined since one has to divide the surface tension term, in which appears a Dirac delta function, by the density, which is a Heaviside function. The KHM equation derived here as many other formulations of scale-by-scale energy budgets \citep[for instance][]{Pandey2020,Narula2025,Freund2019} thus inherit this limitation. However, in most numerical solvers, whether they use front-capturing or diffuse-interface techniques, the discontinuities are regularized: in front-capturing methods at the mesh scale, and in diffuse-interface methods across the prescribed interface thickness. Hence, the one-fluid formulation of the NS equation is solved in a regularized sense. The KHM equation derived here as several other scale-by-scale formulations is thus valid in the same regularized sense. This limitation should be kept in mind when interpreting the results obtained from the present theoretical framework.

The third cautionary note concerns the somewhat arbitrary definition of the scale-by-scale kinetic energy $|\delta \vect{u}|_\rho^2$. As stressed by \cite{Aluie2013, Narula2025}, in the coarse-grained framework, there is no unique way to define the scale-by-scale kinetic energy in flows with significant fluctuations of density. The same applies to the point-splitting methods used here (see Appendix \ref{app:dq2_def}). The choice made here is motivated by the fact that, at least, it recovers the correct one-point scale-integrated kinetic energy when taking the limit to infinite separations. Other definitions filling this condition exists (see Appendix \ref{app:dq2_def}). In absence of consensus on the most appropriate definition, it is perhaps wise to consider that the present definition is just one possible choice among others. At this stage, some comparative analyses using different definitions of the scale-by-scale kinetic energy should be carried out since this arbitrary choice may have consequences on the physical interpretation of the results.

Indeed, the definition proposed here yields the presence of a non-zero pressure term in the budget for the liquid-gas mixture ($\mathbb{C} \equiv \mathbb{T}$). In the lines of \cite{Aluie2011,Aluie2013}, it is referred to as the baropycnal work and interpreted as being responsible for the transfer of kinetic energy among scales. It cancels out when integrating over all scales, and thus disappears from the one-point kinetic energy budget. This phenomenon must not be mistaken for the conversion of kinetic to internal energy by pressure--dilatation, which is a feature unique to compressible flows. As detailed in Appendix \ref{app:dq2_def}, regardless of the definition, when a sort of density weighted (i.e. Favre) averaging is used to define the scale-by-scale kinetic energy, a consistent term emerges, correlating pressure gradient, velocity and density variations, underscoring the baropycnal work's role. In the coarse-grained framework, \cite{Narula2025} showed that this term may however contribute differently to the budget depending on the precise definition of the scale-by-scale kinetic energy. In the spectral energy budget derived by \cite{Ramirez2024}, where the power of each physical process acting on the kinetic energy is expressed as the product of forces by velocity, there is no term similar to the baropycnal work. Therefore, one cannot rule out that the baropycnal term in the KHM equation arises as an artefact of the chosen definition of scale-by-scale kinetic energy, rather than reflecting an unambiguous physical process. This fundamental question calls for additional research.

\section{Numerical simulations of multiphase turbulence} \label{sec:DNS}

\subsection{Numerical methods}

The framework described above is used to analyse data from numerical simulation of turbulent multiphase flows issued from the code \texttt{archer} \citep{Menard2007,Vaudor2017}. This code solves the one-fluid formulation of the NS equation Eq. \eqref{eq:nsa} on a Cartesian staggered grid. The convective term in Eq. \eqref{eq:nsa} is written in conservative form and solved using the improved \cite{Rudman1998} technique \cite[see][for more details]{Vaudor2017}. The latter allows mass and momentum to be transported consistently enabling better accuracy and stability for flows with large liquid/gas density ratios. The viscous term in Eq. \eqref{eq:nsa} is computed following the method presented by \cite{Sussman2007}. The surface tension term is resolved using the Ghost-Fluid method \citep{Fedkiw1999}.

The interface is transported using a coupled level-set and volume-of-fluid (CLSVOF) solver, in which the level-set function accurately describes the geometric features of the interface (its normal and curvature) and the volume-of-fluid (VOF) function ensures mass conservation. For more information about the \texttt{archer} solver, the reader can refer to e.g. \cite{Menard2007, Duret2012, Vaudor2017}.

For time advancement, we use a slightly modified version of the fast three steps Runge-Kutta algorithm (abbreviated fastRK3) proposed by \cite{Aithal2023}. More details on this algorithm are given in Appendix \ref{app:fastRK3}.

\subsection{Numerical configuration}

% Some authors \citep[see e.g][]{Elghobashi2019,TrefftzPosada2023,Desjardin2024} argue against artificially forced turbulence since it may create unphysical interactions between the two-phases thereby precluding drawing any firm conclusions about the features of turbulence by the presence of droplets and/or bubbles. \cite{Elghobashi2019,TrefftzPosada2023} rather advocate for studying decaying turbulence or homogeneous shear turbulence while \cite{Desjardin2024} prefer using a forcing that acts at large scales only, allowing the intermediate and small-scales to evolve naturally. 

% With no will to contest these statements, we want to stress however that recent work on single-phase turbulence reveal that decaying, artificially forced, and shear turbulence are very different classes of turbulent flows \citep[see e.g.][]{Danaila2002,Antonia2006,Thiesset2014}. This is even more perceptible at low Reynolds number where the dynamics of large, intermediate and small scales strongly overlaps. At finite Reynolds number, these different flows are just not comparable and their approach towards a universal behavior at infinite Reynolds number is still an open question. The same is very likely to apply in multiphase turbulence. Therefore, our opinion is that it is perhaps rather hasty to put artificially forced turbulence to the trash. This configuration is worth being analyzed.

The theoretical framework described in \S \ref{sec:KHM} is applied to numerical data of homogeneous isotropic turbulence that is maintained at steady state by adding a forcing term to the NS equation. This configuration has now become a standard for studying multiphase turbulence \citep[see e.g][among others]{Duret2012,McCaslin2014,Loisy2017,Mukherjee2019,Thiesset2020,Boukharfane2021,Riviere2021,Cannon2024}. Here, we use the large-scale stochastic forcing of \cite{Eswaran1988} \citep[see also][]{Cannon2024}. The numerical implementation of this forcing procedure is described in Appendix \ref{app:forcing}.

The initial condition consists in one spherical droplet placed at the centre of the domain, and 6 half-cylinders of same diameter centred on each face of the periodic box. {This configuration is set from the very beginning of the simulation and not released at a given time.} The initial velocity field corresponds to an ABC flow with $u'=1$ {that evolves to finally reach a steady state thanks to the forcing term}. We have checked that the initial condition has no effect on the final statistical properties. This initial condition was chosen as it rapidly achieves the statistically steady state, {hence saving some computational time}. For each simulation, after a transient period of $4s \approx 8T_L$ where $T_L$ is the eddy turn-over time, a steady state is reached. The simulations are then run for 16 additional seconds (physical time), corresponding to roughly $32T_L$. Statistics are gathered throughout the statistically steady period. Statistical convergence is estimated using the 95\% confidence level (assuming Gaussian distributions), and is found to be within few percent. In most figures below, the statistical uncertainty will be displayed as shaded region surrounding the averaged value. 

The relevant non-dimensional parameters for our study are
\begin{subequations}
\begin{eqnarray}
  {\rm the~liquid~volume~fraction~} &\alpha &:= \langle \phi_L \rangle_\mathbb{T}, \\
  {\rm the~density~ratio~} &R_\rho &:= \frac{\rho_L}{\rho_G}, \\
  {\rm the~viscosity~ratio~} &R_\mu &:= \frac{\mu_L}{\mu_G}, \\
  {\rm the~Reynolds~number~of~phase~C~} &Re_C &:= \frac{\rho_C u' L}{\mu_C}, \\
  {\rm the~Weber~number~of~phase~C~} &We_C &:= \frac{\rho_C u'^2 L }{\sigma},
\end{eqnarray}
\end{subequations}
where $\phi_L$ is the liquid phase indicator function defined in \eqref{eq:phiL}. 
Throughout the study we set $L=1$, $\rho_G=1$, $u'^2=  \langle \rho |\vect{u}|^2\rangle / \langle \rho \rangle =1$. In order to simplify the analysis, we decided to use a constant kinematic viscosity $\nu = \mu/\rho$ between the two phases. Hence, $R_\mu^{-1} = R_\rho$, so that only one Reynolds number needs to be defined $Re \equiv \nu^{-1}$ (recall that $u'=1$ and $L=1$). In addition, the effect of surface tension is studied as a function of an averaged Weber number which is defined by $We := \overline{\rho}u'^2L/\sigma$, where $\overline{\rho} = (\rho_L + \rho_G)/2$. Since $u'=1$ and $L=1$, the averaged Weber number reduces to $We \equiv \overline{\rho}/\sigma$.

In summary, our parameter space simplifies to 4 non-dimensional numbers ($\alpha, R_\rho, Re, We$), which can be explored by varying the amount of liquid in the box, the liquid density $\rho_L$, the kinematic viscosity $\nu$ and the surface tension $\sigma$. In the present study, we have performed 14 different numerical simulations covering a range of ($\alpha, R_\rho, Re, We$) as described in Table \ref{tab:dns_params}. This database is also supplemented by three additional simulations of single-phase turbulence covering the same range of Reynolds numbers. For the sake of clarity in Table \ref{tab:dns_params}, we have isolated five different groups from the 14 different simulations. The two first groups corresponds to variations of the density ratio $R_\rho$ for two different volume fractions: $\alpha = 25\%$ corresponding to what is referred to as the 'drop' regime, and $\alpha = 75\%$ corresponding to the 'bubble' regime. Note that the Weber number is fixed which means that the surface tension is adjusted for each $R_\rho$ so that $We = \overline{\rho}/\sigma = 25$. The third group corresponds to variations of the liquid volume fraction $\alpha$ between 12.5\% and 87.5\%. Here, the density ratio is set to $R_\rho = 25$. The case with $R_\rho = 1$ has already been extensively investigated \citep[e.g.][among others]{Mukherjee2019,CrialesiEsposito2022,Cannon2024} and is thus not reproduced here. The fourth and fifth groups correspond to variations of the Reynolds number and the Weber number, respectively. Here again, $R_\rho = 25$ while the liquid volume fraction is $\alpha = 25\%$. The last group of Table \ref{tab:dns_params} corresponds to the single-phase flow simulations.

In Table \ref{tab:dns_params}, we also report the values for the total energy injection rate per unit mass, noted $\varepsilon_f = \langle F \rangle_\mathbb{T}/ \langle \rho \rangle_\mathbb{T}$, which by virtue of Eqs. \eqref{eq:1pt_budget} is also the total kinetic energy dissipation rate. The associated Taylor-based Reynolds number $R_\lambda= u'^2\sqrt{ 5 / (3\nu\varepsilon_f) } $ and the Kolmogorov microscale $\eta = \nu^{3/4} / \varepsilon_f^{1/4}$ are also given. The latter is given in terms of grid cell size $dx$ and is equal to roughly 2 for all cases considered here. Table \ref{tab:dns_params} further reports the surface area of the liquid-gas interface $A_\Gamma$ and some typical scales noted $r_H, r_\mathcal{S}$ and $r_c$ which will de defined and discussed later.

The reader is referred to appendix \ref{app:resolution} for some further discussions about the appropriateness of the numerical resolution. Due to resolution constraints, the Reynolds number achieved in the present numerical database remains moderate, preventing the development of an inertial range and even a restricted scaling range. As a result, the flow statistics reflect a superposition of large, intermediate, and small scales precluding exploring the scaling exponent for kinetic energy at intermediate scales. Nevertheless, the scaling for $\langle |\delta \vect{u}|^2_\rho\rangle_\mathbb{C}$ at small scales remains amenable to exploration. This is discussed in more details in Appendix \ref{app:limit_r0} and in the results' section below. The description of the post-processing procedures for computing the different terms of the scale-by-scale energy budget is given in appendix \ref{app:post_pro}.

\begin{table}
  \begin{center}
  \begin{tabular}{c  c c c c c c  c c c c  c c c }
    studied & \multirow{2}{*}{$\alpha$ [\%]} & \multirow{2}{*}{$R_\rho$} &  \multirow{2}{*}{$\nu$ [m$^2$.s$^{-1}$]} & \multirow{2}{*}{$We$} & \multirow{2}{*}{$Re$}  & \multirow{2}{*}{$N$} & \multirow{2}{*}{$\varepsilon_f$} & \multirow{2}{*}{$R_\lambda$} & \multirow{2}{*}{$\eta/dx$} & \multirow{2}{*}{$A_\Gamma$}  & \multirow{2}{*}{$r_H$}  & \multirow{2}{*}{$r_\mathcal{S}$}  & \multirow{2}{*}{$r_c$} \\
    parameter &&&&&&&&&&    \\
    \hline \\
    liquid & 25  & 1  & 1.4$\times$10$^{-3}$  & 25 & 714 & 256  & 0.99 & 34.6 & 1.86 & 7.10 & 0.100 & 0.061 & 0.010\\
    density, & 25 & 5    & 1.4$\times$10$^{-3}$  & 25  & 714 & 256 & 1.09 & 33.0 & 1.81 & 6.21 & 0.099 & 0.057 & 0.010\\
    drop & 25 & 25   & 1.4$\times$10$^{-3}$  & 25  & 714 & 256 & 1.15 & 32.2 & 1.79 & 5.67 & 0.096 & 0.051 & 0.011\\
    regime & 25 & 125  & 1.4$\times$10$^{-3}$  & 25  & 714 & 256 & 1.23 & 31.1 & 1.76 & 5.76 & 0.095 & 0.049 & 0.012\\
    % \hline
    &&&&&&&&&&   \\
    density, &  75  & 1    & 1.4$\times$10$^{-3}$  & 25  & 714 & 256 & 0.99 & 34.6 & 1.86 & 7.10 & 0.100 & 0.062 & 0.010\\
    bubble & 75  & 5    & 1.4$\times$10$^{-3}$  & 25  & 714 & 256 & 0.99 & 34.6 & 1.86 & 7.10 & 0.099 & 0.057 & 0.011\\
    regime &  75  & 25   & 1.4$\times$10$^{-3}$  & 25 & 714 & 256 & 0.98 & 34.9 & 1.86 & 6.11 & 0.098 & 0.056 & 0.012\\
    % \hline
    &&&&&&&&&& \\
    liquid & 12.5  & 25  & 1.4$\times$10$^{-3}$  & 25  & 714 & 256  &  0.92 & 35.9 & 1.89 & 3.39 & 0.096 & 0.047 & 0.011\\
    volume & 50  &  25  & 1.4$\times$10$^{-3}$ & 25 & 714 & 256 & 1.20 & 31.5 & 1.77 & 7.80 & 0.096 & 0.054& 0.012\\
    fraction & 87.5  &  25  & 1.4$\times$10$^{-3}$ & 25 & 714 & 256 & 0.85 & 37.3 & 1.93 & 3.87 & 0.098 & 0.054& 0.013\\
    % \hline
    &&&&&&&&&& \\
    Reynolds & 25  & 25  & 3.8$\times$10$^{-3}$  & 25  & 263 & 128  & 1.36 & 18.0 & 1.81 & 5.62 & 0.106 & 0.063 & 0.013\\
    number   & 25  & 25  & 5.3$\times$10$^{-4}$  & 25   & 1887 & 512  & 0.99 & 57.3 & 1.81 & 5.55 & 0.089 & 0.042 & 0.109  \\
    % \hline
    &&&&&&&&&& \\
    Weber & 25  & 25  & 1.4$\times$10$^{-3}$ & 12.5 & 714 & 256  & 1.09 & 33.3 & 1.82 & 3.92 & 0.136 & 0.061 & 0.015\\
    number & 25 & 25  & 1.4$\times$10$^{-3}$ & 50 & 714   & 256  & 1.12 & 32.6 & 1.80 & 8.12 & 0.070 & 0.042 & 0.088\\
    &&&&&&&&&& \\
    single & -  & -   & 3.8$\times$10$^{-3}$ & - & 263    & 128  & 0.91 & 21.9 & 2.00 & -    & -       & -     & -    \\
    phase  & -  & -   & 1.4$\times$10$^{-3}$ & -  & 714   & 256  & 0.66 & 42.0 & 2.04 & -    & -       & -     & -    \\
    flows  & -  & -   & 5.3$\times$10$^{-4}$ & -  & 1887   & 512  & 0.60 & 71.3 & 2.02 & -    & -       & -     & - 
  \end{tabular}
  \end{center}
  \caption{List of simulation parameters. The number of simulation points per direction is $N$, hence the resolution $dx = L/N$. The Taylor-scale Reynolds number $R_\lambda = u'^2\sqrt{ 5 / (3\nu\varepsilon_f) } $, where the kinetic energy injection rate per unit mass (= dissipation) $\varepsilon_f = {\langle F \rangle_\mathbb{T}}/{\langle \rho \rangle_\mathbb{T}}$. $\varepsilon_f$ is given in units of $u'^3/L~ (=1)$. The Kolmogorov length scale is defined by $\eta = \nu^{3/4} / \varepsilon_f^{1/4}$. The liquid-gas surface area $A_\Gamma$ is given in units of $L^2~ (=1)$. We also report the Kolmogorov-Hinze scale $r_H$, the scale $r_\mathcal{S}$ and $r_c$ which are defined and discussed later. All are given in units of $L$.} \label{tab:dns_params}
\end{table}

% \subsubsection{Buoyancy-driven bubbly flows}

% The second configuration corresponds to case \texttt{R6} of \cite{Pandey2020}. In this situation, the forcing term $F$ writes $\vect{F} = (\rho - \langle \rho \rangle_\mathbb{T}) \vect{g}$, where $\vect{g}=-g \vect{e_z}$ is the acceleration due to gravity which act only in $z$ direction. The domain is filled with 40 monodispersed and non-overlapping bubbles, randomly located in the domain. The simulation domain is cubic and periodic boundary conditions are used for all three directions. The domain size $L= 256$ and is discretized using 256 grid points in each direction. 

% The relevant parameters for this configuration are:
% \begin{itemize}
%   \item the gas volume fraction $\langle \phi_G \rangle_\mathbb{T} = 1.72\%$. The latter can be deduced from the bubble diameter $d$ that is set to 24 and the number of bubbles. 
%   \item the density ratio between the two phases $\rho_L/\rho_G = (1+At)/(1-At)$, which is related to the Atwood number $At = (\rho_L - \rho_G)/(\rho_L + \rho_G)$. We have set $A_t=0.8$ \citep[same as case \texttt{R6} of][]{Pandey2020} and $\rho_L = 1.0$. This yields $\rho_G = 1/9$. 
%   \item the Galilei number $Ga = \sqrt{\rho_L(\rho_L-\rho_G)gd^3}/\mu_L$. Provided $\mu_L = 0.32$ and $g=1$ \citep[same as case \texttt{R6} of][]{Pandey2020}, one gets $Ga = 346$. We also set $\mu_G = \mu_L$.
%   \item the Bond number $Bo = (\rho_L - \rho_G)g d^2 / \sigma$. For $Bo = 1.9$ \citep[same as case \texttt{R6} of][]{Pandey2020}, one has to set $\sigma = 213.33$
% \end{itemize}

\section{Results} \label{sec:results}

We now present the application of the KHM equation to the numerical data. We sequentially focus on the effect of the liquid-gas density ratio (in subsections \ref{sec:Rrho_drop} and \ref{sec:Rrho_bubble}), the liquid volume fraction (subsection \ref{sec:alpha}), the Reynolds number (subsection \ref{sec:Re}) and Weber number (subsection \ref{sec:We}).

\subsection{Effect of the liquid-gas density ratio in the drop regime} \label{sec:Rrho_drop}

\begin{figure}
  \includegraphics[width=\textwidth]{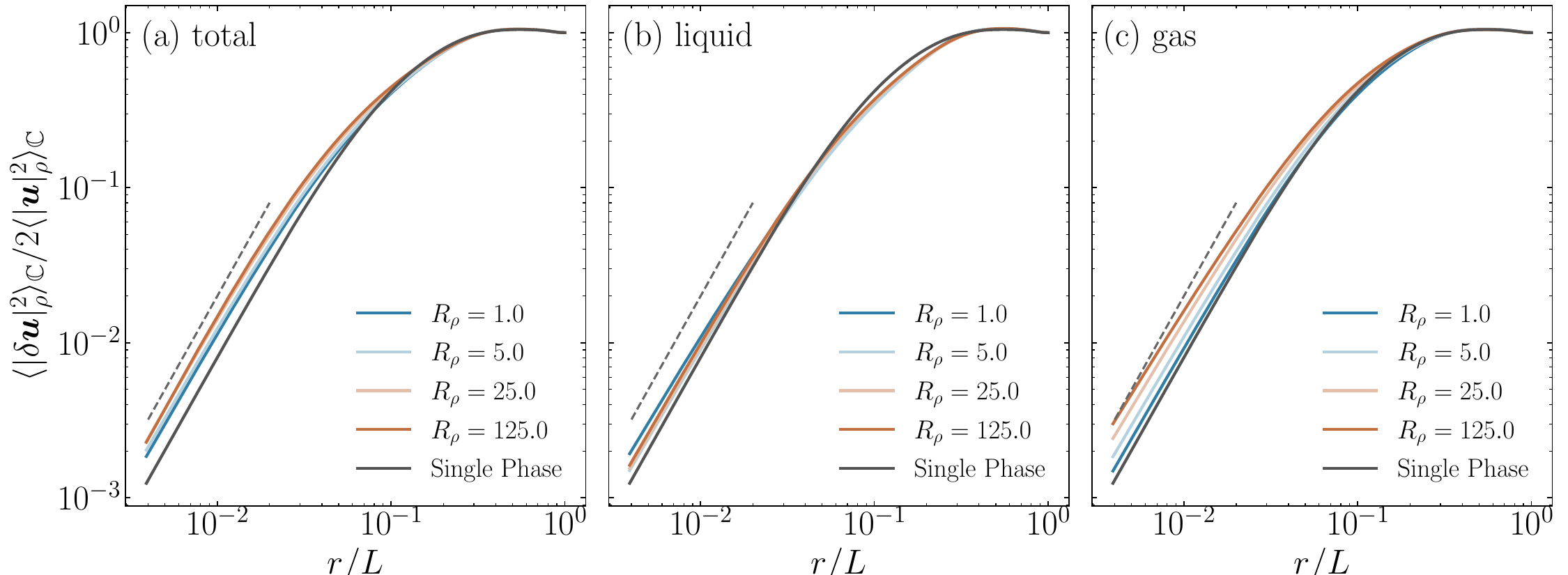}
  \caption{Effect of the density ratio $R_\rho$ on the scale-by-scale kinetic energy for $\alpha = 25\%$. The colours from blue to red correspond to $R_\rho = 1, ~5, ~25, ~125$ while the black curve is for single-phase turbulence. (a) $\mathbb{C} \equiv \mathbb{T}$, (b) $\mathbb{C} \equiv \mathbb{L}$ and (c) $\mathbb{C} \equiv \mathbb{G}$. The dashed line represents the $r^2$ scaling.} \label{fig:rho_phi25_dq2} 
\end{figure}

\begin{figure}
  \begin{tabular}{l l l}
    (a) & (b) & \\
    \includegraphics[trim={2cm 0 2cm 0},clip, width=0.4\textwidth]{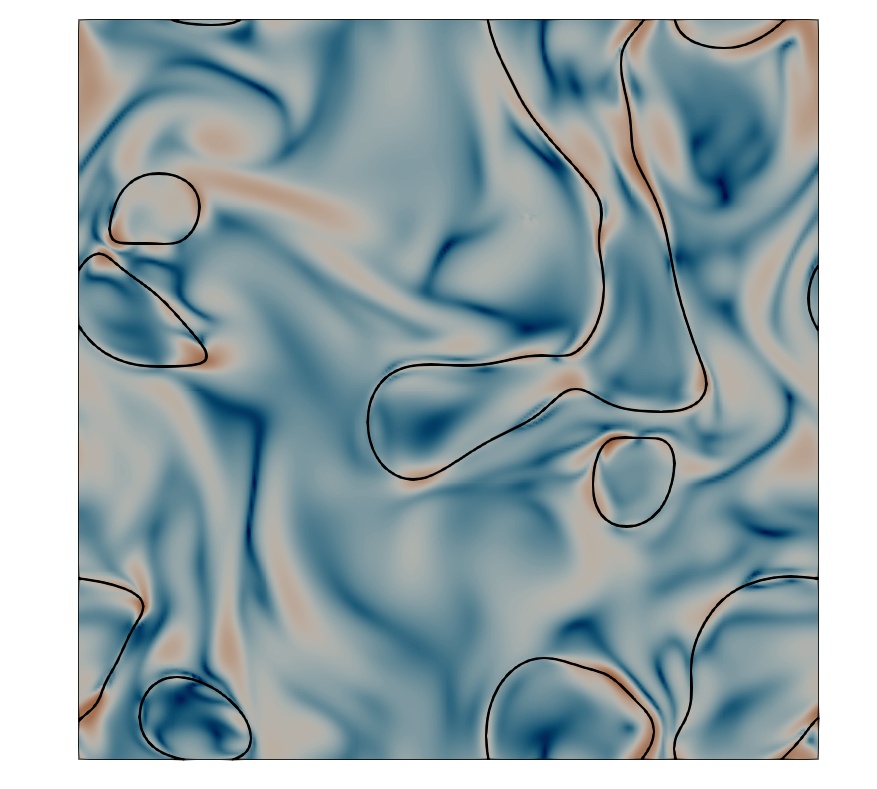}  & 
    \includegraphics[trim={2cm 0 2cm 0},clip, width=0.4\textwidth]{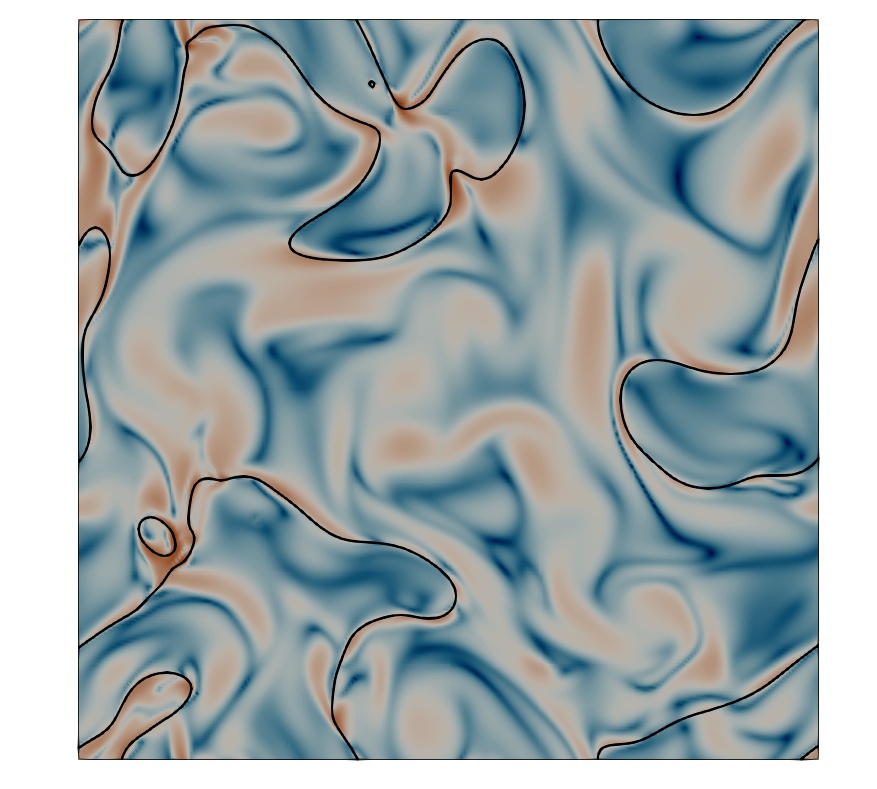} & 
    \includegraphics[width=0.0855\textwidth]{./vert_1-1200_2} \\
    (c) & (d)\\
    \includegraphics[trim={2cm 0 2cm 0},clip, width=0.4\textwidth]{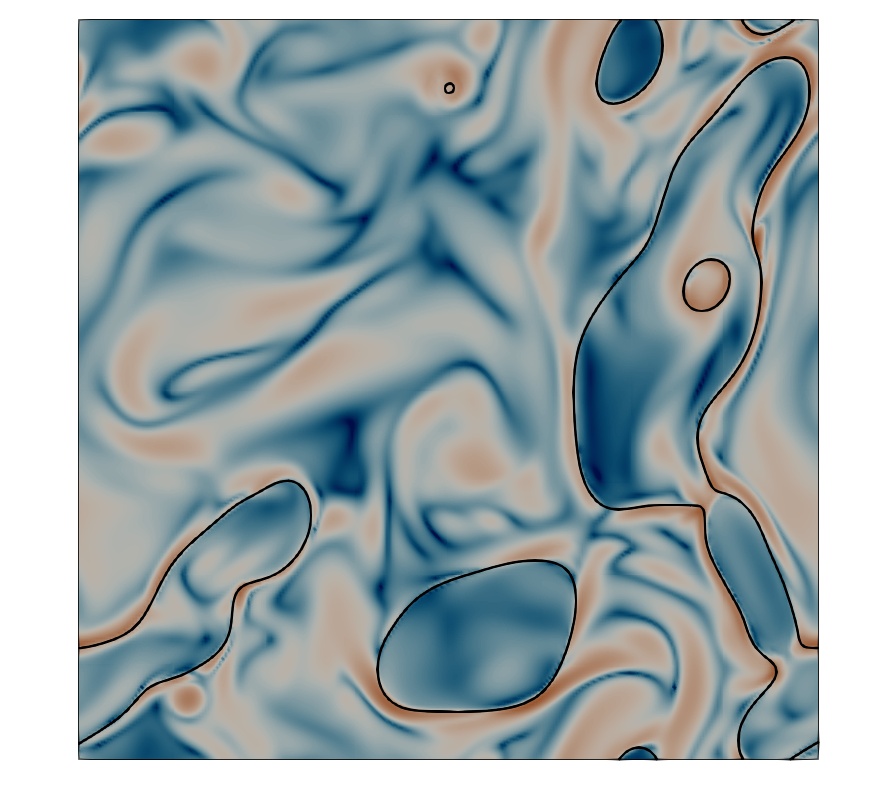} &
    \includegraphics[trim={2cm 0 2cm 0},clip, width=0.4\textwidth]{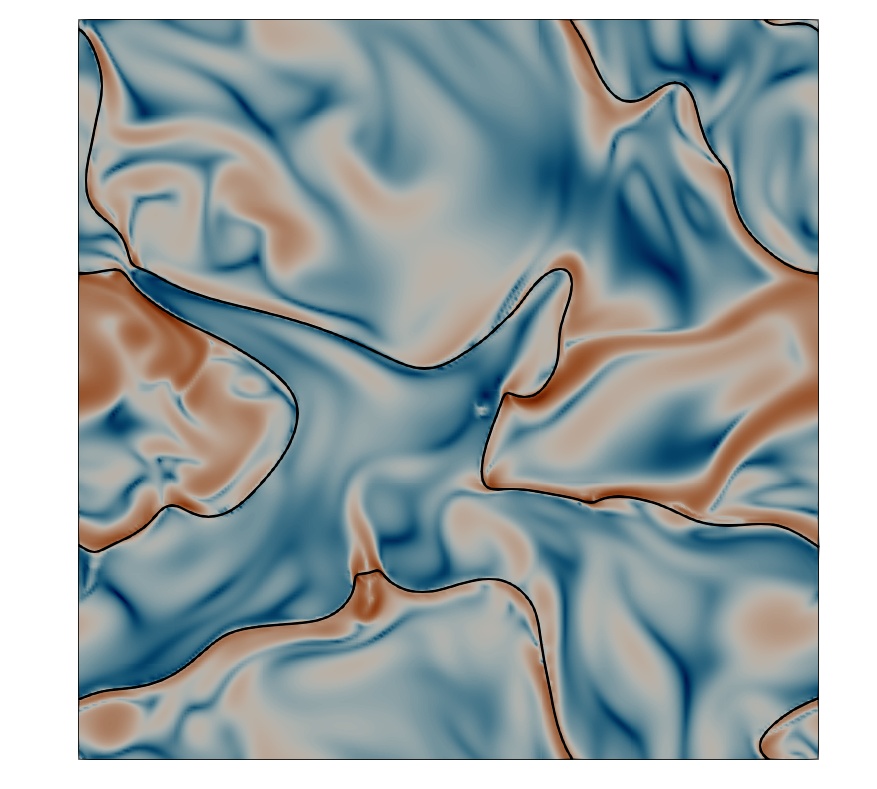} & 
    \includegraphics[width=0.0855\textwidth]{./vert_1-1200_2} 
  \end{tabular}
  \caption{2D slices of the vorticity magnitude (in colour, units of $u'/L$) together with the liquid-gas interface (black curves) for $\alpha=25\%$ and increasing $R_\rho$ from (a) to (d).} \label{fig:rho_phi25_visu}
\end{figure}

The scale distribution of kinetic energy $\langle |\delta \vect{u}|^2_\rho\rangle_\mathbb{C}$ for different density ratio $R_\rho$ in the drop regime ($\alpha = 25\%$) is plotted in Fig. \ref{fig:rho_phi25_dq2}. The scale is normalized by the width of the computational domain $L$ while $\langle |\delta \vect{u}|^2_\rho\rangle_\mathbb{C}$ is normalized by twice the scale-integrated turbulent kinetic energy $\langle |\vect{u}|^2_\rho\rangle_\mathbb{C}$. For the total fluctuating field $\mathbb{C} \equiv \mathbb{T}$ (Fig. \ref{fig:rho_phi25_dq2}(a)), we observe that the small-scale content of kinetic energy is larger in multiphase turbulence than in single phase turbulence. We further note that increasing the density ratio yields an increase of the scale-by-scale kinetic energy at small scales. The activity of turbulence at small scales is linked to the activity of the dissipative scales. Increasing $R_\rho$, and/or moving from single-phase to multiphase flows, thus leads to an increase of the energy dissipation. This is confirmed by the values of $\varepsilon_f$ in the first five lines of Table \ref{tab:dns_params}.

Looking at the curves for the liquid and gas conditional statistics (Figs. \ref{fig:rho_phi25_dq2}(b) and (c) for $\mathbb{C} \equiv \mathbb{L}$ and $\mathbb{C} \equiv \mathbb{G}$, respectively), we note that the scale distribution of kinetic energy in the liquid is mildly changing contrary to the gas phase where kinetic energy substantially increases with $R_{\rho}$ when the separation $r$ decreases. Hence, the increase of $\langle |\delta \vect{u}|^2_\rho\rangle_\mathbb{T}$ at small scales is mostly associated to an increase of $\langle |\delta \vect{u}|^2_\rho\rangle_\mathbb{G}$.

Appendix \ref{app:limit_r0} discusses the small-scale behaviour of the two-point kinetic energy. It is shown that the mixed structure function $\langle |\delta \vect{u}|^2_\rho\rangle_\mathbb{C}$ scales as $r^2$ and is proportional to the enstrophy $\langle |\vect{\omega}|^2\rangle_\mathbb{C}$ averaged over $\mathbb{C} \in \{\mathbb{T,L,G}\}$. This quadratic evolution with respect to $r$ is shown in Fig. \ref{fig:rho_phi25_dq2}(a-c) as gray dashed lines. As $r$ decreases, $\langle |\delta \vect{u}|^2_\rho\rangle_\mathbb{C}$ approaches the expected scaling, although some small deviations are observed, particularly in the gas phase (Fig. \ref{fig:rho_phi25_dq2}(c)). Previous studies \citep{Estivalezes2022,Ling2019} have noted that accurately resolving the enstrophy field in multiphase flow simulations is demanding. Therefore, while mass and kinetic energy are well resolved in our data (see Appendix \ref{app:resolution}), the observed deviations from the $r^2$-scaling might reflect the need for an even finer mesh to fully capture the enstrophy field. Consequently, in the remainder of this paper, the small-scale scaling of $\langle |\delta \vect{u}|^2_\rho\rangle_\mathbb{C}$ will not be further addressed, and the vorticity field will be discussed only qualitatively.

Keeping this consideration in mind, our data indicate that the increase of the small-scale kinetic energy in the gas phase observed in Fig. \ref{fig:rho_phi25_dq2}(c) is related to an increase of the enstrophy field in the gas phase. From $R_\rho =1$ to $R_\rho=125$, the averaged enstrophy in the gas phase is doubled, as shown in Fig. \ref{fig:rho_phi25_dq2}(c). To confirm this, one can scrutinize the visualizations of the flow portrayed in Fig. \ref{fig:rho_phi25_visu}. We observe an increased amplitude of the vorticity magnitude, hence the small scale activity, when one approaches the interface from the gas phase. This behaviour was already reported by \cite{Dodd2016}. The growing production of kinetic energy and vorticity is occurring in the turbulent boundary layers and wakes surrounding the liquid structures. As per \cite{Dodd2016}, this results from an increase of the liquid structures Stokes number which increases with $R_\rho$. The observed increased vorticity magnitude is further consistent with the theory developed by \cite{Terrington2022}. 

\begin{figure}
  \includegraphics[width=\textwidth]{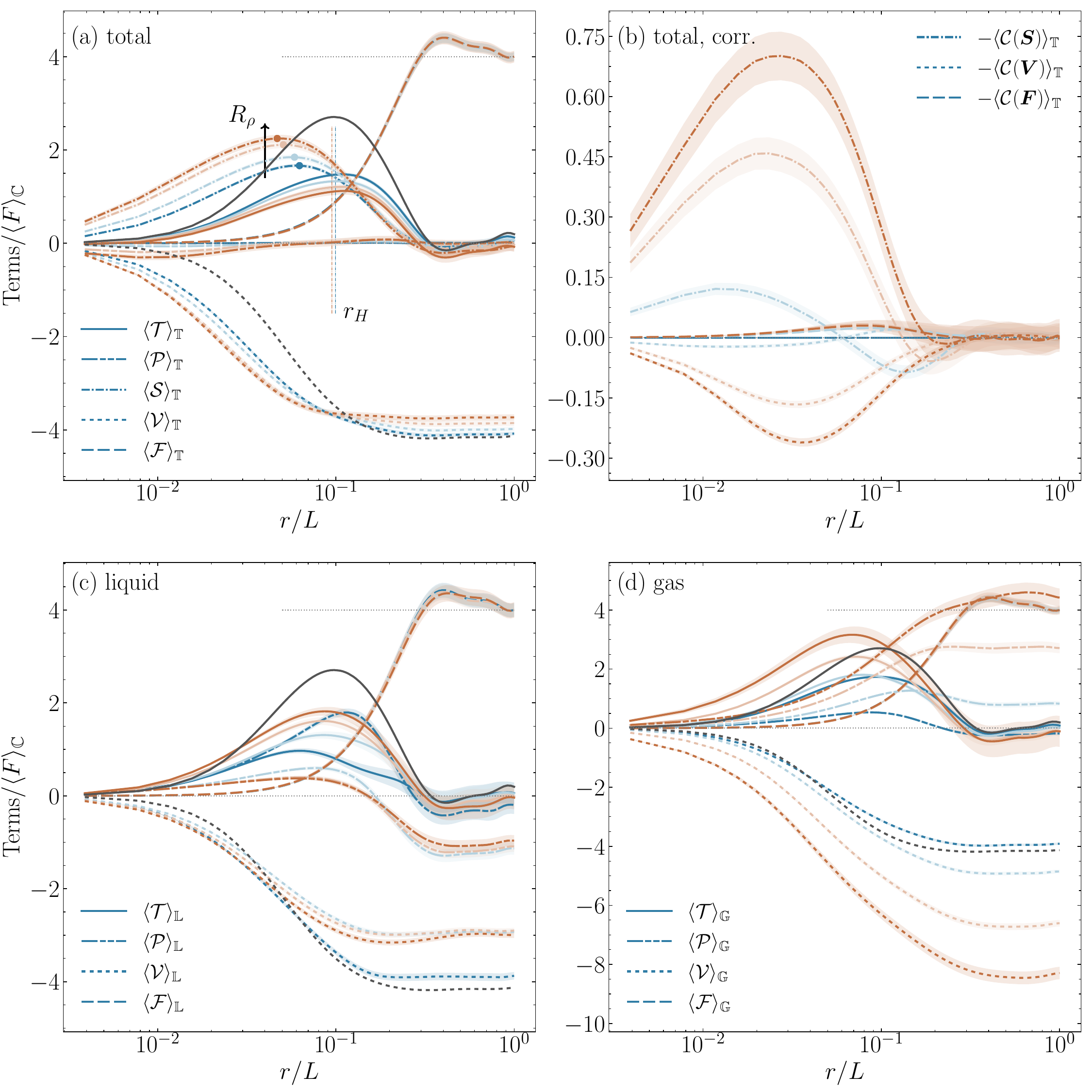}
  \caption{Effect of the density ratio $R_\rho$ on the scale-by-scale kinetic energy budgets for $\alpha = 25\%$. The colours from blue to red correspond to $R_\rho = 1, ~5, ~25, ~125$ as in the legend of Fig. \ref{fig:rho_phi25_dq2}. The black curves are for the single-phase case at same viscosity. The different lines correspond to \Ltransport ~ $\langle \mathcal{T}\rangle_\mathbb{C}$, \Lpressure ~ $\langle \mathcal{P}\rangle_\mathbb{C}$, \Lsurface ~ $\langle \mathcal{S}\rangle_\mathbb{C}$, \Lviscous ~ $\langle \mathcal{V}\rangle_\mathbb{C}$, \Lforcing ~ $\langle \mathcal{F}\rangle_\mathbb{C}$, with (a) $\mathbb{C} = \mathbb{T}$, (c) $\mathbb{C} = \mathbb{L}$ and (d) $\mathbb{C} = \mathbb{G}$. The horizontal dashed line indicates $4\langle F\rangle_\mathbb{C}$, the limit at large separations of the forcing term. Figure (b) represents the density correction terms $\langle \mathcal{C}(\vect{a}) \rangle_\mathbb{T}$. Also represented in (a) are the scales $r_H$ with lines and $r_\mathcal{S}$ with filled circles that will be described later on.}\label{fig:rho_phi25_budget}
\end{figure}

The different terms of the scale-by-scale kinetic energy budget are displayed in Fig. \ref{fig:rho_phi25_budget}. Each contribution is presented in units of $\langle F \rangle_\mathbb{C}$ which is the contribution of the forcing term in the one-point kinetic energy budget. By doing this, the scale distribution of forcing term to the scale-by-scale budget appears to be independent of the investigated physical parameters and of the averaging volume $\mathbb{C} \in \{ \mathbb{T}, \mathbb{L}, \mathbb{G}\}$, as shown in Fig. \ref{fig:rho_phi25_budget}(a-c). The forcing term also appears to be the same as in single-phase flows. Consequently, all the other terms of the scale-by-scale budget can be studied in proportion of an invariant forcing contribution. This significantly eases the interpretation of the results. 

% Furthermore, as it is common in the litterature \citep{Danaila2002,Antonia2006,Lai2018}, instead of plotting directly the viscous term $\langle \mathcal{V} \rangle_\mathbb{C}$, we plot $\langle \mathcal{V} \rangle^*_\mathbb{C}$ which is given by
% \begin{eqnarray}
%   \langle \mathcal{V} \rangle^*_\mathbb{C} = \langle \mathcal{V} \rangle_\mathbb{C} - \lim_{r \to \infty} \langle \mathcal{V} \rangle_\mathbb{C}
% \end{eqnarray}
% By doing this, $\langle \mathcal{V} \rangle^*_\mathbb{C}$ can be viewed as the scale-by-scale contribution of the viscous term to the kinetic energy dissipation rate. 

Let us start by analysing the unconditionally averaged budget, i.e. $\mathbb{C} \equiv \mathbb{T}$, portrayed in Fig. \ref{fig:rho_phi25_budget}(a). We observe that the forcing term $\langle \mathcal{F} \rangle_\mathbb{T}$ contributes positively to the budget and acts at rather large-scales. At intermediate scales, the kinetic energy is transferred from large to small scales through the non-linear energy transport term $\langle \mathcal{T} \rangle_\mathbb{T}$ and the surface tension term $\langle \mathcal{S} \rangle_\mathbb{T}$. The kinetic energy budget is finally equilibrated by the viscous term $\langle \mathcal{V} \rangle_\mathbb{T}$ which is negative. Therefore, our analysis based on the KHM framework confirms previous conclusions using other approaches such as e.g. spectral energy budgets \citep[see][among others]{CrialesiEsposito2022,Cannon2024}. 

We further note that the non-linear transfer term $\langle \mathcal{T} \rangle_\mathbb{T}$ is drastically reduced compared to its single-phase counterpart. The kinetic energy exchange in multiphase turbulence thus appears to comply with the following scenario. \textit{(i)} Part of the kinetic energy injected at large scales is transferred into kinetic energy at smaller scales by the classical non-linear transport process $\langle \mathcal{T} \rangle_\mathbb{T}$. This transfer is however less important than in single-phase flows because \textit{(ii)} part of the kinetic energy injected at large scale is also transferred into surface energy and then released back into kinetic energy, but at smaller scales. Finally, \textit{(iii)} the increase of the overall kinetic energy transfer is compensated by a shift of the viscous term $\langle \mathcal{V} \rangle_\mathbb{T}$ in the direction of smaller scales. The pressure term also contributes as a loss of energy at small scales as soon as $R_\rho \neq 1$. Its contribution is though rather marginal. 

Although not shown in Fig. \ref{fig:rho_phi25_budget}(a), summing up the contribution of the surface tension term and the non-linear transport term indicates that the overall kinetic energy transfer (i.e. $\langle \mathcal{T} \rangle_\mathbb{T} +  \langle \mathcal{S} \rangle_\mathbb{T}$) is larger than the one pertaining to single-phase flows and peaks at smaller scales. It suggests that the overall transfer of kinetic energy in multiphase flows is quite likely to compare to the one observed in single-phase flows but at a larger Reynolds number. This explains that numerical simulations of multiphase turbulence require a much finer resolution compared to single-phase turbulence, even when $R_\rho = 1$.

When the density ratio $R_\rho$ increases, the contribution of the surface tension term $\langle \mathcal{S} \rangle_\mathbb{T}$ gets larger while the one due to the non-linear transport term $\langle \mathcal{T} \rangle_\mathbb{T}$ goes in opposite direction. We note also that $\langle \mathcal{S} \rangle_\mathbb{T}$ peaks at smaller scales compared to $\langle \mathcal{T} \rangle_\mathbb{T}$. The scale at which $\langle \mathcal{S} \rangle_\mathbb{T}$ is maximum is represented by the filled circle symbols in Fig. \ref{fig:rho_phi25_budget}(a). More insights into the behaviour of this scale will be provided later in a dedicated section.

Consistently with our previous conclusions suggesting an increased activity of the smaller scales when $R_\rho$ becomes larger, the contribution of the viscous term starts being visible at smaller scales when $R_\rho$ increases. Finally, the contribution due to pressure $\langle \mathcal{P} \rangle_\mathbb{T} = -\langle \mathcal{C}(-\vect{\nabla}P) \rangle_\mathbb{T}$ is rather small and appears only in the small-scale range where it contributes negatively to the budget. It is zero for $R_\rho = 1$, and increases in amplitude with $R_\rho$.

In Fig. \ref{fig:rho_phi25_budget}(b), we represent the contribution of $-\langle \mathcal{C}(\vect{a}) \rangle_\mathbb{T}$ for the viscous, forcing and surface tension terms. The one corresponding to the pressure term is already plotted and discussed in Fig. \ref{fig:rho_phi25_budget}(a). As expected, $\langle \mathcal{C}(\vect{a}) \rangle_\mathbb{T}$ is zero for $R_\rho = 1$ and its contribution increases when $R_\rho$ grows. We also have $\langle \mathcal{C}(\vect{a}) \rangle_\mathbb{T} \to 0$ when $r \to \infty$ (see appendix \ref{app:limit}). The sign of the density correction for each term is the same as the term itself which means that variations of density tends to amplify the contribution of each process. We note that the density correction term contributes to about 1/3 of the surface tension at $R_\rho = 125$, and about 10\% to the viscous term. 

The different terms of the conditionally averaged KHM equation are presented in Fig. \ref{fig:rho_phi25_budget}(c) and (d) for the liquid and gas phase, respectively. The first striking observation is for the gas phase (Fig. \ref{fig:rho_phi25_budget}(d)), where we note that the pressure term $\langle \mathcal{P} \rangle_\mathbb{G}$ is positive and thus represents a gain for the scale-by-scale kinetic energy budget. Recall that when conditionally averaged, only the pressure transport contributes to the budget while the term $\langle \mathcal{C} \rangle_\mathbb{L,G}$ is zero. Hence the significant contribution attributed to pressure arises solely from the scale-by-scale pressure transport. Its contribution increases greatly with $R_\rho$ and is even exceeding the one of the forcing term $\langle \mathcal{F} \rangle_\mathbb{T}$ at $R_\rho = 125$. This additional amount of kinetic energy injected into the gas phase yields an amplified contribution of the non-linear transport term $\langle \mathcal{T} \rangle_\mathbb{G}$. When the density ratio $R_\rho = 125$, the latter is observed to be larger than the one pertaining to single-phase flows. For $R_\rho = 1$, the sum of the pressure and non-linear transport, i.e. $\langle \mathcal{T} \rangle_\mathbb{G} + \langle \mathcal{P} \rangle_\mathbb{G}$ is roughly similar to the non-linear transport term of single-phase turbulence. The effect of the viscous stress $\langle \mathcal{V} \rangle_\mathbb{G}$ at each scale also becomes larger with $R_\rho$. Here again, this is explained by an increased activity of the small scales in the gas phase. 

In the liquid-phase, our results show that the curves corresponding to $R_\rho = 1$ distinguishes substantially from the others. For instance, it appears that the pressure term $\langle \mathcal{P} \rangle_\mathbb{L}$ (the pressure transport) for $R_\rho = 1$ is very different from the one at $R_\rho = 5$. It seems however to vary only marginally for $R_\rho \geq 25$. The same applies for the viscous term $\langle \mathcal{V} \rangle_\mathbb{L}$ but not for the non-linear transfer term $\langle \mathcal{T} \rangle_\mathbb{L}$ which continues to grow as $R_\rho$ increases. That means that for the liquid phase, the effect of the density contrast between the two phases is not trivial. Some processes seem to saturate while some others continues to evolve. Therefore, the physics observed for $R_\rho = 1$ which is a common simplification in the literature, cannot be readily extrapolated to flows with variations of density, even if the density contrast is moderate.

\subsection{Effect of the liquid-gas density ratio in the bubble regime} \label{sec:Rrho_bubble}

\begin{figure}
  \includegraphics[width=\textwidth]{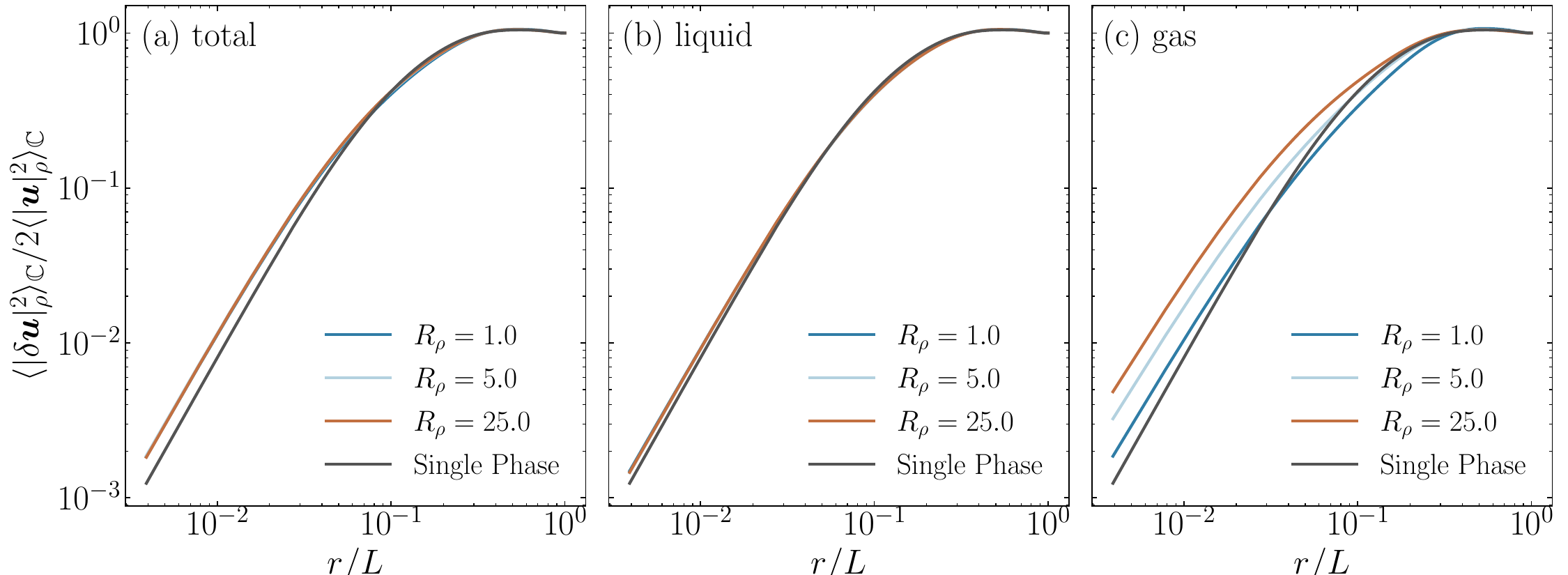}
  \caption{Effect of the density ratio $R_\rho$ on the scale-by-scale kinetic energy for $\alpha = 75\%$. The colours from blue to red correspond to $R_\rho = 1, ~5, ~25$ while the black curve is for single-phase turbulence. (a) $\mathbb{C} \equiv \mathbb{T}$, (b) $\mathbb{C} \equiv \mathbb{L}$ and (c) $\mathbb{C} \equiv \mathbb{G}$. } \label{fig:rho_phi75_dq2} 
\end{figure}
\begin{figure}
  \begin{tabular}{l l l l}
    (a) & (b) & (c)\\
    \includegraphics[trim={2cm 0 2cm 0},clip, width=0.29\textwidth]{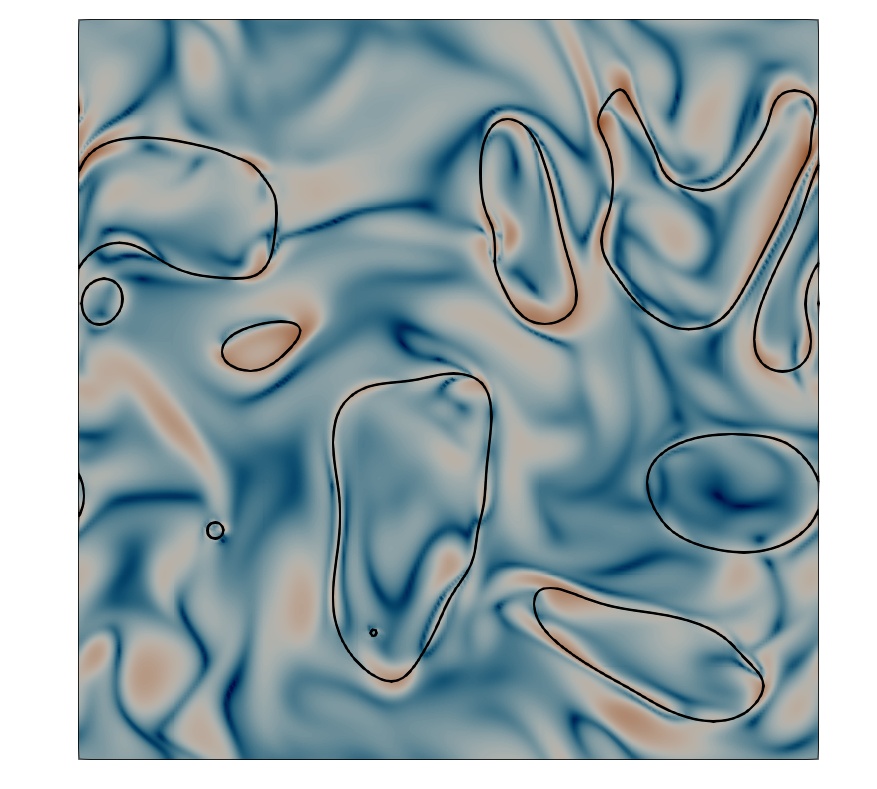}  & 
    \includegraphics[trim={2cm 0 2cm 0},clip, width=0.29\textwidth]{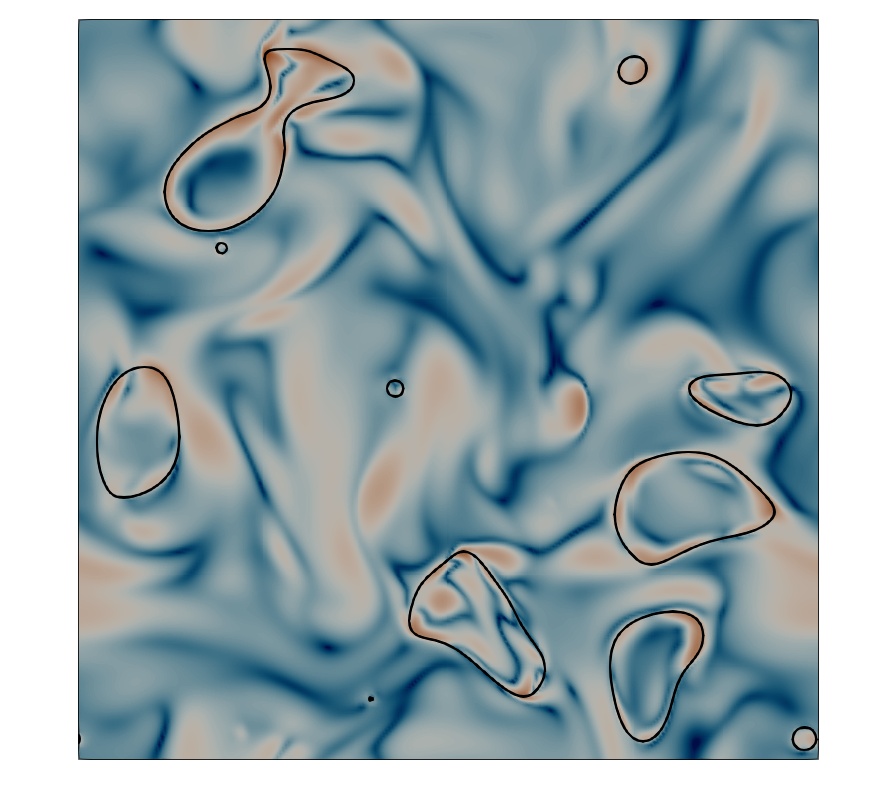} & 
    \includegraphics[trim={2cm 0 2cm 0},clip, width=0.29\textwidth]{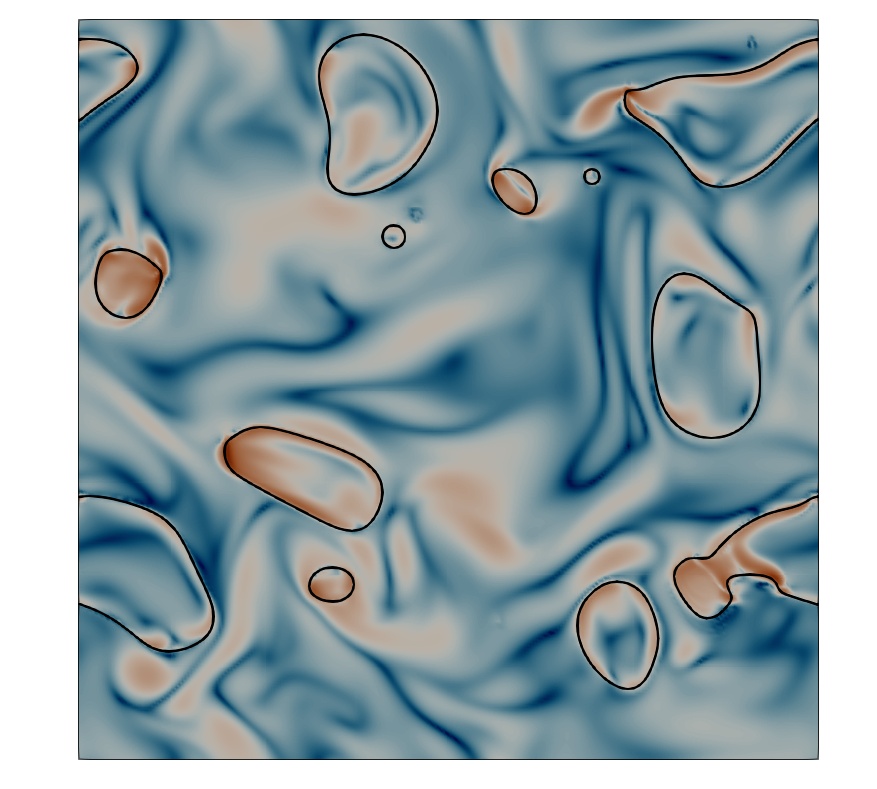} &
    \includegraphics[width=0.062\textwidth]{./vert_1-1200_2} \\
    % (c)\\
    % \includegraphics[width=0.0855\textwidth]{./vert_1-1200_2.jpeg} \\
  \end{tabular}
    \caption{2D slices of the vorticity magnitude (in colour, units of $u'/L$) together with the liquid-gas interface (black curve) at $\alpha = 75\%$ and for increasing $R_\rho$ from (a) to (c).} \label{fig:rho_phi75_visu} 
\end{figure}

The same analysis is performed for $\alpha = 75\%$ and $R_\rho = 1, 5, 25$. The scale distribution of kinetic energy $\langle |\delta \vect{u}|^2_\rho\rangle_\mathbb{C}/2 \langle | \vect{u}|^2_\rho\rangle_\mathbb{C}$ is presented in Fig. \ref{fig:rho_phi75_dq2}(a), (b) and (c) for $\mathbb{C}=\mathbb{T}, \mathbb{L}$ and $\mathbb{G}$, respectively. Here, we note that $\langle |\delta \vect{u}|^2_\rho\rangle_\mathbb{T}/2\langle | \vect{u}|^2_\rho\rangle_\mathbb{C}$ is mostly unchanged when the density ratio $R_\rho$ is increased. Hence, the total dissipation rate $\varepsilon_f$ is found to be constant with respect to $R_\rho$ (see Table \ref{tab:dns_params}). At small scales, the scale-by-scale kinetic energy is slightly larger than that of single-phase turbulence. This is reflected by a larger dissipation rate $\varepsilon_f$ in multiphase turbulence compared to its single-phase counterpart. In the liquid phase, $\langle |\delta \vect{u}|^2_\rho\rangle_\mathbb{L}$ is almost independent of $R_\rho$ and is very comparable to single-phase turbulence. In the bubble regime, the effect of density ratio is only perceptible in the gas phase. We note again a systematic shift of $\langle |\delta \vect{u}|^2_\rho\rangle_\mathbb{G}$ towards smaller scales when the density ratio is increased. Here again, this trend is associated with an increased intensity of the small-scales, which is reflected by a larger dissipation rate and a larger viscous transport in the gas phase. 

This is indeed confirmed by looking at the visualizations of the vorticity magnitude presented in Fig. \ref{fig:rho_phi75_visu}. Although less apparent than in Fig. \ref{fig:rho_phi25_visu} for the drop regime, we still observe larger amplitude of vorticity in the gas phase close to the interface. This is in agreement with the analytical considerations of \cite{Terrington2022}.

\begin{figure}
  \includegraphics[width=\textwidth]{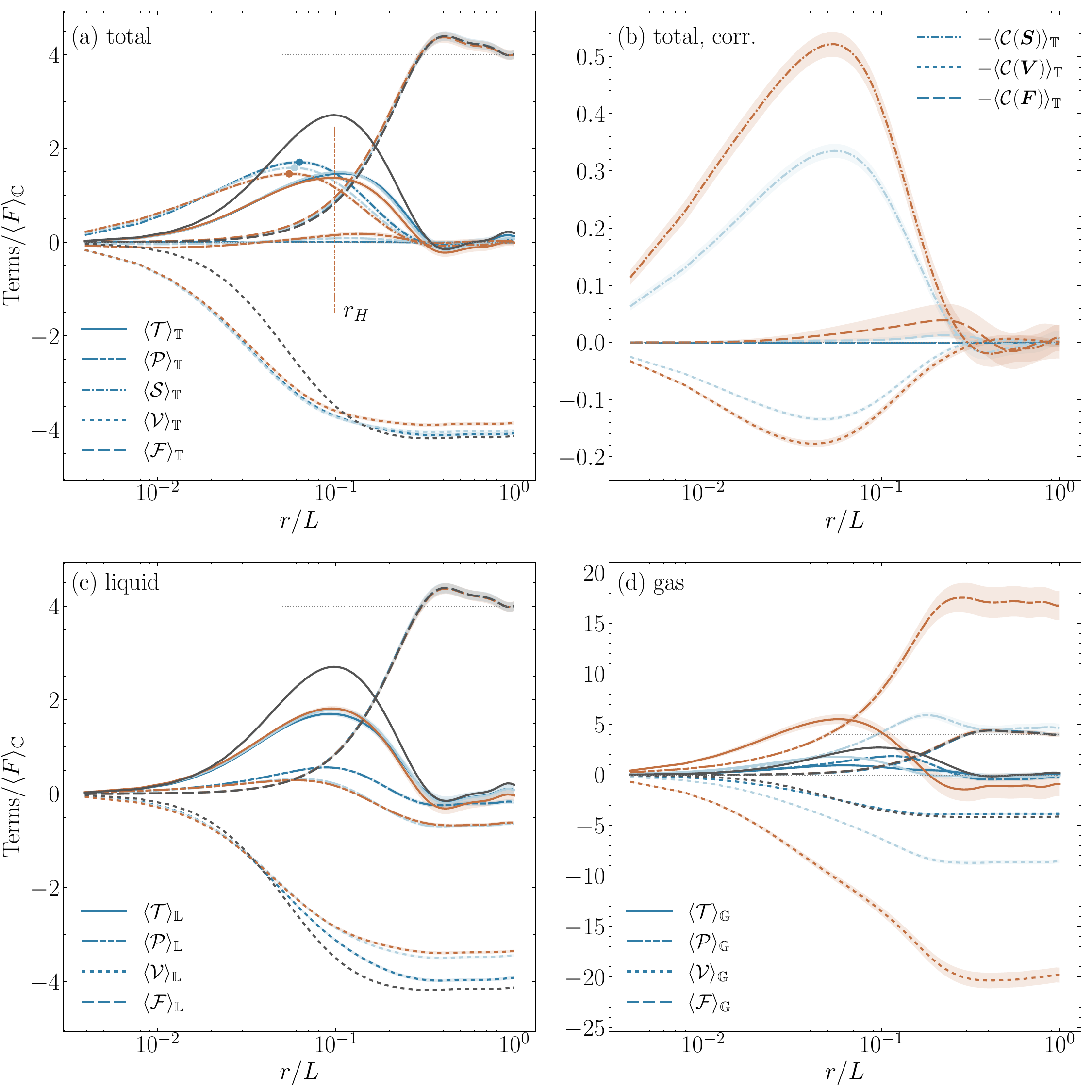}
  \caption{Effect of the density ratio $R_\rho$ on the scale-by-scale kinetic energy budgets for $\alpha = 75\%$. The colours from blue to red correspond to $R_\rho = 1, ~5, ~25, ~125$ as in the legend of Fig. \ref{fig:rho_phi25_dq2}. The black curves are for the single-phase case at same viscosity. The different lines correspond to \Ltransport ~ $\langle \mathcal{T}\rangle_\mathbb{C}$, \Lpressure ~ $\langle \mathcal{P}\rangle_\mathbb{C}$, \Lsurface ~ $\langle \mathcal{S}\rangle_\mathbb{C}$, \Lviscous ~ $\langle \mathcal{V}\rangle_\mathbb{C}$, \Lforcing ~ $\langle \mathcal{F}\rangle_\mathbb{C}$, with (a) $\mathbb{C} = \mathbb{T}$, (c) $\mathbb{C} = \mathbb{L}$ and (d) $\mathbb{C} = \mathbb{G}$. The horizontal dashed line indicates $4\langle F\rangle_\mathbb{C}$, the limit at large separations of the forcing term. Figure (b) represents the density correction terms $\langle \mathcal{C}(\vect{a}) \rangle_\mathbb{T}$. }\label{fig:rho_phi75_budget}
\end{figure}

The different contributions to the KHM equation are portrayed in Fig. \ref{fig:rho_phi75_budget}. For the total fluctuating field (Fig. \ref{fig:rho_phi75_budget}(a)), the effect of the density contrast between the two phases is perceptible on the surface tension term $\langle \mathcal{S}\rangle_\mathbb{T}$ which appears to decrease in amplitude and slides towards smaller scales when $R_\rho$ increases. These variations are quite limited though. We further confirm that the transfer associated to the non-linear transport $\langle \mathcal{T}\rangle_\mathbb{T}$ is much weaker than in single-phase turbulence. The overall transfer of kinetic energy $\langle \mathcal{T}\rangle_\mathbb{T}$ + $\langle \mathcal{S}\rangle_\mathbb{T}$ is substantially larger than single-phase turbulence and is shifted towards smaller scales. This is compensated by a larger contribution of the viscous term in the small scale region. The scale-by-scale contributions of the 'density correction' terms $\langle \mathcal{C}(\vect{a}) \rangle_\mathbb{T}$ (Fig. \ref{fig:rho_phi75_budget}(b)) are similar to what was observed before. They have same sign as the term to which they apply and increase in amplitude with the density contrast $R_\rho$. Hence, these results do not vary substantially from those related to the drop regime.

Scrutinizing the conditionally averaged budget in the liquid or gas phase reveals that $R_\rho$ has a strong effect on the pressure and viscous term (see Fig. \ref{fig:rho_phi75_budget}(d)). In the gas phase, pressure transport appear to act as an additional gain in the budget which is compensated by the viscous term. These processes dominate at $R_\rho = 25$ as they reach values roughly 5 times larger than the forcing term. In the liquid phase (see Fig. \ref{fig:rho_phi75_budget}(c)), we note that the statistics are mostly unchanged between $R_\rho = 5$ and $R_\rho = 25$. For $R_\rho  \geq 5$ and $\alpha = 75\%$, the mass of the liquid represent more than 90\% of the total mass. Hence, in the bubble regime, one can readily assume that the dynamics of the liquid phase dominate the whole flow, and thus increasing further $\rho_L$ has only a limited impact. Only the case with $R_\rho = 1$ differ from the two other set of curves. The transfer of kinetic energy in the liquid phase due to the non-linear transport is systematically smaller than in single-phase turbulence. This effect might be the consequence of the smaller $R_\lambda$ in multiphase turbulence compared to its single-phase counterpart (see Table \ref{tab:dns_params}). The Reynolds number $R_\lambda$ does not vary much when $R_\rho$ varies.

\subsection{Effect of the liquid volume fraction $\alpha$} \label{sec:alpha}

\begin{figure}
  \includegraphics[width=\textwidth]{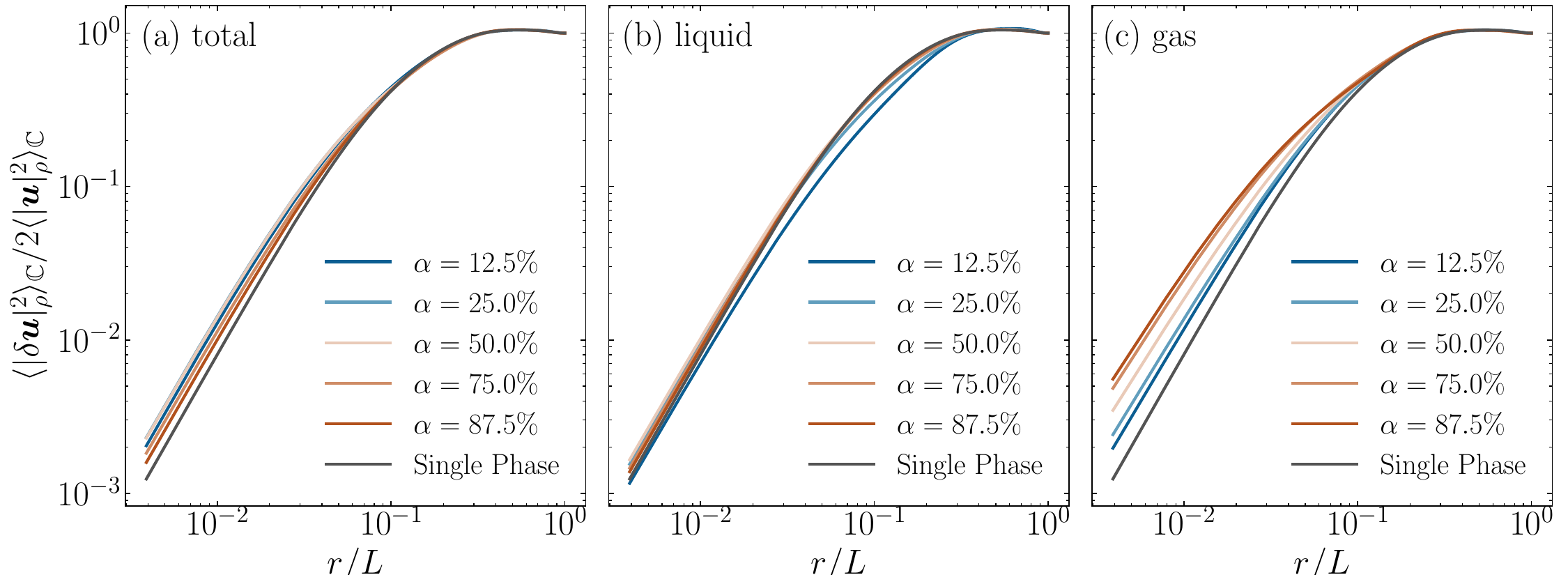}
  \caption{Effect of the liquid volume fraction $\alpha$ on the scale-by-scale kinetic energy. The colours from blue to red correspond to $\alpha = 12.5, ~25, ~50, ~75, ~87.5 \%$. (a) $\mathbb{C} = \mathbb{T}$, (b) $\mathbb{C} = \mathbb{L}$ and (c) $\mathbb{C} = \mathbb{G}$. }\label{fig:phi_rho25_dq2}
\end{figure}

The analysis is now repeated to emphasize the effect of the liquid volume fraction $\alpha$. We start by investigating the scale-by-scale kinetic energy which is plotted in Fig. \ref{fig:phi_rho25_dq2}(a-c) for $\mathbb{C} \equiv \mathbb{T}, ~\mathbb{L} {\rm ~and~} \mathbb{G}$, respectively. We note that the liquid volume fraction has a non-monotonic effect on the total kinetic energy distribution across scales. Indeed, scrutinizing the small-scale region in Fig. \ref{fig:phi_rho25_dq2}(a) reveals that $\langle |\delta \vect{u}|^2_\rho\rangle_\mathbb{T}$ first increase when $\alpha$ grows from 12.5\% to 50\% before decreasing again for $\alpha$ from 50\% to 87.5\%. The scale distribution of kinetic energy in multiphase turbulence tends towards the one of single-phase turbulence when $\alpha \to 1$, i.e. when the liquid drives the whole flow dynamics.

The kinetic energy scale distribution in the liquid (Fig. \ref{fig:phi_rho25_dq2}(b)) shows that the case with $\alpha = 12.5\%$ departs significantly from the different curves. At such a small liquid volume fraction, some confinement effects are likely to be at play which precludes turbulent eddies of size larger than the typical size of liquid structures to develop, thereby impeding the turbulent activity in the liquid phase. The scale distribution of kinetic energy in the liquid for the other values of $\alpha$ is roughly invariant. The scale-by-scale turbulent kinetic energy conditioned by the gas phase is presented in Fig. \ref{fig:phi_rho25_dq2}(c). The contribution of the small-scales to the kinetic energy is observed to increase with $\alpha$. Here again, the scenario proposed above stating that the liquid phase having much larger inertia yields an increased production of vorticity and thus small-scale kinetic energy in the gas phase is likely to hold true. Hence, the more liquid, the more kinetic energy production at small scales in the gas phase. Despite this increased activity of the small scales in the gas phase, table \ref{tab:dns_params} indicates that the total kinetic dissipation rate $\varepsilon_f$ is maximum for $\alpha = 50\%$. This is obviously due to the decreasing proportion in mass and volume of the gas phase when $\alpha$ increases. 

\begin{figure}
  \begin{tabular}{l l l l}
    (a) & (b) & (c)\\
    \includegraphics[trim={2cm 0 2cm 0},clip, width=0.29\textwidth]{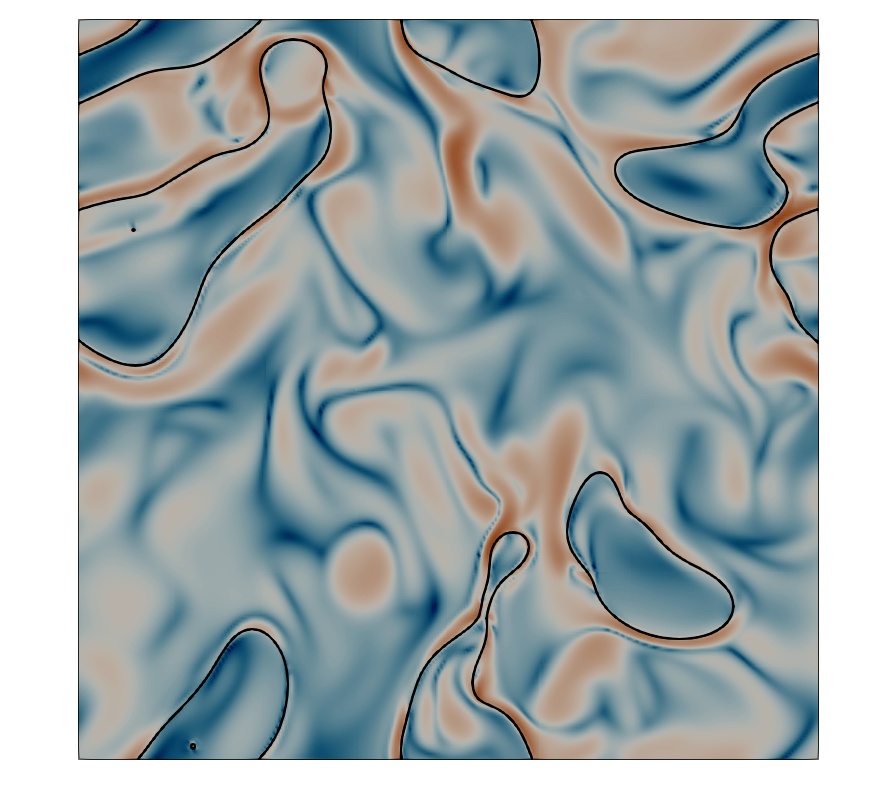}  & 
    \includegraphics[trim={2cm 0 2cm 0},clip, width=0.29\textwidth]{./phi25.0_rho025_sig0.52_mu00140_256_t4} & 
    \includegraphics[trim={2cm 0 2cm 0},clip, width=0.29\textwidth]{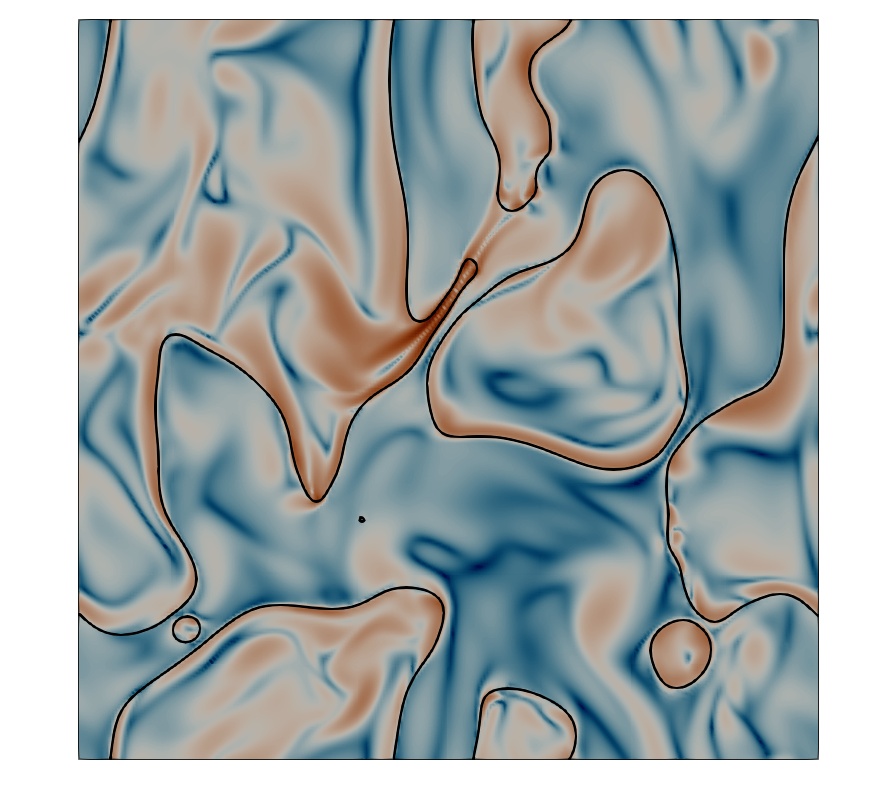} &
    \includegraphics[width=0.062\textwidth]{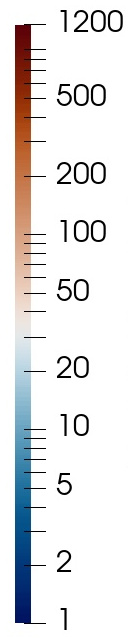} \\
    (d) & (e) \\
    \includegraphics[trim={2cm 0 2cm 0},clip, width=0.29\textwidth]{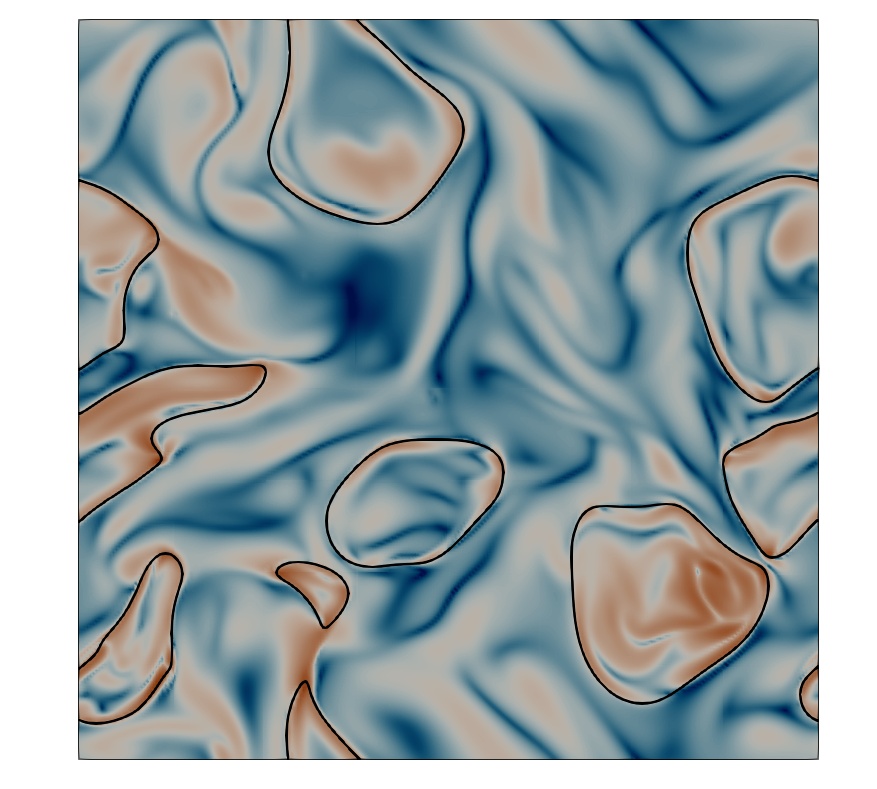} & 
    \includegraphics[trim={2cm 0 2cm 0},clip, width=0.29\textwidth]{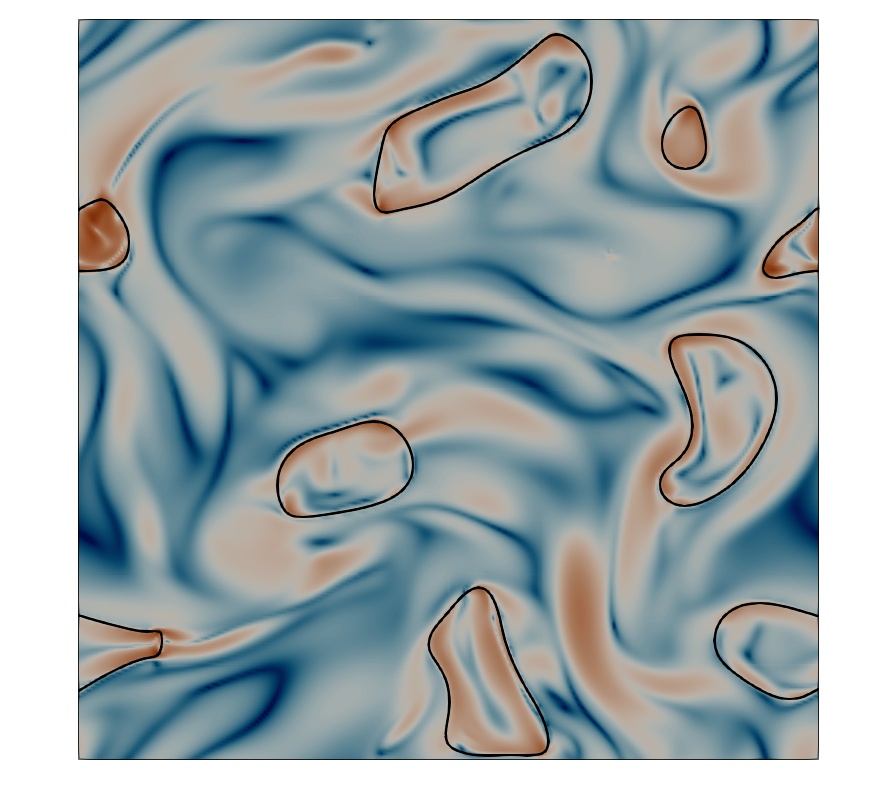}
    & \includegraphics[width=0.062\textwidth]{./vert_1-1200_2.jpeg} \\
  \end{tabular}
  \caption{2D slices of the vorticity magnitude (in colour, units of $u'/L$) together with the liquid-gas interface (black curves) for increasing $\alpha$ from (a) to (e).} \label{fig:phi_rho25_visu}
\end{figure}

The visualizations presented in Fig. \ref{fig:phi_rho25_visu} do not really show any substantial difference of the vorticity magnitude when $\alpha$ changes. However, when $\alpha \to 1$, the volume of the gas phase which lies close to the interface increases relatively to the total gas volume. Hence, the volume where the vorticity magnitude is larger increases in proportion. This is likely to explain why $\langle |\delta \vect{u}|^2_\rho\rangle_\mathbb{G}$ increases in the small-scale range when the liquid volume fraction increases. 

\begin{figure}
  \includegraphics[width=\textwidth]{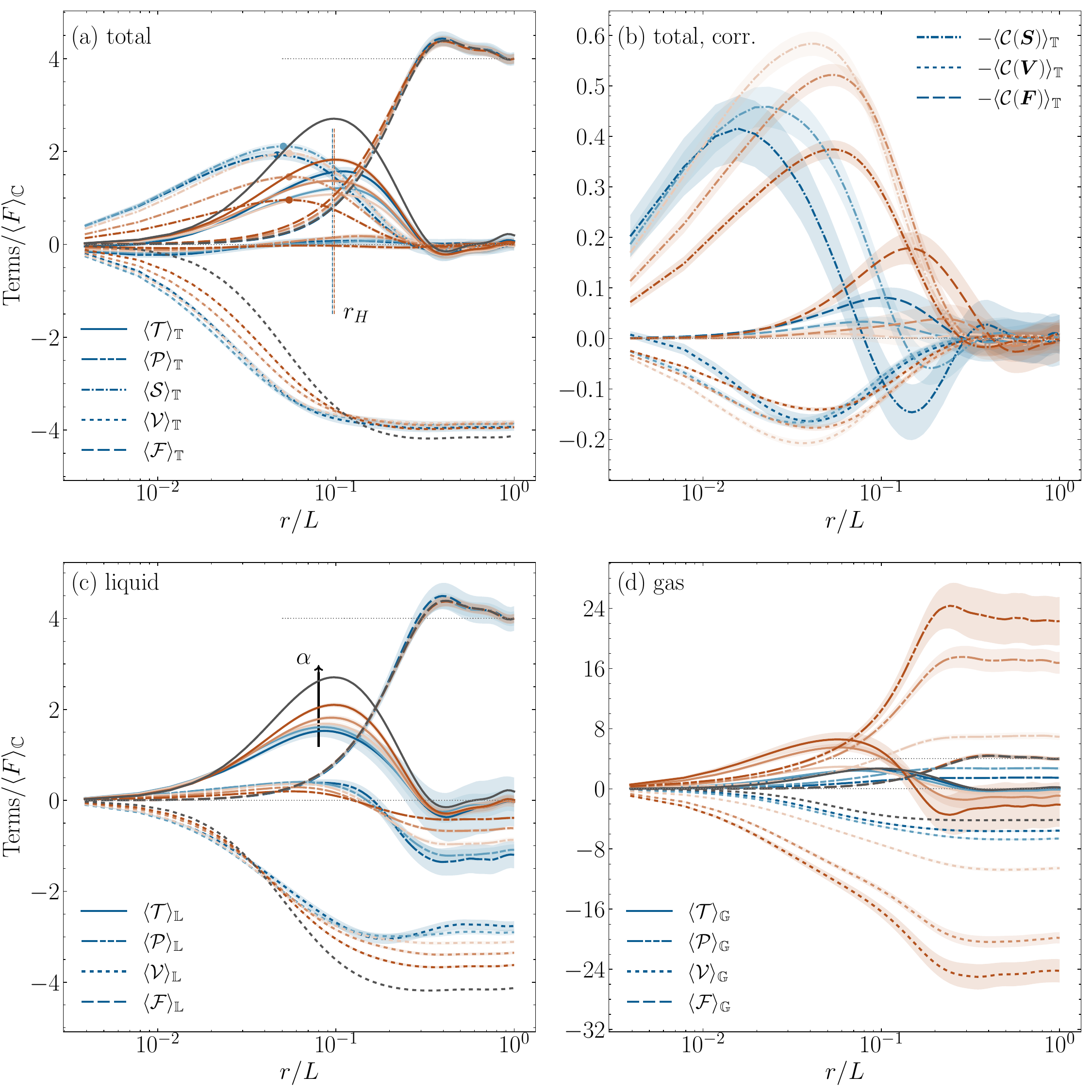}
  \caption{Effect of the liquid volume fraction $\alpha$ on the scale-by-scale kinetic energy budgets. The colours from blue to red correspond to $\alpha = 12.5, ~25, ~50, ~75, ~87.5 \%$ as in the legend of Fig. \ref{fig:phi_rho25_dq2}, while the black curves correspond to the single-phase case. The different lines correspond to \Ltransport~$\langle \mathcal{T}\rangle_\mathbb{C}$, \Lpressure~$\langle \mathcal{P}\rangle_\mathbb{C}$, \Lsurface~$\langle \mathcal{S}\rangle_\mathbb{C}$, \Lviscous~$\langle \mathcal{V}\rangle_\mathbb{C}$, \Lforcing~$\langle \mathcal{F}\rangle_\mathbb{C}$, with (a) $\mathbb{C} = \mathbb{T}$, (c) $\mathbb{C} = \mathbb{L}$ and (d) $\mathbb{C} = \mathbb{G}$. The horizontal dashed line indicates $4\langle F\rangle_\mathbb{C}$, the limit at large separations of the forcing term. Figure (b) represents correction terms $\mathcal{C}(\vect{a})$} \label{fig:phi_rho25_budget}
\end{figure}

The different terms of the KHM equation for different volume fractions are presented in Fig. \ref{fig:phi_rho25_budget}. Figure \ref{fig:phi_rho25_budget}(a) indicates that the unconditionally averaged terms of the KHM equation corresponding to a given $\alpha$ (e.g. 12.5 and 25\%) differ greatly from those corresponding to an equivalent $1-\alpha$ (i.e. 87.5 and 75\%). For instance, the surface tension term $\langle \mathcal{S} \rangle_\mathbb{T}$ for $\alpha = 12.5\%$ appears to be much larger than the one for $\alpha = 87.5\%$, even though the surface area $A_\Gamma$ of the interface is similar (see Table \ref{tab:dns_params}). The transfer due to the non-linear transport $\langle \mathcal{T} \rangle_\mathbb{T}$ has also a non-monotonic evolution with respect to the liquid volume fraction. Nevertheless, it seems to tend towards its single phase counterpart when $\alpha \to 1$. The overall transfer of kinetic energy which is due to both the non-linear transport and surface tension term is compensated by the viscous term $\langle \mathcal{V} \rangle_\mathbb{T}$. The latter is shifted towards larger scales when $\alpha \geq 50 \%$. Here again, the term $\langle \mathcal{V} \rangle_\mathbb{T}$ tends towards its single-phase counterpart when $\alpha \to 1$.

The contribution of the density correction terms $\langle \mathcal{C}(\vect{a})\rangle_\mathbb{T}$ is presented in Fig. \ref{fig:phi_rho25_budget}(b). Except for the forcing term, we note that the correction terms are maximum when $\alpha = 50\%$. This suggests that $\langle \mathcal{C}(\vect{a})\rangle_\mathbb{T}$ should be related to some extent to the surface area of the liquid-gas interface which is minimum for $\alpha = 12.5\% {\rm ~and~} 87.5\%$ and maximum at $\alpha = 50\%$. The contribution of $\langle \mathcal{C}(\vect{S})\rangle_\mathbb{T}$ also peaks at very different scales. Its contribution is perceptible at small scales when $\alpha$ is small and systematically moves towards the large-scales when $\alpha$ becomes larger. For the viscous term, the correction appears to peak around the same scale, independently of $\alpha$.

The terms of the KHM equation conditionally averaged in the liquid and gas phase are presented in Figs. \ref{fig:phi_rho25_budget}(c) and (d), respectively. For the liquid phase, we note that when $\alpha \to 1$, the contribution of the pressure transport term $\langle \mathcal{P} \rangle_\mathbb{L}$ tends to zero and the other terms (the viscous $\langle \mathcal{V} \rangle_\mathbb{L}$ and non-linear transport term $\langle \mathcal{T} \rangle_\mathbb{L}$) tends towards those obtained in single-phase flows. A careful examination of Fig. \ref{fig:phi_rho25_budget}(d) reveals that the same holds true for the gas phase when $\alpha \to 0$. For small values of $\alpha$ (i.e. the drop regime), the pressure term in the liquid phase $\langle \mathcal{P} \rangle_\mathbb{L}$ contributes negatively at large scales which means that it represents a loss for the liquid kinetic energy budget at those scales. On the other hand, the budget in the gas phase (Fig. \ref{fig:phi_rho25_budget}(d)) indicates that the pressure term $\langle \mathcal{P} \rangle_\mathbb{G}$ contributes positively and very substantially when $\alpha \to 1$, i.e. in the bubble regime. Note the amplitude of the pressure term $\langle \mathcal{P} \rangle_\mathbb{G}$ for $\alpha = 87.5\%$ which is roughly 6 times larger than the forcing term $\langle \mathcal{F} \rangle_\mathbb{G}$. This means that when the liquid-volume fraction is large, most of the kinetic energy production in the gas phase comes from pressure transport. This process is compensated by an increased contribution from the viscous term $\langle \mathcal{V} \rangle_\mathbb{G}$.

In summary, varying the liquid volume fraction reveals that the physics of the drop regime ($\alpha < 50\%$) differs significantly from the one of the bubble regime ($\alpha > 50\%$). This stresses again that the results obtained by setting $R_\rho = 1$ can very hardly be representative of what occurs in 'real' multiphase flows characterized by fluctuations, even moderate, of the fluid density. Our results also reveal that the budget in the most abundant phase, when the minority phase represents less than say 5\%, could possibly correspond to the one observed in single phase turbulence.

\subsection{Effect of the Reynolds number $Re$} \label{sec:Re}

\begin{figure}
  \includegraphics[width=\textwidth]{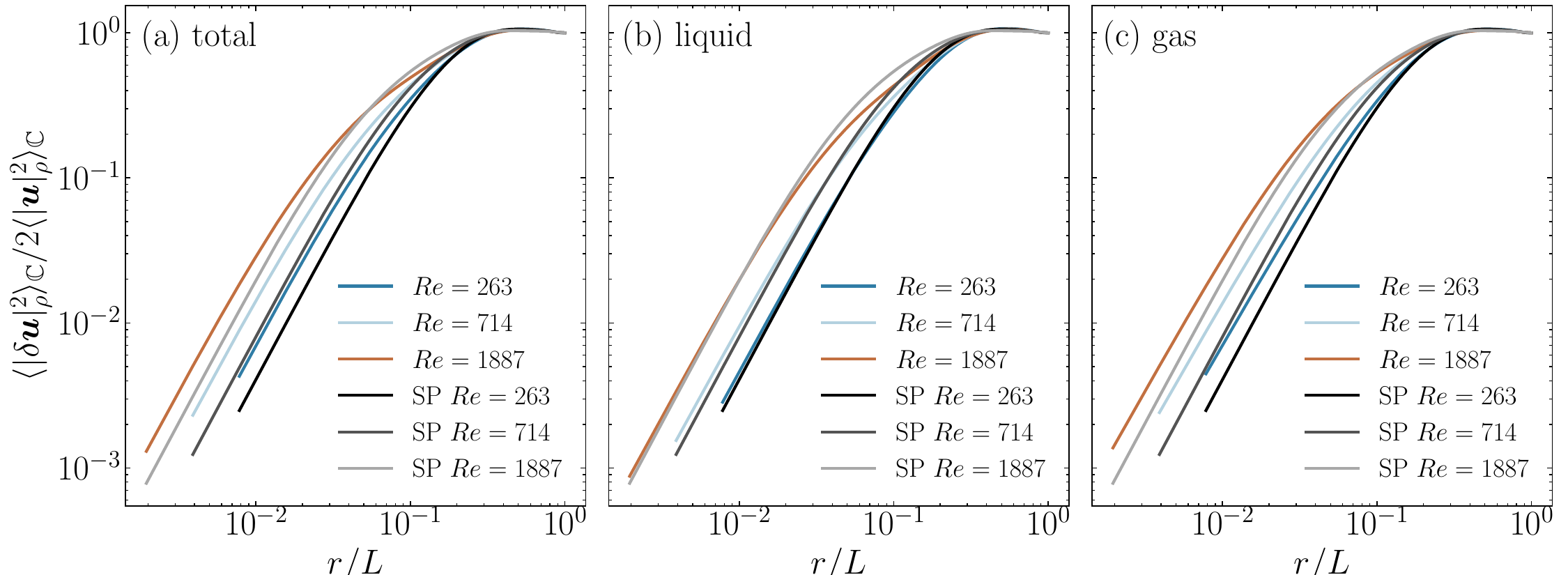}
  \caption{Effect of the Reynolds number $Re  \equiv 1/\nu$ on the scale-by-scale kinetic energy. The colours from blue to red correspond to $Re = 263, 714, 1887$. The grey curves correspond to the single-phase (SP) data at same Reynolds number. (a) $\mathbb{C} = \mathbb{T}$, (b) $\mathbb{C} = \mathbb{L}$ and (c) $\mathbb{C} = \mathbb{G}$. } \label{fig:nu_rho25_dq2}
\end{figure}

\begin{figure}
  \begin{tabular}{l l l l}
    (a) & (b) & (c)\\
    \includegraphics[trim={2cm 0 2cm 0},clip, width=0.29\textwidth]{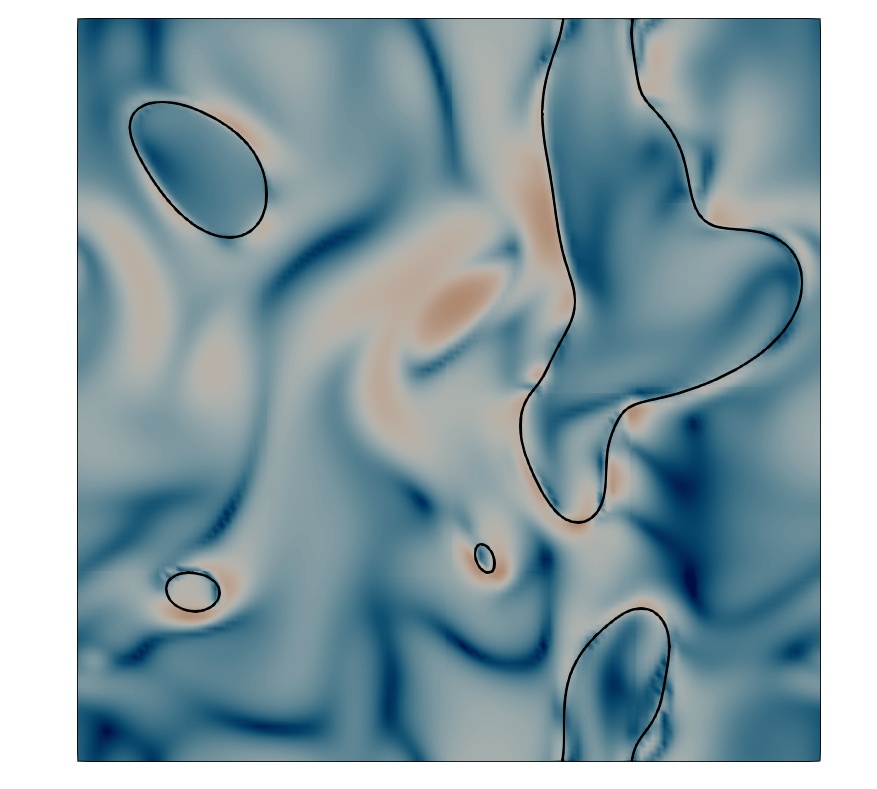}  & 
    \includegraphics[trim={2cm 0 2cm 0},clip, width=0.29\textwidth]{./phi25.0_rho025_sig0.52_mu00140_256_t4} & 
    \includegraphics[trim={2cm 0 2cm 0},clip, width=0.29\textwidth]{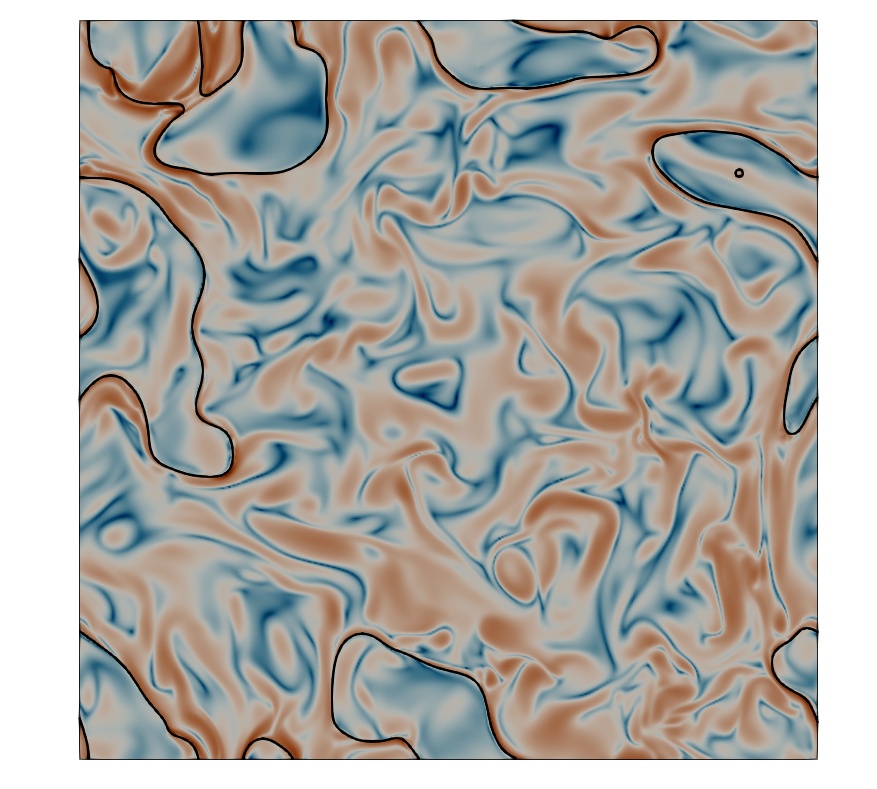} &
    \includegraphics[width=0.062\textwidth]{./vert_1-1200_2.jpeg} \\
  \end{tabular}
    \caption{2D slices of the vorticity magnitude (in colour, units of $u'/L$) together with the liquid-gas interface (black curves) for increasing $Re$ from (a) to (c).} \label{fig:nu_rho25_visu} 
\end{figure}

The effect of Reynolds number $Re$ is studied independently of the other parameters by varying the kinematic viscosity (see Table \ref{tab:dns_params}). The kinetic energy distribution among the different scales is plotted in Fig. \ref{fig:nu_rho25_dq2}. Decreasing the value of the kinematic viscosity shifts the viscous cut-off towards the small scales. The distribution of kinetic energy is thus spread over a wider range of scales and the small scales are more energetic when the Reynolds number increases. Here again, we further note that for $\mathbb{C}=\mathbb{T}$ (Fig. \ref{fig:nu_rho25_dq2}(a)), the small-scale kinetic energy of multiphase turbulence is larger than its single-phase counterpart at the same Reynolds number. As with Fig. \ref{fig:rho_phi25_dq2}(a), the distribution conditioned in the liquid is very comparable to single-phase turbulence, only the distribution in the gas phase deviates significantly. Hence, the larger kinetic energy observed for $\mathbb{C}=\mathbb{T}$ at small scales is due to a larger kinetic energy in the gas phase.

The visualizations of the vorticity field presented in Fig. \ref{fig:nu_rho25_visu} indicate again that these small scales are produced in the gas phase in close vicinity of the interface. When the Reynolds number is increased, there are more vortical structures, they are of smaller size and their magnitude increases. This is reflected in the increase of $\langle |\delta \vect{u}|^2_\rho \rangle_\mathbb{T}$ and $\langle |\delta \vect{u}|^2_\rho \rangle_\mathbb{G}$ at small-scales. Despite this, the amount of surface area of the two-fluid interface appears to be independent of the Reynolds number (see Table \ref{tab:dns_params}). 

\begin{figure}
  \includegraphics[width=\textwidth]{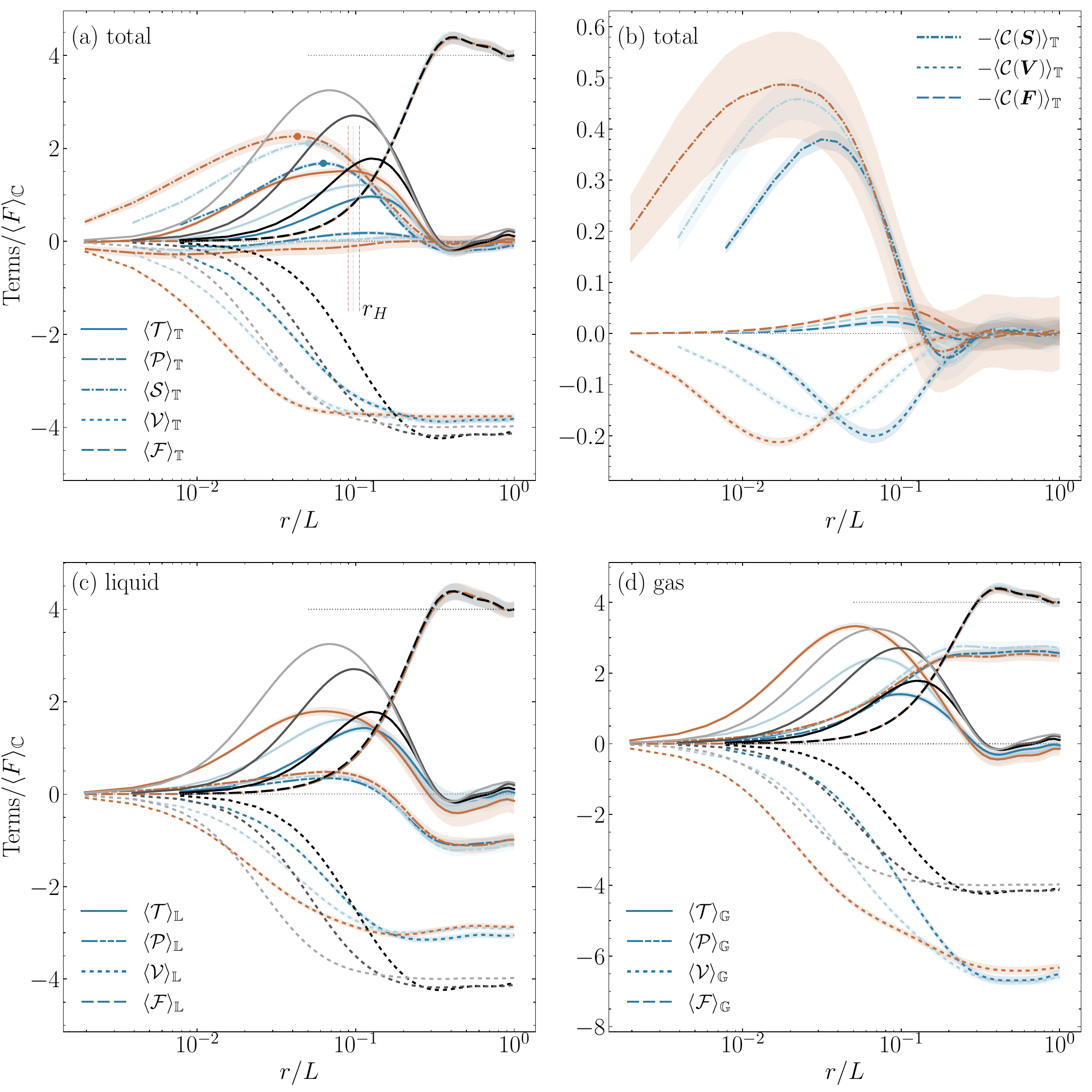}
  \caption{Effect of the Reynolds number $Re \equiv 1/\nu$ on the scale-by-scale kinetic energy budgets. The colours from blue to red correspond to $Re = 263, 714, 1887$ as in the legend of Fig. \ref{fig:nu_rho25_dq2}. The grey curves correspond to the single-phase data at same Reynolds number. The grey curves correspond to the single-phase data at same Reynolds number. The different lines correspond to \Ltransport~$\langle \mathcal{T}\rangle_\mathbb{C}$, \Lpressure~ $\langle \mathcal{P}\rangle_\mathbb{C}$, \Lsurface~ $\langle \mathcal{S}\rangle_\mathbb{C}$, \Lviscous~ $\langle \mathcal{V}\rangle_\mathbb{C}$, \Lforcing~ $\langle \mathcal{F}\rangle_\mathbb{C}$, with (a) $\mathbb{C} = \mathbb{T}$, (c) $\mathbb{C} = \mathbb{L}$ and (d) $\mathbb{C} = \mathbb{G}$. The horizontal dashed line indicates $4\langle F\rangle_\mathbb{C}$, the limit at large separations of the forcing term. Figure (b) represents the density correction terms $\mathcal{C}(\vect{a})$} \label{fig:nu_rho25_budget}
\end{figure}

The terms of KHM equation are displayed in Fig. \ref{fig:nu_rho25_budget} for (a,b) the total fluctuating field, (c) and (d) the liquid and gas phase, respectively. For the total fluctuating field (Fig. \ref{fig:nu_rho25_budget}(a)), decreasing the viscosity increases the separation between large and small scales and thus the intermediate scales develop. The viscous term is shifted towards smaller scales and the terms acting at the intermediate range, i.e. $\langle \mathcal{T}\rangle_\mathbb{T}$ and $\langle \mathcal{S}\rangle_\mathbb{T}$ widen and increase in amplitude. This is qualitatively equivalent to what occurs in single-phase turbulence. The same observation stands true for the additional term due to density variations $\langle \mathcal{C}(\vect{a})\rangle_\mathbb{T}$ presented in Fig. \ref{fig:nu_rho25_budget}(b). The overall transfer of kinetic energy which is composed of both the non-linear transport and the surface tension term, exceeds the one of single-phase turbulence at the same Reynolds number. As observed before, this is compensated by the viscous term being active at smaller scales. The pressure term does not contribute much to the budget of $\langle |\delta \vect{u}|^2_\rho \rangle_\mathbb{T}$.

Looking at the terms of the KHM equation in liquid and gas phase (Fig. \ref{fig:nu_rho25_budget}(c,d)), it is noted that $\langle \mathcal{T}\rangle_\mathbb{L,G}$ increases in amplitude and peaks at smaller scales when the Reynolds increases. Its contribution is quite smaller to that of single-phase turbulence. Surprisingly, the pressure term appears to be independent of the Reynolds number. This applies to both the liquid and gas phase. Pressure transport is again found to contribute positively in the gas phase. For the liquid phase, it is a gain at small scales and a loss at large scales.

\subsection{Effect of the Weber number $We$} \label{sec:We}

\begin{figure}
  \includegraphics[width=\textwidth]{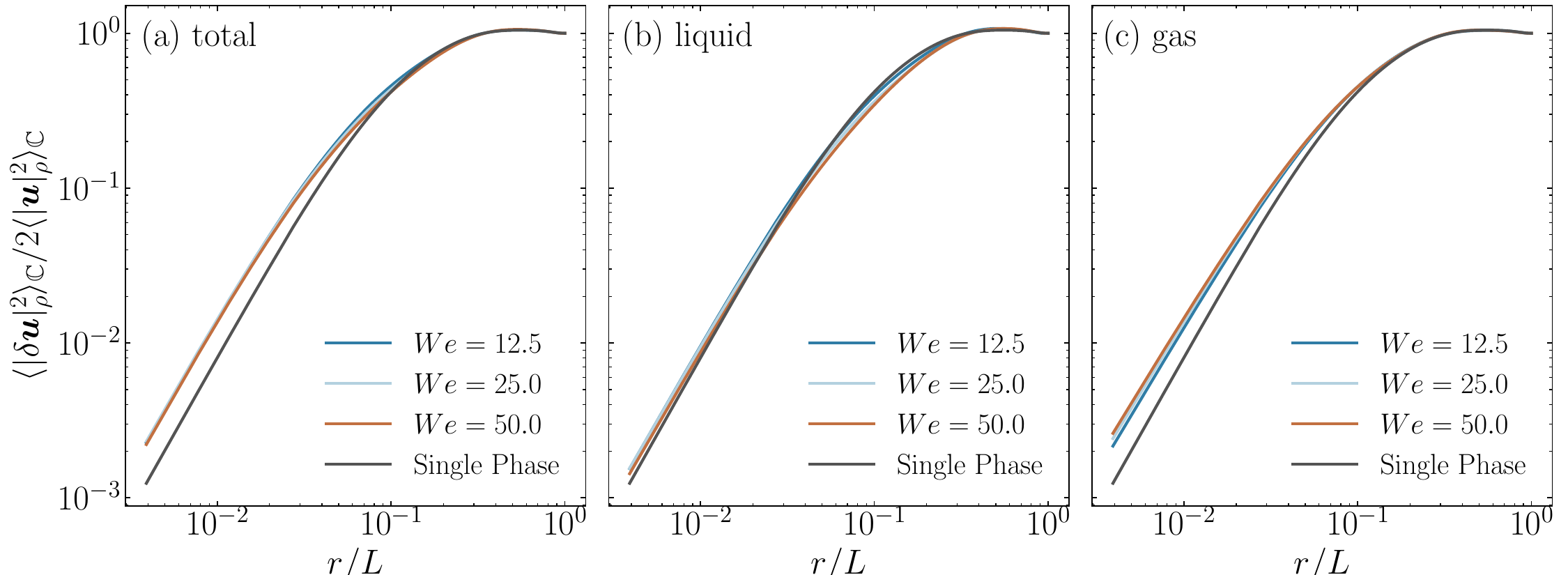}
  \caption{Effect of the Weber number $We  \equiv \overline{\rho}/\sigma$ on the scale-by-scale kinetic energy. The colours from blue to red correspond to $We = 12.5, ~25, ~50$, while the black curves correspond to the single-phase case. (a) $\mathbb{C} = \mathbb{T}$, (b) $\mathbb{C} = \mathbb{L}$ and (c) $\mathbb{C} = \mathbb{G}$. } \label{fig:sigma_rho25_dq2}
\end{figure}

\begin{figure}
  \begin{tabular}{l l l l}
    (a) & (b) & (c)\\
    \includegraphics[trim={2cm 0 2cm 0},clip, width=0.29\textwidth]{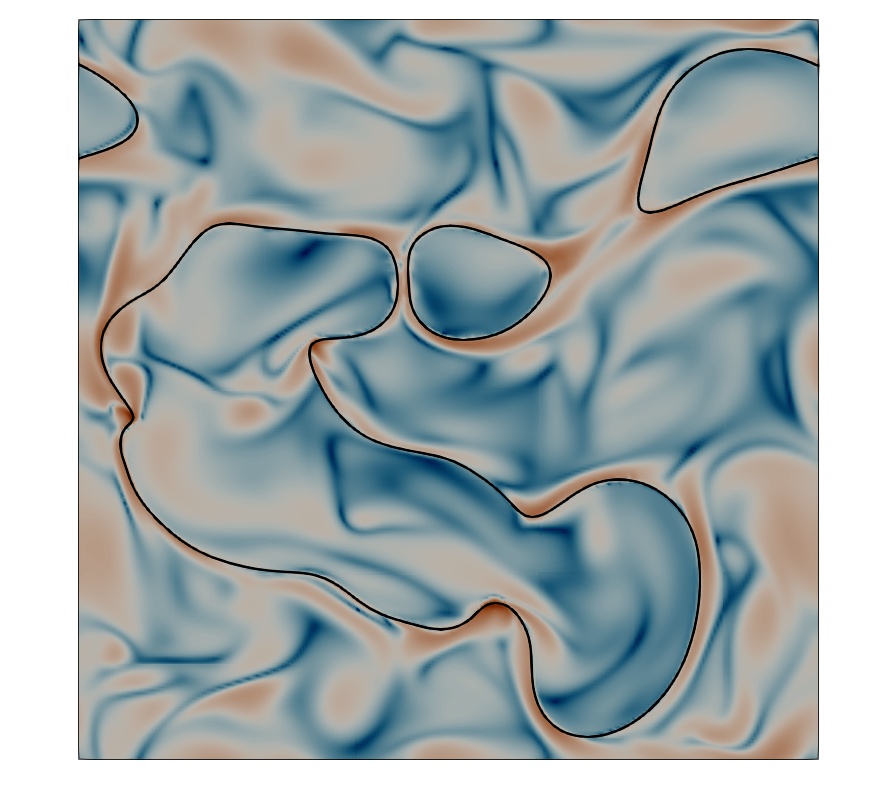}  & 
    \includegraphics[trim={2cm 0 2cm 0},clip, width=0.29\textwidth]{./phi25.0_rho025_sig0.52_mu00140_256_t4} & 
    \includegraphics[trim={2cm 0 2cm 0},clip, width=0.29\textwidth]{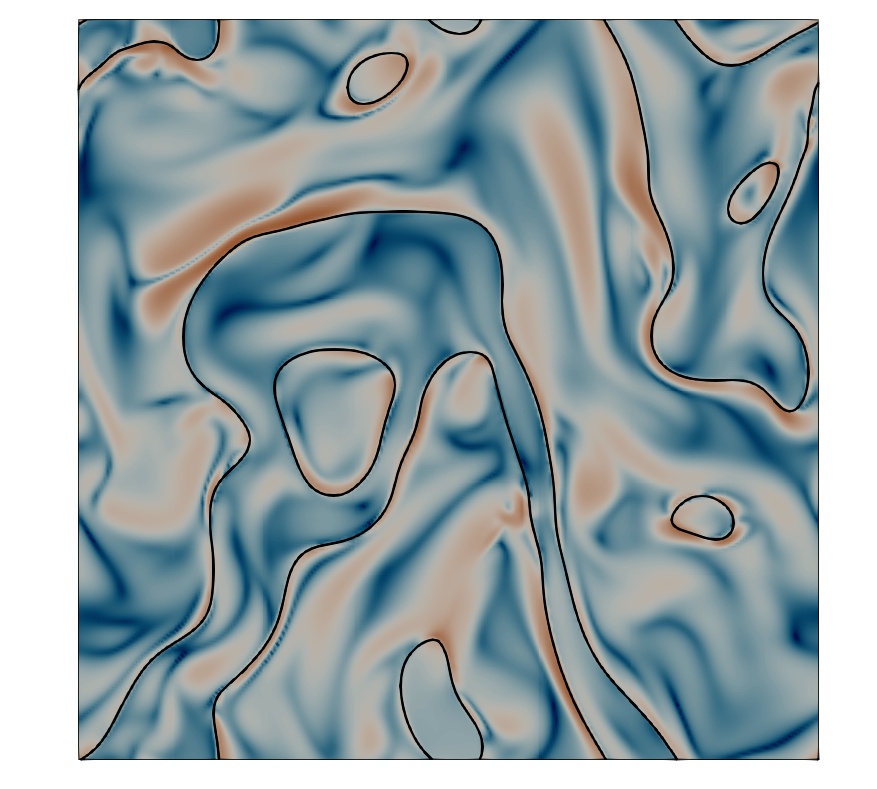} &
    \includegraphics[width=0.062\textwidth]{./vert_1-1200_2.jpeg} \\
    % (c)\\
    % \includegraphics[width=0.0855\textwidth]{./vert_1-1200_2.jpeg} \\
  \end{tabular}
    \caption{2D slices of the vorticity magnitude (in colour, units of $u'/L$) together with the liquid-gas interface (black curves) for increasing $We$ from (a) to (c).} \label{fig:sigma_rho25_visu} 
\end{figure}

We now investigate the effect of Weber number $We$ with $R_\rho = 25$, $\alpha = 25\%$ and $Re = 714$. The scale-by-scale kinetic energy $\langle |\delta \vect{u}|^2_\rho\rangle_\mathbb{C}$ is displayed in Fig. \ref{fig:sigma_rho25_dq2}(a), (b) and (c) for $\mathbb{C} \equiv \mathbb{T}, ~\mathbb{L}, ~{\rm and}~ \mathbb{G}$, respectively.

As an overall observation, the scale-by-scale kinetic energy appears to vary only marginally when the surface tension is varied. One observes though some very slight modifications in the intermediate range of scales for the unconditionally averaged kinetic energy $\langle |\delta \vect{u}|^2_\rho\rangle_\mathbb{T}$ (Fig. \ref{fig:sigma_rho25_dq2}(a)). The small-scales remain more or less unchanged consistently with the values of $\varepsilon_f$ reported in Table \ref{tab:dns_params}. The same small variations in the intermediate range of scales are also visible in the liquid phase in Fig. \ref{fig:sigma_rho25_dq2}(b). In the gas phase (Fig. \ref{fig:sigma_rho25_dq2})(c), the distribution of kinetic energy remain independent of the Weber number in the large and intermediate range of scales. Some marginal differences appear when one travels towards smaller scales where $\langle |\delta \vect{u}|^2_\rho\rangle_\mathbb{G}$ appears to increase very slightly with $We$. Looking at $\langle |\delta \vect{u}|^2_\rho\rangle_\mathbb{T}$ in Fig. \ref{fig:sigma_rho25_dq2}(a), the small-scale kinetic energy of multiphase turbulence is again observed to be larger than the one pertaining to single-phase turbulence. This results from an increased small-scale activity in the gas phase (see Fig. \ref{fig:sigma_rho25_dq2}(c)).

The direct visualizations of the flow presented in Fig. \ref{fig:sigma_rho25_visu} indicate that the magnitude of the vorticity field does not vary substantially when $We$ is varied. The increase of the Weber yields more pronounced corrugations of the interface and hence a larger surface area (see also Table \ref{tab:dns_params}). The theory of \cite{Terrington2022} indicates that one particular term of the vorticity production is proportional to the surface tension $\sigma$ and the gradient along the interface of the mean curvature. Hence, when the Weber number increases, the decrease of surface tension is likely compensated by larger gradients of curvature (more corrugations), leading to roughly the same  vorticity magnitude, and hence a comparable small-scale kinetic energy distribution in the gas phase.

\begin{figure}
  \includegraphics[width=\textwidth]{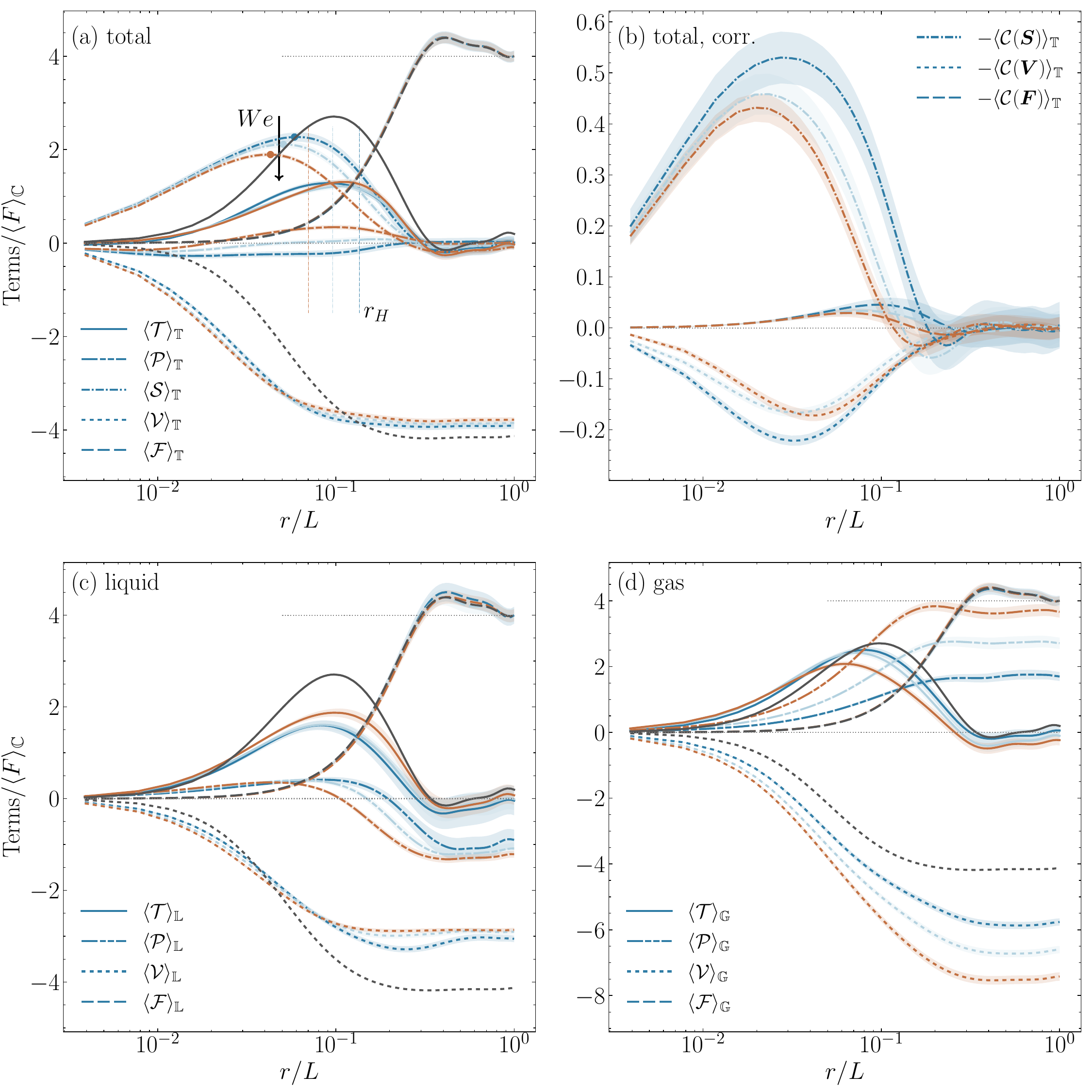}
  \caption{Effect of the Weber number $We  = \overline{\rho}/\sigma$ on the scale-by-scale kinetic energy budgets. The colours from blue to red correspond to $We = 12.5, ~25, ~50$ as in the legend of Fig. \ref{fig:sigma_rho25_dq2}, while the black curves correspond to the single-phase case. The different lines correspond to 
  \Ltransport~$\langle \mathcal{T}\rangle_\mathbb{C}$, \Lpressure~ $\langle \mathcal{P}\rangle_\mathbb{C}$, \Lsurface~ $\langle \mathcal{S}\rangle_\mathbb{C}$, \Lviscous~ $\langle \mathcal{V}\rangle_\mathbb{C}$, \Lforcing~ $\langle \mathcal{F}\rangle_\mathbb{C}$, with (a) $\mathbb{C} = \mathbb{T}$, (c) $\mathbb{C} = \mathbb{L}$ and (d) $\mathbb{C} = \mathbb{G}$. The horizontal dashed line indicates $4\langle F\rangle_\mathbb{C}$, the limit at large separations of the forcing term. Figure (b) represents the density correction terms $\mathcal{C}(\vect{a})$} \label{fig:sigma_rho25_budget}
\end{figure}

The effect of surface tension on the scale-by-scale kinetic energy budget is presented in Fig. \ref{fig:sigma_rho25_budget}. Scrutinizing the unconditionally averaged terms (Fig. \ref{fig:sigma_rho25_budget}(a)), we note that the surface tension term $\langle \mathcal{S}\rangle_\mathbb{T}$ drops when surface tension decreases. It is thus plausible that the scale-by-scale contribution of the surface tension term will tend to zero at infinite Weber number. The Weber number further appears to have an influence on the pressure term $\langle \mathcal{P}\rangle_\mathbb{T}$, whose contribution changes sign when the Weber changes. This behaviour is rather surprising. The two other terms of Eq. \eqref{eq:KHM_general} due to the non-linear transport $\langle \mathcal{T}\rangle_\mathbb{T}$ and the viscous term $\langle \mathcal{V}\rangle_\mathbb{T}$ are only marginally influenced by variations of the surface tension. However, they are substantially different from the ones obtained in single-phase flows. For instance, the viscous term starts to contribute at much smaller scales to compensate an increased overall transfer of kinetic energy, which is again much larger in multiphase than in single-phase flows.

The density correction terms are portrayed in Fig. \ref{fig:sigma_rho25_budget}(b). As observed previously, we note that $-\langle \mathcal{C}(\vect{S})\rangle_\mathbb{T}$, $-\langle \mathcal{C}(\vect{V})\rangle_\mathbb{T}$ and $-\langle \mathcal{C}(\vect{F})\rangle_\mathbb{T}$ have same sign as $\langle \mathcal{S}\rangle_\mathbb{T}$, $\langle \mathcal{V}\rangle_\mathbb{T}$ and $\langle \mathcal{F}\rangle_\mathbb{T}$, respectively. The contribution of these terms seem to decrease when the Weber number increases. 

The terms of Eq. \eqref{eq:KHM_general} conditionally averaged budget in the liquid and gas phase are displayed in Fig. \ref{fig:sigma_rho25_budget}(c) and (d), respectively. Noticeable is the evolution of the pressure transport term in the liquid phase $\langle \mathcal{P}\rangle_\mathbb{L}$ when the surface tension is varied. The latter is positive at small scales (a gain) and negative (a loss) at large scales. The scale at which the change of sign occurs decreases with surface tension. The effect of surface tension is also clearly visible on the pressure term in the gas phase, $\langle \mathcal{P}\rangle_\mathbb{G}$, which appears to increase when the Weber number increases. Its contribution to the scale-by-scale budget is positive (a gain) and is compensated mainly by an increased contribution of the viscous term $\langle \mathcal{V}\rangle_\mathbb{G}$.

\subsection{Phenomenological predictions for the pivotal scale}

So far, we have shown that surface tension yields an additional transfer term of kinetic energy between scales. Of particular interest is the scale noted $r_\mathcal{S}$ at which the term $\langle \mathcal{S} \rangle_\mathbb{T}$ is maximum. Indeed, for scales larger than $r_\mathcal{S}$, the interface pumps kinetic energy, stores it as surface energy before releasing it back into kinetic energy but at scales smaller than $r_\mathcal{S}$. 

As already mentioned in the introduction, \citep{CrialesiEsposito2023a} interpreted $r_\mathcal{S}$ as the pivotal scale between, at large-scales, a range dominated by breakup, and at small scales, a range dominated by coalescence. They thus suggested that $r_\mathcal{S}$ should coincide with the Kolmogorov-Hinze scale, noted $r_H$, which is known as the largest fluid parcel which would resist breakup.

Going back to the original idea of \cite{Hinze1955}, \cite{Mukherjee2019} extracted the Kolmogorov-Hinze scale from a local, i.e. scale dependent Weber number \citep[see also][]{Perlekar2012}. We proceed similarly here, and the local Weber number is defined from the scale-by-scale kinetic energy, viz.
\begin{eqnarray}
  We(r) := \frac{1}{2} \frac{\overline{\rho}}{\langle \rho \rangle_\mathbb{T}} \frac{ \langle |\delta \vect{u}|_\rho^2\rangle_\mathbb{T} r}{\sigma}
\end{eqnarray}
Since $\langle |\delta \vect{u}|_\rho^2\rangle_\mathbb{T} / \langle \rho \rangle_\mathbb{T} \to 2u'^2$ when $r \to L$, this definition of the local Weber number is such that $We(r) \to We$ as $r \to L$. The Kolmogorov-Hinze scale is then simply given as $We(r=r_H) =1$. 

Note that the classical definition for the Kolmogorov-Hinze scale can be recovered if the Reynolds number is sufficiently large for an inertial range to exist and for $r_H$ to fall in such an inertial range. In this situation, Kolmogorov scaling might apply for the Favre-averaged structure functions, viz.
\begin{eqnarray}
  \frac{\langle |\delta \vect{u}|_\rho^2 \rangle_\mathbb{T}}{ \langle \rho \rangle_\mathbb{T}} = C_K(\epsilon_f r )^{{2}/{3}}, \label{eq:We_r}
\end{eqnarray}
where $C_K$ is known as the Kolmogorov constant. This yields the following relation for Kolmogorov-Hinze scale:
\begin{eqnarray}
  r_H = \left(\frac{2}{C_K}\right)^{3/5} \left(\frac{\sigma}{\overline{\rho}}\right)^{3/5} \varepsilon_f^{-2/5} \label{eq:Hinze},
\end{eqnarray}
which is very similar to the classical definition of the Kolmogorov-Hinze scale \citep[see e.g.][]{Hinze1955,Perlekar2012,CrialesiEsposito2022,CrialesiEsposito2023a,Cannon2024}. Defining $r_H$ from Eq. \eqref{eq:We_r} instead of Eq. \eqref{eq:Hinze} applies to cases where the Reynolds number is not sufficiently large for an inertial range scaling to be invoked. Note that \cite{Hinze1955} introduced an {\it ad hoc} critical Weber number, $We_c$, which initially was aimed at accounting for the variations of viscosity. This can readily be done here as well by defining $r_H$ as $We(r=r_H) = We_c$. Note also that Eq. \eqref{eq:Hinze} differs from the classical definition of the Kolmogorov-Hinze scale through the appearance of the mean density $\overline{\rho}$ instead of the density of the most abundant fluid. One can thus think of Eq. \eqref{eq:Hinze} as a likely more general expression for $r_H$ that applies to the dense regime where the notion of carrier versus dispersed phase is irrelevant.

The values for $r_\mathcal{S}$ and $r_H$ are reported in Table \ref{tab:dns_params}. They are also represented in Figs. \ref{fig:rho_phi25_budget}(a), \ref{fig:rho_phi75_budget}(a), \ref{fig:phi_rho25_budget}(a), \ref{fig:nu_rho25_budget}(a), \ref{fig:sigma_rho25_budget}(a) as filled circles and vertical lines, respectively. Comparing the values of $r_\mathcal{S}$ and $r_H$ reveals that the Kolmogorov-Hinze scale $r_H$ does not systematically scale with $r_\mathcal{S}$. This is particularly visible when the density ratio between the two phases is varied. In this situation, $r_H$ is roughly constant while $r_\mathcal{S}$ decreases when $R_\rho$ is varied from 1 to 125. It is also visible when the Weber number is varied. In this situation, both $r_H$ and $r_\mathcal{S}$ decrease when $We$ increases. However, $r_H$ does not vary in proportion of $r_\mathcal{S}$.

The discrepancy between $r_H$ and $r_\mathcal{S}$ was already noted by \citep{CrialesiEsposito2023a} at $R_\rho = 1$ but for varying viscosity ratio $R_\mu$. This difference could possibly be taken into account by using an {\it ad hoc} critical Weber number function of $R_\rho$. One could otherwise seek for another characteristic scale as done below. 

Using some different forms of the scale-by-scale kinetic energy budget equations in gravity driven bubbly flows, \cite{Pandey2020,Pandey2023,Ramirez2024} have shown that the surface tension term has qualitatively the same scale distribution as the one reported here or in \cite{CrialesiEsposito2022,CrialesiEsposito2023a,Cannon2024}. However, the regime of bubbly flows investigated by \cite{Pandey2020,Pandey2023,Ramirez2024} does not reveal any breakup nor coalescence, and thus does not fall into the Kolmogorov-Hinze framework. This means that breakup/coalescence may not be required to explain the behaviour of the surface tension term.

Here, we extend somewhat the phenomenology proposed by \cite{CrialesiEsposito2023a} and replace breakup/coalescence events by the more general mechanisms of interface deformation/restoration. We hypothesize that scales larger than $r_\mathcal{S}$ contribute to the interface \textit{deformation}, thereby loosing kinetic energy which is converted into surface energy. The mechanism of interface \textit{deformation} can, but not necessarily includes breakup. On the other hand, for scales smaller than $r_\mathcal{S}$ surface energy is released back into kinetic energy through interface \textit{restoration}. By \textit{restoration}, we mean that any corrugations of the interface smaller than $r_\mathcal{S}$ are flattened through the action of surface tension. Coalescence can also contribute to the disappearance of scales smaller than $r_\mathcal{S}$, thereby releasing kinetic energy at those scales. Breakup can also contribute to restoration. This is true for instance during the Rayleigh-Plateau instability where the deformed ligament have larger surface area than the formed droplets after breakup. For simplicity, the range of scales $r>r_\mathcal{S}$ will be now referred to as the deformation range, and the range of scales $r<r_\mathcal{S}$, the restoration range. 

In recent work \citep{Thiesset2021,Gauding2022,Thiesset2023}, we introduced a general morphological analysis of corrugated interfaces. It was shown in particular in \cite{Gauding2022,Thiesset2023} that when observed at a sufficiently small scale, an interface appears flat. If observed at larger scales, the interface is seen to be curved though smooth, and can eventually be observed to be fractal at even larger scales. Building on the exact analytical results of \cite{Kirste1962,Frisch1963} \citep[see also][]{Gauding2022,Thiesset2023} a typical scale below which an interface is observed to be flat can be obtained. The latter writes:
\begin{eqnarray}
  r_c = \sqrt{\frac{8}{3\langle H^2 \rangle_s - \langle G \rangle_s}}, \label{eq:dphi2_rc}
\end{eqnarray}
where $H$ and $G$ represent mean curvature and Gaussian curvature, respectively, while $\langle \bullet \rangle_s$ stands for a surface area weighted average. For iso-surfaces formed by passive scalar \cite{Gauding2022} proved that $r_c$ scales with the Batchelor length-scale.

To say it differently, an observer zooming on the interface down to scale $r_c$ will see a flat surface. The curvature of the interface is visible when it is observed with a field of view larger than $r_c$. In other words, the smallest interface corrugation is related to $r_c$. This scale is thus likely to be a good candidate to represent $r_\mathcal{S}$, the pivotal scale between the deformation and the restoration range. The values for $r_c$ computed from Eq. \eqref{eq:dphi2_rc} are reported in Table \ref{tab:dns_params}. Unfortunately, although the above reasoning is appealing on the paper, we find that the variations of $r_c$ do not coincide with the variations $r_\mathcal{S}$. The scale $r_\mathcal{S}$ is systematically smaller and does not vary in proportion of $r_c$. Nevertheless, it is found that $r_c$ coincides rather satisfactorily with $r_H$ so that one can conclude that the Kolmogorov-Hinze framework provides a rather suitable prediction for the smallest corrugation scales of the interface.

\section{Conclusions} \label{sec:conclusions}

In the present study, a KHM equation is derived and used to explore the scale-by-scale kinetic energy exchanges in multiphase turbulence. We consider both classical spatial averages and phase conditional averages. This allows the kinetic energy budget to be inferred in each phase separately. This framework is applied to numerical data of forced homogeneous isotropic turbulence simulated using the \texttt{archer} code. %Particular attention has been paid to the adequate resolution of the simulations which is achieved at the sacrifice of lower Reynolds and Weber numbers compared to what can be found in the literature. 
The database does not restrict to cases with $\rho_L = \rho_G$ and considers different values for the liquid/gas density ratio. The liquid volume fraction, Weber and Reynolds numbers were also varied. The statistics of multiphase turbulence are systematically compared to those of single-phase turbulence at the same Reynolds number. 

From this methodology yield several outcomes. The most important are listed below. 

Firstly, we confirm that the interface modifies the overall energy transfer between scales. The term associated with the surface tension is found to act similarly to an additional transfer term, although it should be more accurately interpreted as the energy conversion between kinetic and surface energy. The physical picture is then that part of the kinetic energy injected into the system is pumped from the large scales to deform the interface. Surface energy stored by the interface is then released back into kinetic energy but at smaller scales. In the meantime, kinetic energy is also transferred to the smaller scales by the classical non-linear transport, although at smaller rates when compared to its single-phase counterpart. Summing these two contributions, the overall rate of kinetic energy transfer/conversion is larger than in single-phase flows and is compensated by a larger contribution of the viscous term at small scales. In this respect, the visualizations of the vorticity magnitude have revealed that the presence of the interface creates more intense small-scale structures in the close vicinity of the interface in the gas phase.

Secondly, conditioning the KHM equation in either the liquid and gas phase have revealed the role played by pressure transport. The latter was found to act as a gain (loss) of scale-by-scale kinetic energy for the lighter (heavier) fluid. In the gas phase, the contribution due to pressure can also severely exceed the one of the forcing mechanism which is used to maintain the flow at statistically steady state. This additional pressure power is compensated by the viscous term. For the liquid-gas mixture also i.e. $\mathbb{C}\equiv\mathbb{T}$, a non-zero contribution arises from the pressure term. We refer to this term as the baropycnal work, following \cite{Aluie2011,Aluie2013}. It is not clear however at this stage whether it is merely an artefact of the Favre averaging or whether it has a genuine physical significance.

Thirdly, we have shown that the physics of bubbles are very different to that of drops. This conclusion arises as soon as one considers different densities between the liquid and gas phases. Hence, the case with $R_\rho=1$ is a very special case that can hardly be extrapolated to situations with some, even moderate, density jumps.

Fourthly, building on previous work, we hypothesize the existence of a pivotal scale, noted $r_\mathcal{S}$, between surface deformation (kinetic energy pumping) at large scales and surface restoration (kinetic energy release) at small scales. These two processes are believed to be slightly more general than those invoked by \cite{CrialesiEsposito2023a} as they may apply to situations with no breakup nor coalescence, e.g. gravity driven bubbly flows. We have sought a suitable prediction for this scale. It appears that the Kolmogorov-Hinze scale $r_H$ may not scale as $r_\mathcal{S}$. Using a morphological analysis of corrugated interfaces, we extract a new characteristic scale, noted $r_c$. The latter is argued to be the smallest scale below which the interface appears flat to an observer. Unfortunately, $r_\mathcal{S}$ and $r_c$ do not scale with each other. However, the Kolmogorov-Hinze scale $r_H$ provides satisfactory prediction for $r_c$.

The present work may also inspire future dedicated studies. For instance, in the present database $r_H$, $r_\mathcal{S}$ and $r_c$ do not vary much, except maybe when the Weber number is varied. A wider range of parameters should thus be considered to provide further evidence on how these different scales evolve. At this stage, our hypothesis stating that $r_\mathcal{S}$ separates the deformation range and restoration range still remains plausible, although no definite proofs are provided yet. Another open question is that of the asymptotic behaviour of the different terms of the KHM equation in the limit of large Reynolds and Weber numbers. Dedicated studies are required to fully capture the behaviour that emerges at higher Reynolds regimes.

The discrepancy between $r_c$ and $r_\mathcal{S}$ recalls that the scale-by-scale contribution of the surface tension term is obviously not only related to the geometrical features of the interface. It depends also on the coupling between the dynamical field (the velocity field), the interface geometry and its kinematics. Consequently, more insights could be provided by looking at the alignment between the interface normal, the velocity vector and/or the strain eigenvectors. Further, because the $\langle\mathcal{S}\rangle_\mathbb{T}$ term is assumed to depend on the creation or destruction of surface area at a given scale, one should probably look at some scale-by-scale (e.g. coarse grained) versions of the transport equation for the surface density.

Our data have also indicated an increased activity of the small scales associated with an increase of $R_\rho$ or $\alpha$. This was reflected in particular in larger vorticity amplitudes in the gas phase close to the interface. More insight could be provided by looking at the vorticity production mechanisms in the line of \cite{Terrington2022}, and references therein. In Appendix \ref{app:limit_r0}, we provide first insights into the relation between the KHM equation and the transport equation for the enstrophy. This could be continued in future analysis in order to shed light onto the vorticity dynamics in multiphase systems.

Finally, we have heretofore investigated the KHM equation for the total, liquid, and gas phase. The case referred to as the mixed case ($\mathbb{C}=\mathbb{M}$) was not considered. This could be done in follow-up study with the aim of shedding light into the energy exchanges between the two phases at a given scale $r$.

\backsection[Acknowledgements]{The present work would not have been feasible without the contribution of the members of the \texttt{archer} team, viz. J-C. Br{\"a}ndle de Motta, B. Duret, T. M{\'e}nard, A. Poux who are gratefully acknowledged. We are very grateful to P. Bousquet-Melou from CRIANN for helping us optimising our GPU-processing code and R. Zamansky for providing us a first version of the forcing routine. This work has benefited from discussions with C. Federrath, R. Zamansky, A. Burlot, V. Boniou, J-L. Pierson, M. Crialesi-Esposito. Calculations were performed using the computing resources of CRIANN (Normandy, France), under the project 2018002.}

\backsection[Funding]{This research received funding from the french research national agency (ANR) under the contract ANR-22-CE51-0005.}

\backsection[Declaration of interests]{The authors report no conflict of interest.}

\backsection[Data availability statement]{The data and the code can be made available upon reasonable request.}

\backsection[Author ORCIDs]{F. Thiesset, https://orcid.org/0000-0001-7303-5106}

\backsection[Author contributions]{Both authors contributed equally to all aspects of the present work.}

\appendix

\section{Detailed derivation of the scale-by-scale budgets} \label{app:derivation}

\subsection{Unaveraged equation}

The mathematical machinery for deriving the different terms of the K{\'a}rm{\'a}n-Howarth-Monin for incompressible two-phase flows is same as already presented elsewhere by e.g. \cite{Galtier2011,Lai2018,Hellinger2021a}, initially developed in the context of either compressible flows and/or variable density miscible mixtures. The reader is referred to these papers for more details on the algebra. Here, we followed the exact same method as \cite{Hellinger2021a} but apply it to the one-fluid formulation to the two-phase incompressible NS-equation.

We start by the convective term, noted $\mathcal{T}$, which writes:
\begin{eqnarray}
  \mathcal{T} &:=& -(\delta \vect{u}) \cdot (\delta (\vect{\nabla} \cdot \rho \vect{u} \vect{u}) ) - (\delta ( \rho \vect{u})) \cdot (\delta ((\vect{u} \cdot \vect{\nabla}) \vect{u})) \nonumber \\
  &=& -(\delta \vect{u}) \cdot (\delta (\vect{u} \cdot ( \vect{\nabla} \rho \vect{u})))  - (\delta ( \rho \vect{u})) \cdot (\delta ((\vect{u} \cdot \vect{\nabla}) \vect{u})), \label{eq:Tgen}
\end{eqnarray}
where we have used the incompressibility condition $\vect{\nabla} \cdot \vect{u} = 0$. It is then useful to define the derivatives w.r.t the separation $\vect{r}$ and the mid-point $\vect{X} := (\vect{x^+} + \vect{x}^-)/2$ \citep[see e.g.][]{Hill2002,Danaila2012}.
\begin{subequations}
  \begin{eqnarray}
    &\vect{\nabla_r} = \frac{1}{2} &\left(\vect{\nabla^+} -\vect{\nabla^-} \right), \\
    &\vect{\nabla_X} = &\left(\vect{\nabla^+} +\vect{\nabla^-}\right).
  \end{eqnarray}
\end{subequations}
For any quantity $\bullet$, we also have $\vect{\nabla^+} \bullet^- = \vect{\nabla^-} \bullet^+ = 0$. After some manipulations, Eq. \eqref{eq:Tgen} can be rewritten as \citep[see e.g.][]{Hellinger2021a}
\begin{eqnarray}
  \mathcal{T} = - \vect{\nabla_X} \cdot (\overline{\delta} \vect{u}) |\delta \vect{u}|^2_\rho - \vect{\nabla_r} \cdot (\delta \vect{u}) |\delta \vect{u}|^2_\rho, \label{eq:T}
\end{eqnarray}
where $\overline{\delta} \bullet := (\bullet^+ + \bullet^-)/2$ is the arithmetic mean of any quantity $\bullet$ between point $\vect{x^+}$ and $\vect{x^-}$. Equation  \eqref{eq:T} reveals that the two-point convective term writes as the sum of two divergence terms, with respect to the geometrical space $\vect{X}$ and scale space $\vect{r}$, respectively.

For the other terms, we use the ruse of \cite{Hellinger2021a} who noticed that for any quantity $\vect{a}$, each term of the KHM equation has a generic form $\mathcal{D}(\vect{a})$ in the form
\begin{eqnarray}
  \mathcal{D}(\vect{a}) &:=&  (\delta \vect{u}) \cdot (\delta \vect{a}) + (\delta (\rho \vect{u})) \cdot (\delta v\vect{a}) \nonumber \\
  &=& 2(\delta \vect{u}) \cdot (\delta \vect{a}) - \mathcal{C}(\vect{a}), \label{eq:Dab}
\end{eqnarray}
where 
\begin{eqnarray}
  \mathcal{C}(\vect{a}) &:=& (\delta \vect{u}) \cdot (\delta \vect{a}) - (\delta (\rho \vect{u})) \cdot (\delta v\vect{a}) \nonumber \\
  &=& (\rho^+ v^- - 1) \vect{u}^+ \cdot \vect{a}^- + (\rho^- v^+ - 1) \vect{u}^- \cdot \vect{a}^+, \label{eq:Cab}
\end{eqnarray}
is a correction term which accounts explicitly for the variation of density. It is indeed easy to show that in constant density flows, we have $\rho^+ v^- = \rho^- v^+ = 1$ and hence $\mathcal{C}(\vect{a}) = 0$ .

For the term due to the pressure gradient, which is noted $\mathcal{P}$, we set $\vect{a} = - \vect{\nabla} P$ in eq. \eqref{eq:Dab}, and obtain
\begin{eqnarray}
  \mathcal{P} &:=& (\delta \vect{u}) \cdot (\delta (-\vect{\nabla} P) ) + (\delta ( \rho \vect{u})) \cdot (\delta (-v \vect{\nabla} P) \nonumber \\
  &=&  2(\delta \vect{u}) \cdot (\delta (-\vect{\nabla} P) ) - \mathcal{C}(-\vect{\nabla}P). \label{eq:Pgen}
\end{eqnarray}
The first term on RHS of Eq. \eqref{eq:Pgen} can be written as
\begin{eqnarray}
  -2(\delta \vect{u}) \cdot (\delta (\vect{\nabla} P) ) = - 2 \vect{\nabla_X} \cdot (\delta \vect{u}) (\delta P), \label{eq:Pdec}
\end{eqnarray}
where again we have used $\vect{\nabla} \cdot \vect{u}=0$. The term on RHS of Eq. \eqref{eq:Pdec} is generally referred to as the pressure transport. Using Eq. \eqref{eq:Pdec}, the pressure term (Eq. \ref{eq:Pgen}) can finally be recast in the form 
\begin{eqnarray}
  \mathcal{P} = - 2 \vect{\nabla_X} \cdot (\delta \vect{u}) (\delta P) - \mathcal{C}(-\vect{\nabla}P).
\end{eqnarray}
Proceeding similarly for the surface tension term noted $\mathcal{S}$, the viscous term noted $\mathcal{V}$, and the forcing term $\mathcal{F}$, yields
\begin{subequations}
  \begin{eqnarray}
    \mathcal{V} &:=& 2 (\delta \vect{u}) \cdot (\delta (\vect{\nabla} \cdot \mathsfbi{t})) - \mathcal{C}(\vect{\nabla} \cdot \mathsfbi{t}) \\
    \mathcal{S} &:=& 2 (\delta \vect{u}) \cdot (\delta \vect{S}) - \mathcal{C}(\vect{S}) \\
    \mathcal{F} &:=& 2 (\delta \vect{u}) \cdot (\delta \vect{F}) - \mathcal{C}(\vect{F}).
  \end{eqnarray} \label{eq:VST}
\end{subequations}
These terms can hardly be recast in a more compact or illustrative notation and are thus kept in this form. Note however that invoking additional hypothesis could lead to further simplifications. 
% For instance, one may assume a constant dynamic viscosity to end up with \citep{Hellinger2021a} 
% \begin{eqnarray}
%   \mathcal{V} = - 2 (\delta (\vect{\nabla}\vect{u})) : (\delta \mathsfbi{t}) - C(\vect{u},\vect{\nabla} \cdot \mathsfbi{t}),
% \end{eqnarray}
% where $\mathsfbi{a}:\mathsfbi{b} = a_{ij}b_{ij}$ (summation applies to repeated indices) denote the double contraction of second order tensors.
For instance, if a linear forcing is used, i.e. $\vect{F} := A \rho \vect{u}$ \citep{Lundgren2003,Rosales2005}, with a forcing amplitude $A$ that does not depend on space, one ends up with
\begin{eqnarray}
  \mathcal{F} = 2 A |\delta \vect{u}|^2_\rho
\end{eqnarray}
Lumping together the general formulation of the different terms leads to
\begin{eqnarray}
  \mathcal{D}_t &=& \mathcal{T} + \mathcal{P} + \mathcal{V} + \mathcal{S} + \mathcal{F} \nonumber \\
  \partial_t |\delta \vect{u}|^2_\rho &=&  - \vect{\nabla_X} \cdot (\overline{\delta} \vect{u}) |\delta \vect{u}|^2_\rho - \vect{\nabla_r} \cdot (\delta \vect{u}) |\delta \vect{u}|^2_\rho  \nonumber \\
  && - 2 \vect{\nabla_X} \cdot (\delta \vect{u}) (\delta P) - \mathcal{C}(-\vect{\nabla}P) \nonumber \\
  && + 2 (\delta \vect{u}) \cdot (\delta (\vect{\nabla} \cdot \mathsfbi{t})) - \mathcal{C}(\vect{\nabla} \cdot \mathsfbi{t}) \nonumber \\
  && + 2 (\delta \vect{u}) \cdot (\delta \vect{S}) - \mathcal{C}(\vect{S}) \nonumber \\
  && + 2 (\delta \vect{u}) \cdot (\delta \vect{F}) - \mathcal{C}(\vect{F}) \label{eq:KHM_unav}
\end{eqnarray}
Equation \eqref{eq:KHM_unav} is the general unaveraged KHM equation applying to incompressible two-phase flows. In this formulation, Eq. \eqref{eq:KHM_unav} reveals that the effect of density contrasts are accounted for through the different correction terms $\mathcal{C}(\vect{a})$. These terms will be zero when the two-phases are assumed to have same density, which is a reasonable approximation in case of emulsions \citep[e.g.][]{CrialesiEsposito2022}. The final key distinction between Eq. \eqref{eq:KHM_unav} and its single-phase variable-density analogue \citep[see e.g.][]{Galtier2011,Lai2018,Hellinger2021a} lies in an additional term arising from surface tension, inherited from the surface tension contribution in the one-fluid formulation of the two-phase incompressible Navier--Stokes equations.

\subsection{Averaged equations}

Eq. \eqref{eq:KHM_unav} should be supplemented by some averaging operations that depend on the flow situations. Here, the flow under consideration is stationary and homogeneous, thereby allowing to employ both temporal (over time $t$) and spatial (over geometrical positions in flow $\vect{X}$) averages. We will also employ angular averages over all orientations of the separation vector $\vect{r}$. These averages will be denoted by $\langle \bullet \rangle_\mathbb{T}$. In this situation, Eq. \eqref{eq:KHM_unav} simplifies to:
\begin{eqnarray}
  \underbrace{\langle \partial_t |\delta \vect{u}|^2_\rho \rangle_\mathbb{T}}_{{\rm Time~deriv}~\mathcal{D}_t} &=& \underbrace{- \langle \vect{\nabla_r} \cdot (\delta \vect{u}) |\delta \vect{u}|^2_\rho \rangle_\mathbb{T}}_{{\rm Transport}~\mathcal{T}} ~ \underbrace{- \langle \mathcal{C}(-\vect{\nabla}P) \rangle_\mathbb{T}}_{{\rm Pressure}~\mathcal{P}} \nonumber \\
  && \underbrace{+ 2\langle (\delta \vect{u}) \cdot (\delta (\vect{\nabla} \cdot 2\mu \mathsfbi{S}))\rangle_\mathbb{T} - \langle \mathcal{C}(\vect{\nabla} \cdot 2\mu \mathsfbi{S})\rangle_\mathbb{T}}_{{\rm Viscous}~\mathcal{V}} \nonumber \\
  && \underbrace{+ 2\langle (\delta \vect{u}) \cdot (\delta \vect{S})\rangle_\mathbb{T} - \langle \mathcal{C}(\vect{S})\rangle_\mathbb{T}}_{{\rm Surface~tension}~\mathcal{S}} \nonumber \\
  &&~ \underbrace{+ 2\langle (\delta \vect{u}) \cdot (\delta \vect{F})\rangle_\mathbb{T} - \langle \mathcal{C}(\vect{F})\rangle_\mathbb{T}}_{{\rm Forcing}~\mathcal{F}}. \label{eq:KHM_av_T}
\end{eqnarray}
For the sake of completeness, we have kept the time derivative term in Eq. \eqref{eq:KHM_av_T}. It will however be zero in the sequel because the flow is at statistically steady state and the averaging volume has fixed, non-moving boundaries. The spatially averaged $\vect{X}$-divergence terms is not included in Eq. \eqref{eq:KHM_av_T} as it vanishes due to periodicity \citep{Hill2002}. The first term on RHS of Eq. \eqref{eq:KHM_av_T} is a transfer term and writes as the divergence in scale-space of the flux $(\delta \vect{u}) |\delta \vect{u}|^2_\rho$. As for its single-phase counterpart, this term represents the cascade process of turbulent kinetic energy across the different scales. The other terms represent the scale-by-scale contribution of pressure, viscous stress, surface tension and forcing. 

In two-phase flows, it is also convenient to define conditional averages where statistics are gathered only within a given phase \citep{Dodd2016,Rosti2019,CrialesiEsposito2022,TrefftzPosada2023}. When applied to two-point statistics, one has to consider three different situations \citep{Yao2020,Yao2023}: 
\begin{itemize}
  \item the two points $\vect{x}^+$ and $\vect{x}^-$ lie in the liquid phase, such conditional averages will be denoted with the subscript $\mathbb{L}$ for “liquid”.
  \item the two points $\vect{x}^+$ and $\vect{x}^-$ lie in the gas phase, such conditional averages will be denoted with the subscript $\mathbb{G}$ for “gas”.
  \item the two points $\vect{x}^+$ and $\vect{x}^-$ lie in different phases, such conditional averages will be denoted with the subscript $\mathbb{M}$ for “mixed”.
\end{itemize}
For computing such conditional averages \citep{Yao2020,Gauding2021,Yao2023} one needs to define a phase indicator function which for the liquid phase can be defined by
\begin{eqnarray}
  \phi_L(\vect{x},t):=\begin{cases}
    1 ~~ {\rm ~if~} \vect{x} {\rm ~lies~in ~the ~liquid ~phase}  \\
    0 ~~ {\rm elsewhere},
 \end{cases} 
\end{eqnarray} \label{eq:phiL}
while for the gas phase,
\begin{eqnarray}
  \phi_G(\vect{x},t):=\begin{cases}
    1 ~~ {\rm ~if~} \vect{x} {\rm ~lies~in ~the ~gas ~phase}  \\
    0 ~~ {\rm elsewhere}. 
 \end{cases}
\end{eqnarray}

Then, for any two-point quantity $\delta\bullet$ one may define the conditional averages as \citep{Yao2020,Gauding2021,Yao2023}
\begin{subequations}
\begin{eqnarray}
  &\langle \delta \bullet \rangle_\mathbb{L} &= \frac{\langle \phi_L^+ \phi_L^- \delta \bullet \rangle_\mathbb{T}}{\gamma_L} \\
  &\langle \delta \bullet \rangle_\mathbb{G} &= \frac{\langle \phi_G^+\phi_G^-\delta \bullet \rangle_\mathbb{T}}{\gamma_G} \\
  &\langle \delta\bullet \rangle_\mathbb{M} &= \frac{\langle \phi_L^+ \phi_G^- \delta\bullet \rangle_\mathbb{T}}{\gamma_{M}},
\end{eqnarray} \label{eq:conditional_average}
\end{subequations}
where,
\begin{subequations}
  \begin{eqnarray}
    \gamma_L &:=& \langle \phi_L^+ \phi_L^-\rangle_\mathbb{T} \\
    \gamma_G &:=& \langle \phi_G^+ \phi_G^-\rangle_\mathbb{T} \\
    \gamma_M &:=& \langle \phi_L^+ \phi_G^-\rangle_\mathbb{T}.
  \end{eqnarray} \label{eq:gamma1}
\end{subequations}
We thus have $\gamma_L + \gamma_G + 2 \gamma_M=1$, and for any two-point statistics $\delta \bullet$ \citep{Yao2023}
\begin{eqnarray}
  \langle \delta\bullet \rangle_\mathbb{T} = \gamma_L \langle \delta\bullet \rangle_\mathbb{L} + \gamma_G \langle \delta\bullet \rangle_\mathbb{G} + 2\gamma_M \langle \delta\bullet \rangle_\mathbb{M}. \label{eq:gamma2}
\end{eqnarray}
Therefore, the conditionally averaged KHM equation is simply obtained by pre-multiplying each term by a correlation of the phase indicator function before averaging over all space, time, and orientations. Any two-point statistics averaged over the whole domain can be retrieved from the sum of the conditional ones weighted by their corresponding phase indicator correlation function $\gamma$ \citep{Yao2023}. These correlations are linked together by \citep[see e.g.][p 27-28]{Torquato2002} 
\begin{subequations}
\begin{eqnarray}
  \gamma_G &=& 1-2\langle \phi_L \rangle_\mathbb{T} + \gamma_L \\
  \gamma_L &=& 1-2\langle \phi_G \rangle_\mathbb{T} +\gamma_G \\
  \gamma_M &=& \langle \phi_L \rangle_\mathbb{T} - \gamma_L = \langle \phi_G \rangle_\mathbb{T} - \gamma_G 
\end{eqnarray} \label{eq:gamma3}
\end{subequations}
Although rather far from the scope of the present work, it is worth noting the correlation functions of the phase indicator field contains interesting information about the morphology of the liquid and gas phase interface. They can notably be used to infer the interface surface area, its mean and Gaussian curvatures, together with some fractal characteristics, if any \citep{Gauding2022,Thiesset2023,Thiesset2020,Thiesset2021}.

Applying conditional averages has some consequences in the final form of the KHM equation. For instance, the $\vect{X}$-divergence terms do not vanish {\it a priori}, even when averaged over a periodic domain and/or homogeneous flow. The time derivative term should also be retained in the final formulation of the conditionally averaged KHM equation since the conditional averaging volume has non-fixed boundaries. Indeed, let us define the set $\mathbb{C}(\vect{r}) := \{\vect{X}: \phi_C^+ \phi_{C'}^- =1\}$, where $C \in \{L,G\}$ and $C' \in \{L,G\}$. The conditional volume average of the time derivative term can then be rewritten as
\begin{eqnarray}
  \langle \partial_t |\delta \vect{u}|^2_\rho \rangle_\mathbb{C} = \frac{1}{\gamma_C} \int_\mathbb{T} \phi_C^+ \phi_C^- ~ \partial_t |\delta \vect{u}|^2_\rho \textrm{d} \vect{X} = \frac{1}{\gamma_C}  \int_{\mathbb{C}} \partial_t |\delta \vect{u}|^2_\rho \textrm{d} \vect{X}. \nonumber 
\end{eqnarray}
Then, by virtue of the Reynolds transport theorem, 
\begin{eqnarray}
  \langle \partial_t |\delta \vect{u}|^2_\rho \rangle_\mathbb{C} = \frac{1}{\gamma_C} \frac{d}{dt} \int_{\mathbb{C}} |\delta \vect{u}|^2_\rho \textrm{d} \vect{X} - \frac{1}{\gamma_C} \int_{\partial \mathbb{C}} |\delta \vect{u}|^2_\rho ~\vect{u}_b \cdot \vect{n} \textrm{d}S, \label{eq:RTT}
\end{eqnarray}
where $\partial \mathbb{C}$ is the boundary of the control volume $\mathbb{C}$, $\vect{n}$ is the outwardly oriented normal vector to $\partial \mathbb{C}$ and $\vect{u}_b$ the velocity of the control volume boundary in the laboratory frame of reference. For immiscible fluids, the velocity of the control surface, $\vect{u}_b$, is the fluid velocity, $\vect{u}$, at the control surface $\partial \mathbb{C}$. While the leftmost term on RHS of Eq. \eqref{eq:RTT} is zero when the flow is at statistically steady state, the rightmost term cannot however be dropped as it represents the surface averaged flux of $|\delta \vect{u}|^2_\rho$ across the control surface. Consequently, even in statistically steady flows, the conditionally averaged time derivative of the quantity $|\delta \vect{u}|^2_\rho$ term cannot be neglected. Since the rightmost term of Eq. \eqref{eq:RTT} writes as a convective flux across the averaging volume boundary, it is worth being added to the convective term. This is what has been done in Eq. \eqref{eq:KHM_symbolic} to form the “transport” term $ \langle \mathcal{T} \rangle_\mathbb{C}$.

Since the conditional averages are performed in the bulk of each phase which does not comprise the interface, the contribution of the surface tension term in the conditionally averaged two-point kinetic energy budget is zero. The same applied to the one-point kinetic energy budget \citep{Dodd2016}. 

In summary, the general conditionally averaged KHM equation reads
\begin{eqnarray}
  \underbrace{\langle \partial_t |\delta \vect{u}|^2_\rho \rangle_\mathbb{C}}_{{\rm Time~deriv.} ~ \mathcal{D}_t} &=& \underbrace{- \langle \vect{\nabla_r} \cdot (\delta \vect{u}) |\delta \vect{u}|^2_\rho \rangle_\mathbb{C} - \langle \vect{\nabla_X} \cdot (\overline{\delta} \vect{u}) |\delta \vect{u}|^2_\rho \rangle_\mathbb{C}}_{{\rm Transport}~\mathcal{T}}  \nonumber \\
  && \underbrace{-2 \langle \vect{\nabla_X} \cdot (\delta \vect{u}) (\delta P) \rangle_\mathbb{C} - \langle \mathcal{C}(-\vect{\nabla}P) \rangle_\mathbb{C}}_{{\rm Pressure}~\mathcal{P}} \nonumber \\
  && \underbrace{+ 2 \langle (\delta \vect{u}) \cdot (\delta (\vect{\nabla} \cdot 2\mu\mathsfbi{S}))\rangle_\mathbb{C} - \langle \mathcal{C}(\vect{\nabla}\cdot 2\mu \mathsfbi{S})\rangle_\mathbb{C}}_{{\rm Viscous}~\mathcal{V}} \nonumber \\
  && \underbrace{+ 2\langle (\delta \vect{u}) \cdot (\delta \vect{F})\rangle_\mathbb{C} - \langle \mathcal{C}(\vect{F})\rangle_\mathbb{C} }_{{\rm Forcing}~\mathcal{F}} \label{eq:KHM_av_LGLG},
\end{eqnarray}
where the subscript $\mathbb{C} \in \{\mathbb{L}, \mathbb{G}, \mathbb{M}\}$. 

Since the density is constant in each phase separately (and equals either $\rho_L$ or $\rho_G$), all variable-density correction terms $\mathcal{C}(\vect{a})$ should be zero when conditionally averaged within either the liquid or gas phase. Eq. \eqref{eq:KHM_av_LGLG} can then be simplified in the special case $\mathbb{C} \in \{\mathbb{L}, \mathbb{G}\}$, viz.
\begin{eqnarray}
  \underbrace{\langle \partial_t |\delta \vect{u}|^2_\rho \rangle_\mathbb{C}}_{{\rm Time~deriv.}~\mathcal{D}_t} &=& \underbrace{- \langle \vect{\nabla_r} \cdot (\delta \vect{u}) |\delta \vect{u}|^2_\rho \rangle_\mathbb{C} - \langle \vect{\nabla_X} \cdot (\overline{\delta} \vect{u}) |\delta \vect{u}|^2_\rho \rangle_\mathbb{C}}_{{\rm Transport}~\mathcal{T}} ~ \underbrace{-2 \langle \vect{\nabla_X} \cdot (\delta \vect{u}) (\delta P) \rangle_\mathbb{C}}_{{\rm Pressure}~\mathcal{P}} ~ \nonumber \\
  && \underbrace{+ 2 \langle(\delta \vect{u}) \cdot (\delta (\vect{\nabla }\cdot 2 \mu \mathsfbi{S}))\rangle_\mathbb{C} }_{{\rm Viscous}~\mathcal{V}} ~ \underbrace{+ 2 \langle (\delta \vect{u}) \cdot (\delta \vect{F})\rangle_\mathbb{C}}_{{\rm Forcing}~\mathcal{F}} \label{eq:KHM_av_LG}.
\end{eqnarray}
Equation \eqref{eq:KHM_av_LG} applies true only if $\mathbb{C} \in \{\mathbb{L},\mathbb{G}\}$. It is also valid for $\mathbb{C}=\mathbb{M}$ only if $\rho_L = \rho_G$. In Eq. \eqref{eq:KHM_av_LG}, since the density is constant in each phase, it can be dropped from the average, and one recovers the classical single-phase constant-density KHM equation.

\subsection{Different definitions of the two-point kinetic energy} \label{app:dq2_def}

In variable density flows, the definition of the two-point kinetic energy is not unique \citep{Aluie2013,Narula2025}. In the present work, we have used Eq. \eqref{eq:dq2} following \cite{Galtier2011,Hellinger2021a,Lai2018}. Different definitions can however be found in the literature. For instance, \cite{Hellinger2021} define the kinetic energy as $|\delta \vect{w}|^2$ where $\vect{w} = \rho^{1/2}\vect{u}$ as in \cite{Kida1990}. \cite{Hellinger2021} found that the KHM equation for this definition of the kinetic energy is virtually the same as Eq. \eqref{eq:KHM_av_T}. There appears a term analogous to the $\mathcal{C}$-term for the pressure which now writes
\begin{eqnarray}
    \mathcal{C}_{\sqrt{\rho}}(\vect{\nabla}P) = ((\rho^+ v^-)^{1/2} - 1) \vect{u}^+ \cdot (\vect{\nabla}P)^- + ((\rho^- v^+)^{1/2} - 1) \vect{u}^- \cdot (\vect{\nabla}P)^+. \label{eq:Cabrho}
\end{eqnarray}
The latter can also be interpreted as the contribution of the baropycnal work. Similar density variation correction terms also arise from the viscous and forcing terms of the Navier-Stokes equation \citep{Hellinger2021}. More generally, one could use any $0\leq\beta\leq1$ such that $\delta (\rho^\beta \vect{u}) \cdot \delta (\rho^{1-\beta} \vect{u})$ is a possible definition for the scale-by-scale turbulent kinetic energy \citep{Aluie2013}. Based on this definition for the scale-by-scale kinetic energy, the same kind of terms due to density variations are likely to operate in the KHM equation.

Another choice is made by \cite{Ferrand2020} and \cite{Brahami2020} who define the two-point kinetic energy as $\overline{\delta}\rho |\delta \vect{u}|^2$. In this case, \cite{Brahami2020} derived an additional term due to density variations in the KHM equation (see their Eq. 2.39),
% \begin{eqnarray}
%     (\delta \vect{u}) \cdot \left[ \left(1-\rho^+v^- \right) (\vect{\nabla}P)^- + \left(1-\rho^-v^+ \right) (\vect{\nabla}P)^+\right],
% \end{eqnarray}
which can also be associated to the effect of the baropycnal work. Some additional contributions due to density variations emerge from the viscous (and the forcing) term of the Navier-Stokes equations.

{Because in multiphase flows, the density is piecewise constant, one could also have used the single-phase definition for the structure function, i.e. $\langle |\delta \vect{u}|^2\rangle_\mathbb{C}$, i.e. without any sort of density weighted increments. The latter could therefore be interpreted as a Reynolds-averaged version for the velocity fluctuations. This is indeed a possible definition. Although it should better be referred to as the “agitation” rather than the kinetic energy, it may solve the artefact associated with the Favre average (see \S\ref{sec:discussion}). Note that this definition would lead to the exact same KHM equation as Eq. \eqref{eq:KHM_symbolic} for the case $\mathbb{C} \in \mathbb{\{L,G\}}$ since the density is constant per phase and could thus be dropped from the structure function.}

In summary, a consensus on the optimal definition of two-point kinetic energy in variable density flows remains elusive. \cite{Narula2025} explored various definitions for scale-by-scale turbulent kinetic energy within the coarse-grained framework. Similar research is required to determine the most suitable definition when a point-splitting method is used.

\subsection{Limit at large separations} \label{app:limit}

It is worth noting that for homogeneous flows, the scale-by-scale kinetic energy \citep{Lai2018},
\begin{eqnarray}
  \langle |\delta \vect{u}|^2_\rho \rangle_\mathbb{T} = 2 \langle \rho |\vect{u}|^2\rangle_\mathbb{T} - 4\langle (\overline{\delta} \rho) \vect{u}^+ \cdot \vect{u}^-\rangle_\mathbb{T} = 4 \langle k \rangle_\mathbb{T} - 4\langle (\overline{\delta} \rho) \vect{u}^+ \cdot \vect{u}^-\rangle_\mathbb{T},
\end{eqnarray}
where the turbulent kinetic energy is defined as $k = \frac{1}{2}\rho |\vect{u}|^2$. Since, in the limit $|\vect{r}| \to \infty$, the correlation $\langle (\overline{\delta} \rho) \vect{u}^+ \cdot \vect{u}^-\rangle_\mathbb{T} \to 0$, the time derivative term in Eq. \eqref{eq:KHM_av_T} has the following limit
\begin{eqnarray}
  \lim_{r\to \infty}  \langle \partial_t |\delta \vect{u}|^2_\rho \rangle_\mathbb{T} = 4 \frac{d}{dt} \langle k \rangle_\mathbb{T}.
\end{eqnarray}
Proceeding similarly for the other terms of the KHM equation, it can be shown that, in the limit of large separations, each term of the total KHM equation Eq. \eqref{eq:KHM_av_T} tends towards four times its counterpart in the one-point total kinetic energy budget. It is also easy to show that the variable-density corrections $\mathcal{C}(\vect{a}) \to 0 $ when $r \to \infty$. As an example, the two-point viscous term can be proven to asymptote the kinetic energy dissipation rate \citep{Hellinger2021a}, viz.
\begin{eqnarray}
  \lim_{r \to \infty} + 2 \langle (\delta \vect{u}) \cdot (\delta (\vect{\nabla} \cdot 2 \mu \mathsfbi{S}))\rangle_\mathbb{T} &-& \langle \mathcal{C}(\vect{\nabla}\cdot 2 \mu \mathsfbi{S})\rangle_\mathbb{T} \nonumber \\
  &=& 4 \langle \vect{u} \cdot (\vect{\nabla} \cdot 2 \mu \mathsfbi{S})\rangle_\mathbb{T} \nonumber - 0\\
  &=& 4 \langle\vect{\nabla} \cdot 2 \mu ( \vect{u}  \mathsfbi{S})\rangle_\mathbb{T}  - 4 \langle \vect{\nabla u} : 2 \mu \mathsfbi{S}\rangle_\mathbb{T} \label{eq:lim_V_LS},
\end{eqnarray}
By periodicity, the first term on RHS of Eq. \eqref{eq:lim_V_LS} cancels out, and remains only the kinetic energy dissipation rate which reads $\langle \epsilon \rangle_\mathbb{T} := \langle \vect{\nabla u} : 2 \mu \mathsfbi{S}\rangle_\mathbb{T} = \langle 2 \mu \mathsfbi{S} : \mathsfbi{S}\rangle_\mathbb{T}$ \citep[see e.g.][p 251]{Wilcox1998}. The other terms of the total kinetic energy transport equation can be retrieved similarly to finally obtain the one-point total kinetic energy budget, which for the present configuration writes:
\begin{eqnarray}
  \frac{d}{dt} \langle k \rangle_\mathbb{T}=\langle F \rangle_\mathbb{T}  -  \langle \epsilon \rangle_\mathbb{T} + \langle S \rangle_\mathbb{T},
\end{eqnarray} 
where $\langle F \rangle_\mathbb{T} := \langle \vect{u} \cdot \vect{F}\rangle_\mathbb{T}$ and $\langle S \rangle_\mathbb{T} := \langle \vect{u} \cdot \vect{S}\rangle_\mathbb{T}$ represent the contribution of forcing and surface tension to the kinetic energy budget, respectively. Note that, in statistically steady flows, the time derivative cancels out together with surface tension term. The latter can indeed be written as \citep{Dodd2016,TrefftzPosada2023}
\begin{eqnarray}
  \langle S \rangle_\mathbb{T} \sim -\sigma d_t A_\Gamma ,
\end{eqnarray}
which is zero since the surface area of the liquid/gas interface $A_\Gamma$ is on average constant. The other terms vanish because the flow under consideration is periodic.

The same reasoning can be applied to the different terms of the conditionally averaged KHM equation Eq. \eqref{eq:KHM_av_LG}. Here again, in the limit of large separations, each term of Eq. \eqref{eq:KHM_av_LG} tends towards four times their counterpart in the one-point conditionally averaged kinetic energy budget derived by \cite{Dodd2016}. For instance, the two-point viscous term conditionally averaged in either the liquid or gas phase has the following limit,
\begin{eqnarray}
  \lim_{r \to \infty} 2 \langle (\delta \vect{u}) \cdot (\delta (\vect{\nabla} \cdot 2 \mu \mathsfbi{S})) \rangle_\mathbb{C} & = &
  4 \langle \vect{u} \cdot (\vect{\nabla} \cdot 2 \mu \mathsfbi{S}) \rangle_\mathbb{C}\nonumber \\
  &=& - 8 \mu_C \langle \vect{\nabla u} : \mathsfbi{S} \rangle_\mathbb{C} + 8 \mu_C \langle \vect{\nabla} \cdot (\vect{u} \mathsfbi{S}) \rangle_\mathbb{C}
  \nonumber \\
  &=&  -8 \mu_C \langle \mathsfbi{S}:\mathsfbi{S} \rangle_\mathbb{C} \nonumber + 8 \mu_C \langle \vect{\nabla} \cdot (\vect{u} \mathsfbi{S}) \rangle_\mathbb{C} \\
  &:=& -4 \langle \epsilon \rangle_\mathbb{C} + 4\langle T_\nu \rangle_\mathbb{C},
\end{eqnarray}
where the kinetic energy dissipation rate in phase $\mathbb{C} \in \{\mathbb{L},\mathbb{G}\}$ is $\langle \epsilon \rangle_\mathbb{C} := 2 \mu_C \langle \mathsfbi{S}:\mathsfbi{S} \rangle_\mathbb{C}$ and the viscous transport is $\langle T_\nu \rangle_\mathbb{C} := 2 \mu_C  \langle \vect{\nabla} \cdot (\vect{u} \mathsfbi{S}) \rangle_\mathbb{C}$. The conditional one-point average of a quantity $\bullet$ is defined by:
\begin{eqnarray}
  \langle \bullet \rangle_\mathbb{C} := \frac{\langle \phi_C \bullet \rangle_\mathbb{T} }{ \langle \phi_C \rangle_\mathbb{T}}.
\end{eqnarray}  
Proceeding with the pressure gradient term in Eq. \eqref{eq:KHM_av_LG} yields
\begin{eqnarray}
  \lim_{r \to \infty} -2 \langle \vect{\nabla_X} \cdot (\delta \vect{u}) (\delta P) \rangle_\mathbb{C} = -4 \langle \vect{\nabla} \cdot \vect{u}P\rangle_\mathbb{C} := 4\langle T_p \rangle_\mathbb{C},
\end{eqnarray}
again with $\mathbb{C} \in \{\mathbb{L},\mathbb{G}\}$. These terms are constitutive of the one-point conditional kinetic energy budget \citep[see Eqs. (3.8) to (3.11) in][]{Dodd2016} and are recovered here from the KHM equation in the limit of large separations. In presence of a forcing term, the latter writes:
\begin{eqnarray}
  \frac{d}{dt} \langle k \rangle_\mathbb{C} = \langle  F\rangle_\mathbb{C} - \langle \epsilon \rangle_\mathbb{C} + \langle T_\nu \rangle_\mathbb{C} + \langle T_p \rangle_\mathbb{C},
\end{eqnarray}
For obtaining the above equation, it was necessary to note that,
\begin{eqnarray}
  \langle \partial_t k\rangle_\mathbb{C} + \langle \vect{\nabla} \cdot \vect{u} k \rangle_\mathbb{C} = \frac{d}{dt} \langle k \rangle_\mathbb{C} - \int_\Gamma k \vect{u}_b\cdot \vect{n} {\rm d}S + \int_\Gamma k \vect{u}\cdot \vect{n} {\rm d}S, 
\end{eqnarray}
where we have employed the Reynolds transport theorem for the time derivative term and the Green-Ostogradski theorem for the transport of kinetic energy. In case of immiscible fluids with no phase change, one has $\vect{u}_b = \vect{u}$ at the interface between the two phases, and thus the sum of the time derivative term and the transport term reduces to $d \langle k \rangle_\mathbb{C}/dt$. 

Let us now focus on the large-scales asymptotic behaviour when $\mathbb{C} = \mathbb{M}$. For this, it is first worth noting that for any one-point statistics $\bullet$, one has \citep{Dodd2016}
\begin{eqnarray}
  \langle \bullet \rangle_\mathbb{T} = \langle \phi_L \rangle_\mathbb{T} \langle \bullet \rangle_\mathbb{L} + \langle \phi_G \rangle_\mathbb{T} \langle \bullet \rangle_\mathbb{G}.
\end{eqnarray}
In addition, when the separation $r$ goes to infinity \citep{Torquato2002},
\begin{subequations}
  \begin{eqnarray}
    \lim_{r \to \infty} \gamma_L &=& \langle \phi_L \rangle_\mathbb{T}^2 \\
    \lim_{r \to \infty} \gamma_G &=& \langle \phi_G \rangle_\mathbb{T}^2 \\
    \lim_{r \to \infty} \gamma_{M} &=& \langle \phi_L \rangle_\mathbb{T}\langle \phi_G \rangle_\mathbb{T}
  \end{eqnarray}\label{eq:gamma4}
\end{subequations}
Then using Eq. \eqref{eq:KHM_av_LGLG} (with $\mathbb{C}=\mathbb{M}$), Eqs. \eqref{eq:gamma1}, \eqref{eq:gamma2}, \eqref{eq:gamma3}, \eqref{eq:gamma4}, and after some straightforward manipulations, it can be obtained that for e.g. the forcing term:
\begin{eqnarray}
  \lim_{r \to \infty} + 2\langle (\delta \vect{u}) \cdot (\delta \vect{F})\rangle_\mathbb{M} - \langle \mathcal{C}(\vect{F})\rangle_\mathbb{M} =  4\frac{\langle F \rangle_\mathbb{L} + \langle F \rangle_\mathbb{G}}{2}
\end{eqnarray} 
which is thus simply (four times) the average of $\langle F \rangle$ in the liquid and gas phase. The same applies to the other terms. 

\subsection{Limit at small separations} \label{app:limit_r0}

We are interested in the limit of the two-point kinetic energy at small separations. For this purpose, we use the following decomposition:
\begin{eqnarray}
  \langle (\delta \rho \vect{u}) \cdot (\delta \vect{u}) \rangle_\mathbb{C} = 
\underbrace{\langle \delta \rho ~ \vect{u}(\vect{x}+\vect{r}) \cdot \delta \vect{u}\rangle_\mathbb{C}}_{T_1} + \underbrace{\langle \rho(\vect{x}) |\delta \vect{u}|^2\rangle_\mathbb{C}}_{T_2}. \label{eq:dq2_decomp}
\end{eqnarray}
Let us start with term $T_2$. In the limit of small separations, we can use Taylor expansions to write
\begin{eqnarray}
  |\delta \vect{u}|^2 \sim {r^2} \left(\vect{\nabla u} : \vect{\nabla u}\right) + \mathcal{O}(r^3).
\end{eqnarray}
Decomposing the velocity gradient tensor into symmetric and antisymmetric components, it is then possible to obtain
\begin{eqnarray}
   T_2 = \frac{r^2}{3} \left\langle \rho \left(\mathsfbi{S}:\mathsfbi{S} + \frac{1}{2}|\vect{\omega}|^2\right)\right\rangle_\mathbb{C} + \mathcal{O}(r^3).
\end{eqnarray}
where $\vect \omega$ is the vorticity vector. The 1/3 coefficient is obtained under the assumption of statistical isotropy.

For term $T_1$, we start by recalling that the increment $\delta \rho$ is active only when the two points cross the interface, i.e. only when $\mathbb{C}=\mathbb{T}$. The density increment $\delta \rho$ is given by:
\begin{eqnarray}
  \delta \rho = (\rho_L - \rho_G) \delta_\Gamma(\vec{x}-\vec{x_s}) |\vec{r} \cdot \vec{n}(\vec{x_s})|.
\end{eqnarray}
Here, $\delta_\Gamma(\vec{x}-\vec{x_s}) |\vec{r} \cdot \vec{n}(\vec{x_s})|$ is the probability that vector $\vec{r}$ crosses the interface. This would lead to 
\begin{eqnarray}
  \langle \delta \rho \rangle = \frac{\Delta \rho}{V} \int_\Gamma |\vec{r} \cdot \vec{n}| dA = \Delta \rho \frac{A_\Gamma}{V} \frac{r}{2} + \mathcal{O}(r^3),
\end{eqnarray}
where $\Delta \rho = \rho_L - \rho_G$ and $V$ is the averaging volume for $\mathbb{C}=\mathbb{T}$. The factor $1/2$ comes from the isotropic assumption for the orientation of the surface. This result is a generalization in 3D-space of the Buffon's needle problem \citep{Torquato2002,Thiesset2021,Thiesset2023,Gauding2022}. Higher order terms would include the effect of mean and Gaussian curvature.

Decomposing further $T_1$ as:
\begin{eqnarray}
  T_1 = \left\langle \delta \rho \left( \vect{u}(\vect{x}) \cdot \delta \vect{u} +  |\delta \vect{u}|^2 \right) \right\rangle_\mathbb{C},
\end{eqnarray}
we realize that $ \delta \rho |\delta \vect{u}|^2 $ is third-order in $r$ since $\delta \rho \sim r$ while $|\delta \vect{u}|^2\sim r^2$. Therefore, only $\vect{u}(\vect{x}) \cdot \delta \vect{u}$ contributes to leading order. Taylor series expansion for $\vect{u}\cdot\delta \vect{u}$ yields:
\begin{eqnarray}
    T_1 = \Delta \rho \langle \delta(\vec{x}-\vec{x_s}) |\vec{r} \cdot \vec{n}(\vec{x_s})| \vec{u} \cdot (\vec{r} \cdot \vec{\nabla}) \vec{u} \rangle_\mathbb{T} + \mathcal{O}(r^3).
\end{eqnarray}
The term $\vec{u} \cdot (\vec{r} \cdot \vec{\nabla}) \vec{u}$ can further be rewritten as $\frac{1}{2}\vect{\nabla_r} |\vect{u}|^2$ and therefore, $T_1$ can be expressed as:
\begin{eqnarray}
  T_1 = \frac{\Delta \rho }{V} \frac{r^2}{2} \int_\Gamma |\vec{\hat{r}} \cdot \vect{n}| ~ \vec{\nabla}_{\vec{\hat{r}}} |\vec{u}|^2 dA + \mathcal{O}(r^3),
\end{eqnarray}
where $\vect{\hat{r}} = \vect{r}/r$. Further, it is reasonable (although not proven rigorously) to consider that $\vec{\nabla}_{\vec{\hat{r}}} |\vec{u}|^2$ is independent of $\vect{n}$. With this assumption, one gets:
\begin{eqnarray}
  T_1 = \frac{\Delta \rho}{V} \frac{r^2}{4} \int_\Gamma \vec{\nabla}_{\vec{\hat{r}}} |\vec{u}|^2 dA + \mathcal{O}(r^3).
\end{eqnarray}
Because $\int_\Gamma |\vec{\hat{r}} \cdot \vect{n}| dA = 1/2$. By use of isotropy, the gradient $\vec{\nabla}_{\vec{\hat{r}}} |\vec{u}|^2$ does not depend on the orientation of $\vect{\hat{r}}$. Therefore, there is no preferred direction of $|\vec{u}|^2$ with respect to $\vect{\hat{r}}$. In addition, in statistically homogeneous flows with no mean flow, the mean velocity and the mean gradients should be zero. Furthermore, for immiscible fluids, the velocity $\vect{u}$ is continuous across the interface so that there is no net interfacial exchanges between phases. These arguments constitute a sufficient body of evidence to state that $T_1$ vanishes. Pending a more rigorous proof, we will continue with $T_1 = 0$ for $\mathbb{C}\equiv\mathbb{T}$ as well.

In summary, the small scale limit of the mixed structure function writes:
\begin{eqnarray}
  \langle (\delta \rho \vect{u}) \cdot (\delta \vect{u}) \rangle_\mathbb{C} = \frac{r^2}{3} \left( \left\langle \frac{\rho \epsilon}{2\mu} \right\rangle_\mathbb{C} + \left\langle \frac{\rho |\vect{\omega}|^2}{2} \right\rangle_\mathbb{C} \right) + \mathcal{O}(r^3),
\end{eqnarray}
for any $\mathbb{C} \in \{\mathbb{L,G,T}\}$. Recall that $\epsilon/2\mu = \mathsfbi{S}:\mathsfbi{S}$. Note also that for $\mathbb{C} \in \{\mathbb{L,G}\}$, the fluid physical properties are constant per phase. Under the assumption of statistical homogeneity, one then has:
\begin{eqnarray}
   \left\langle \frac{\rho \epsilon}{\mu} \right\rangle_\mathbb{C} =  \langle \rho  |\vect{\omega}|^2\rangle_\mathbb{C}.
\end{eqnarray}
Because this relation holds true per phase, it should also apply to the total field by simple addition of the statistics in the liquid and gas phase. The small-scale limit of the two-point kinetic energy can thus be rewritten as:
\begin{eqnarray}
  \langle (\delta \rho \vect{u}) \cdot (\delta \vect{u}) \rangle_\mathbb{C} &=& \frac{r^2}{3} \left\langle \rho |\vect{\omega}|^2 \right\rangle_\mathbb{C}  + \mathcal{O}(r^3) \\
  &=& \frac{r^2}{3} \left\langle \frac{\rho \epsilon}{\mu} \right\rangle_\mathbb{C}  + \mathcal{O}(r^3),\label{eq:dq2_omega}
\end{eqnarray}
for any $\mathbb{C}\in\{\mathbb{L,G,T}\}$. Eq. \eqref{eq:dq2_omega} reveals that, in the limit of small separations, the unconditional and conditional two-point kinetic energy is quadratic in $r$ and is proportional to the enstrophy (or kinetic energy dissipation rate).

In single phase flows in absence of solid boundaries, the evolution equation for the enstrophy field is obtained from the Navier-Stokes equation. When averaged over a domain $\mathbb{C}$, one obtains 
\begin{align}
    {d_t} \frac{\langle |\vect{\omega}|^2\rangle_\mathbb{C}}{2} 
    &=& \langle \vect{\omega} \vect{\omega} : \vect{\nabla}\vect{u} \rangle_\mathbb{C}
    + \nu   \left\langle \vect{\nabla}^2 \frac{ |\vect{\omega}|^2}{2} -\vect{\nabla} \vect{\omega} : \vect{\nabla} \vect{\omega}  \right\rangle_\mathbb{C}
    + \langle \vect{\omega} \cdot (\vect{\nabla} \times \vect{F})  \rangle_\mathbb{C} \label{eq:omega2_av}
\end{align}
Equation \eqref{eq:omega2_av} reveals that the time variation of the averaged enstrophy depends on a production term due to vortex stretching (the first term on RHS of Eq \ref{eq:omega2_av}), on the enstrophy viscous diffusion and viscous dissipation (sometimes referred to as the palinstrophy) and on the effect of artificial forcing as represented by the rightmost term on RHS of Eq. \eqref{eq:omega2_av}. 

Given Eq. \eqref{eq:dq2_omega}, the limit at small scales of the KHM equation per phase should tend towards the transport equation for the enstrophy, Eq. \eqref{eq:omega2_av}. In single-phase flows, this was done for instance by \cite{Antonia2000}. Proceeding similarly here, the transport term in scale-space in the KHM equation (Eq. \ref{eq:KHM_av_LG}) tends towards the production by vortex stretching in Eq. \eqref{eq:omega2_av}, the viscous term in Eq. \eqref{eq:KHM_av_LG} tends towards the viscous terms in Eq. \eqref{eq:omega2_av}, while the forcing term in \eqref{eq:omega2_av} is related to its counterpart in Eq. \eqref{eq:KHM_av_LG}.

For the case $\mathbb{C}=\mathbb{T}$, the dynamics of vorticity presents a considerably higher level of complexity. \cite{Terrington2022} showed that conservation laws for the volume integrated vorticity reveal some additional source terms which come from surface tension, viscous stresses and pressure gradients. This was achieved through sophisticated algebraic manipulations, which we have not yet been able to leverage to derive an evolution equation for the enstrophy in multiphase systems. Deriving an equation for the enstrophy starting from the KHM equation in the limit of small separations is not an easy task either. Indeed, as for $\delta (\rho \vect{u})$, some terms are discontinuous at the interface (e.g. the pressure and viscous terms), which require the same treatment as term $T_1$ above. Handling the derivation for all these terms is left for future analysis.

\section{Numerical tools and post-processing procedures} \label{app:dns}

\subsection{Time advancement algorithm} \label{app:fastRK3}

We here describe the fastRK3 algorithm used for the time advancement of the numerical solutions. The first step of the algorithm is to transport the coupled level-set and VOF fields at time $t+\Delta t$ and compute the related density $\rho_{n+1}$ and viscosity $\mu_{n+1}$ at time $t+\Delta t$. For ensuring consistency between the transport of VOF and momentum, the convective term in Eq. \eqref{eq:nsa} is also computed.

Let us now note $\vect{R}$ the right-hand side of Eq. \eqref{eq:nsb}. Formally, the fastRK3 algorithm is similar to the classical RK3 method. It consists in advancing from the velocity at time $t$ noted $\vect{u}_n$ to the velocity at time $t+\Delta t$ noted $\vect{u}_{n+1}$, by calculating three intermediate velocity fields, viz
\begin{subequations}
  \begin{eqnarray}
    \vect{u}_1^* &=& \vect{u}_n + \frac{\Delta t}{3} \vect{R}(\vect{u}_n) \\
    \vect{u}_1 &=& \vect{u}_1^* - \frac{\Delta t}{3} \frac{\vect{\nabla}P_1}{\rho_{n+1}} \\
    \vect{u}_2^* &=& \vect{u}_n + \Delta t \left[ -\vect{R}\left(\vect{u}_n\right) + 2 \vect{R} \left(\vect{u}_1 \right)\right] \\
    \vect{u}_2 &=& \vect{u}_2^* - \Delta t \frac{\vect{\nabla}P_2}{\rho_{n+1}} \\
    \vect{u}_3^* &=& \vect{u}_n + \Delta t \left[\frac{3}{4} \vect{R} \left(\vect{u}_1 \right) + \frac{1}{4} \vect{R} \left(\vect{u}_2 \right) \right] \\
    \vect{u}_{n+1} &=& \vect{u}_3 = \vect{u}_3^* - \Delta t  \frac{\vect{\nabla}P_3}{\rho_{n+1}}
  \end{eqnarray}
\end{subequations}
In the classical RK3 method, the intermediate pressure fields $P_i \in \{P_1, P_2, P_3\}$ are obtained by solving the Poisson equation:
\begin{eqnarray}
  \frac{\nabla^2 P_i}{\rho_{n+1}} = \vect{\nabla} \cdot \vect{u}_i^*
\end{eqnarray}
at all 3 steps of the algorithm. Instead, the fastRK3 method expresses the intermediate pressures $P_1$ and $P_2$ as extrapolation of the pressure at time $t+\Delta t /3$ and $t+\Delta t$, respectively, using $P_{n}$ and $P_{n-1}$, viz.
\begin{subequations}
  \begin{eqnarray}
    P_1 &=& \frac{5}{3}P_n - \frac{2}{3}P_{n-1} \\
    P_2 &=& 2P_n - P_{n-1}.
  \end{eqnarray}
\end{subequations}
This is referred to as the mid-point extrapolation in \cite{Aithal2023}. By doing this, only one Poisson equation is solved in order to compute the pressure at time $t+\Delta t$, noted $P_3 = P_{n+1}$, while 3 are required for the classical RK3 algorithm (one per sub-step of the Runge-Kutta scheme). Here, $P_3$ is obtained by using a multi-grid preconditioned Conjugate Gradient algorithm (MGCG) \citep{Zhang1996}. We do not use the decomposition proposed by \cite{Dodd2016,TrefftzPosada2023} in order to solve a Poisson equation with constant coefficients. Note that the fastRK3 is not self-starting. Hence, in order to get $P_n$ and $P_{n-1}$, the first 3 steps of time advancement are done using the classical RK3 scheme before switching to fastRK3 for the rest of the simulation. %In the present implementation, it was found that the fastRK3 scheme was computationally comparable to the explicit Euler scheme, while allowing for temporal accuracy between first and third-order depending on the physical parameters that are used. 

\subsection{Forcing procedure} \label{app:forcing}

We here provide the details of the forcing procedure used to maintain turbulence at steady state. The forcing acts only at large scales, in a spectral band $\left[ k_{\min} = 2\pi / L, k_{\max} = 2\pi / (L / 3) \right]$ where $L$ is the width of the simulation domain. We use the same forcing spectrum as \cite{Federrath2010}, noted $W(k)$, which corresponds to a paraboloid peaking around $k_f = (k_{\max}+k_{\min})/2$ , viz.
\begin{eqnarray}
  W(k) &=	\begin{cases}
    \frac{k_f}{k}\left[1 - K \left(k - k_f\right)^2\right]~~ {\rm where~} k_{\min}< k < k_{\max}  \\
    0 ~~ {\rm elsewhere},
          \end{cases} 
\end{eqnarray}
where $K = 4/(k_{\max} - k_{\min})^2$. In order to maintain a statistically steady state with a prescribed kinetic energy, the amplitude of the forcing is adjusted using a PID controller, where the amplification factor of the controller, noted $A$, is given by 
\begin{eqnarray}
  A = G \left(\Upsilon(t) + \frac{1}{\tau_i} \int_0^\tau \Upsilon(t) {\rm d}t + \tau_d \frac{d}{dt}\Upsilon(t)\right),
\end{eqnarray}
where $\Upsilon := \widetilde{k}(t) - \widetilde{k}_t$ is the error between the kinetic energy at time $t$, i.e. $\widetilde{k}(t) = \langle k \rangle_\mathbb{T}/ \langle \rho \rangle_\mathbb{T}$ and the target kinetic energy $\widetilde{k_t}$. The controller parameters were set to $G=200$, $\tau_i = 0.25 s$ and $\tau_d = 0.01 s$. The correlation time of the stochastic forcing is equal to $L_f/u'$, where $L_f = 2 \pi / k_f = L/2$ and $u'=(2 \widetilde{k_t})^{1/2}$.

This forcing was chosen because it has some interesting features which are detailed below. First, because it acts at large scales, {\it (i)} the intermediate and small scales are believed to behave more naturally, and {\it (ii)} the approach towards the asymptotic behaviour at infinite Reynolds number is faster than decaying turbulence or by using a linear forcing \citep{Antonia2006}. Further, because this forcing is based on a stochastic process, its contribution to the kinetic energy budget is expected {\it (iii)} to be independent of the chosen flow physical parameters and {\it (iv)} to be the same in the liquid and gas phase. This was confirmed later by analysing the scale-by-scale budgets. Thirdly, by the use of a PID controller, we minimize the time fluctuations of energy injection, thereby {\it (v)}  mitigating what is referred to as non-equilibrium effects \citep{Goto2015,Fang2023}. 

\subsection{Numerical resolution} \label{app:resolution}

To appraise the appropriateness of our numerical solutions, we have monitored two metrics. The first is the difference between the total time-averaged injection and dissipation of kinetic energy. Given Eq. \eqref{eq:1pt_budget}, injection and dissipation should be on average equal, except if numerical errors (numerical dissipation or dispersion) are at play. In case of insufficient resolutions we have noted that both the transport term and the surface tension term were negative, thereby contributing to the kinetic energy budget as non-physical numerical dissipation. Their contribution were though converging to zero when increasing the resolution. The second metric for inferring the appropriate resolution is the conservation of mass. In case of insufficient resolution, there could be some inconsistency between the geometrical properties obtained from the VOF field and the level-set field. In our solver, if such situations occur, the VOF is readjusted to match the geometrical properties of the level-set, at the price of small mass variations. 

In order to keep these metrics of the order of a percent, the viscosity $\nu$ and surface tension $\sigma$ were chosen so that the ratio of the Kolmogorov length-scale $\eta$ to the mesh size $dx$ is about 2 (see Table \ref{tab:dns_params}) and the Weber number is not more than 50. While half this resolution is likely sufficient for low density ratios (say $R_\rho \leq$ 5), we have found that such a fine resolution was necessary for larger density ratio. This results in significantly smaller Taylor scale Reynolds number $R_\lambda$ compared to what can be found in the literature using the same number of simulation points \citep{CrialesiEsposito2022,Cannon2024}. 

\subsection{Post-processing} \label{app:post_pro}

Statistics are computed using the post-processing library \texttt{PyArcher} which comes as a submodule of the \texttt{archer} code. Instead of computing the explicit form of the different terms of the KHM equation which are given in Appendix (Eqs. \eqref{eq:KHM_av_T}, \eqref{eq:KHM_av_LGLG} and Eqs. \eqref{eq:KHM_av_LG}), we use Eq. \eqref{eq:KHM_general} where all terms of the NS equation are extracted directly from the code. This was found to be the most computationally efficient and accurate. The two-point statistics are computed using homemade routines which work on GPU using OpenACC directives. 

In previous studies from the literature \citep{Dodd2016,CrialesiEsposito2022,TrefftzPosada2023}, it is not mentioned how conditional averages are computed. Here, the phase indicator is defined from the distance function used in the level-set method which we note $\Psi$. However, since the code is based on a staggered Cartesian grid, we have decided to interpolate the level-set field (a cell-centred quantity) at the same position as the velocity components, i.e. at the faces of each computational cell. Three different face-centred level-set fields are thus obtained, one for each velocity component. These are noted $\Psi_s^i$ for each direction $i$. The phase indicator function for the liquid phase can subsequently be defined as $\phi_L = 1$ where $\Psi_s^i > 0.5dx$ and for the gas phase, $\phi_G=1$ where $\Psi_s^i < -0.5dx$. This shift of $0.5dx$ either in the liquid or gas phase was imposed in order to exclude the cells where the interface (and hence the surface tension term) is present. 

The budget of the KHM equation is closed up to about $10^{-3} \varepsilon_f$. 
As a projection method is used to enforce the incompressibility condition, the resolution of the NS equation is bounded by the residual of the Poisson equation, which is of the order of $\mathcal{L}_2(\nabla\cdot\vect{u})\sim 10^{-9}$. This constitutes a small yet finite source of error.
The second source of error arises from the transport term in the NS and therefore the KHM equations:  in our code, we solve the transport term for $\rho \vect{u}$ using the Rudman scheme, which needs to be converted to get the transport term for  $\vect{u}$. This operation is likely to come with some error which remains complicated to evaluate. 
The last, and probably the most dominant, loss of precision can be rooted to the angular averaging operation which requires to interpolate the Cartesian  $(r_x, r_y, r_y)$ coordinates to the spherical coordinates $(r, \theta, \varphi)$ as described in \eqref{eq:angle_av}. More details on this operation can be found in \cite{Thiesset2020a}.

\bibliographystyle{jfm}
\bibliography{sbs_budget}

\end{document}